\documentclass[12pt,reqno]{amsart}
\usepackage{graphics,amsmath,amssymb,epsfig,color,amsaddr}
\usepackage{amsfonts,geometry,array}
\usepackage{graphicx}
\graphicspath{ {./images/} }
\usepackage{amsthm}
\usepackage{bm}
\usepackage{dsfont}
\usepackage{mathabx} 
\usepackage{flafter}
\usepackage[svgnames]{xcolor}
\usepackage{bbm}
\usepackage{booktabs}
\usepackage{float}
\newcommand{\indep}{\rotatebox[origin=c]{90}{$\models$}}
\usepackage{natbib}
\usepackage{ulem}
\usepackage{enumerate}
\usepackage{tcolorbox}
\usepackage[ruled,vlined,linesnumbered]{algorithm2e}
\usepackage{placeins}
\SetKwInOut{Input}{Input}
\SetKwInOut{Output}{Output}

\makeatletter
\renewcommand\@endtheorem{\vvv@endmarker\endtrivlist\@endpefalse}
\newcommand\vvv@endmarker{  {\nobreak\hfil\penalty50
  \hskip2em\vadjust{}\nobreak\hfil\openbox
  \parfillskip=0pt \finalhyphendemerits=0 \par
  \penalty 10000 \parskip=0pt\noindent}\ignorespaces}
\makeatother

\theoremstyle{plain}

\newtheorem*{assumption*}{Assumption}
\newtheorem{assumption}{Assumption}[section]

\newtheorem{definition}{Definition}[section]
\newtheorem{example}{Example}[section]

\newtheorem{lemma}{Lemma}[section]

\newtheorem{proposition}{Proposition}[section]
\newtheorem{remark}{Remark}[section]

\numberwithin{equation}{section}

\usepackage[nodisplayskipstretch]{setspace}
\usepackage[hypertexnames=false]{hyperref}
\hypersetup{hidelinks,breaklinks}
\usepackage{threeparttable}
\usepackage{threeparttablex} \usepackage{multirow}
\usepackage[font=bf, justification=centering, width = \textwidth]{caption} \usepackage{subcaption}
\usepackage{array}
\newcolumntype{H}{>{\setbox0=\hbox\bgroup}c<{\egroup}@{}}
\newcolumntype{P}[1]{>{\centering\arraybackslash}p{#1}}
\usepackage{tabularx}
\newcolumntype{Y}{>{\centering\arraybackslash}X}

\newcounter{mysubtable}

\usepackage{pgfplots} 
\pgfplotsset{compat=newest}
\usepgfplotslibrary{fillbetween} 
\usepgfplotslibrary{polar} \usepgflibrary{shapes.geometric}  
\usetikzlibrary{calc}  
\pgfplotsset{my style/.append style={axis x line=middle, axis y line= middle, xlabel={$\theta_{DID}$}, ylabel={$ATT$}, 
		every axis x label/.style={
			at={(ticklabel* cs:1.00)},
			anchor=west,
		},
		every axis y label/.style={
			at={(ticklabel* cs:1.00)},
			anchor=south,
		},
		axis equal}}

\let\origmaketitle\maketitle
\def\maketitle{
  \begingroup
  \def\uppercasenonmath##1{}   \let\MakeUppercase\relax   \origmaketitle
  \endgroup
}

\begin{document}
\nonstopmode

\title[From Pre-Trends to Bounds: Robust Difference-in-Differences]{\Large From Pre-Trends to Bounds: Robust Difference-in-Differences}
\author{Kyunghoon Ban, D\'esir\'e K\'edagni, Zhengrun Chen, and Augustine Denteh}
\address{Rochester Institute of Technology, UNC-Chapel Hill, and Davidson College}
\noindent \date{\scriptsize{The present version is as of \today. This paper supersedes two papers ``Robust Difference-in-differences Models'' by Kyunghoon Ban and D\'esir\'e K\'edagni and ``Anticipation Effects in Difference-in-Differences Models'' by Zhengrun Chen, Augustine Denteh, and D\'esir\'e K\'edagni. We thank participants at various seminars and conferences. All errors are ours.
Kyunghoon Ban (kban@saunders.rit.edu): 3314 Max Lowenthal Hall, 12 Lomb Memorial Drive, Rochester, NY 14623, USA; 
D\'esir\'e K\'edagni (dkedagni@unc.edu): 102 Gardner Hall, CB 3305 University of North Carolina Chapel Hill, NC 27599, USA;
Zhengrun Chen (zhengrun@unc.edu): 208 Swain Hall, 131 S Columbia St, Chapel Hill, NC 27514, USA;
Augustine Denteh (audenteh@davidson.edu): 209 Ridge Rd Box 5000, Davidson, NC 28035, USA}}

\begin{abstract}
Difference-in-differences is a widely used method for estimating causal effects in non-experimental settings, relying on two key assumptions: the Parallel Trends (PT) assumption and the No-Anticipation Effects (No-AE) assumption. However, the validity of these assumptions are challenging to demonstrate empirically because they rely on unobservable quantities. This paper presents a unified framework that is robust to violations of either or both of these assumptions. By reinterpreting the PT assumption through the lens of selection bias, we introduce the concept of a robust difference-in-differences (RDID) estimand. We derive partial identification results for the Average Treatment Effect on the Treated under relaxed assumptions and characterize a misspecification-robust region based on minimum data-consistent relaxations of the standard model. Furthermore, our framework is extended to accommodate covariates and multiple post-treatment periods. We demonstrate the empirical importance of relaxing parallel trends and no anticipation assumptions with our proposed methods in three applications. 
\end{abstract}

\maketitle

\markboth{From Pre-Trends to Bounds: Robust Difference-in-Differences}{From Pre-Trends to Bounds: Robust Difference-in-Differences}

{\footnotesize \textbf{Keywords}: Difference-in-differences; parallel trends; anticipation effects; partial identification; selection bias; misspecification-robust bounds

}
\newpage

\section{Introduction}

Difference-in-differences (DID) is one of the most widely used research designs for estimating causal effects in non-experimental settings. One of the parameters of interest in this context is the average treatment effect on the treated (ATT), which estimates the treatment effect for treated units. The DID design relies on two assumptions to identify the ATT: the ``parallel trends'' (PT) and the ``no-anticipation effects'' (No-AE) assumptions. The parallel trends assumption means that the average outcome among the treated and untreated populations would have followed the same trends in the absence of treatment. The No-AE assumption posits that, in the pre-treatment period, the outcome does not depend on future treatment. Equivalently, it rules out anticipation, which is any behavioral response to foreknowledge of future interventions. However, it is difficult to empirically verify both assumptions because they restrict quantities the researcher does not observe in their data.

Researchers using DID designs typically devote significant attention to justifying the validity of the PT and No-AE assumptions, especially the former. Naturally, such indirect justifications are often controversial and met with skepticism. For example, \citeauthor{kearney2015mtv}'s (\citeyear{kearney2015mtv}) ambitious study the effects of an MTV reality show on teen childbearing provided insightful findings, but there have been heated debates on the validity of its main PT assumption \citep{jaeger2018, kahnlang2019}. \citet{kearney2015mtv} estimate that MTV's \textit{16 and Pregnant} reduced teen birth rates by 4.3 percent, accounting for roughly one-quarter of the contemporaneous decline. \citet{jaeger2018} challenge this causal interpretation: the estimated effect becomes statistically insignificant after allowing trends to vary with pre-treatment demographic composition, similarly sized ``effects'' appear in placebo periods before the show aired, and extending the event-study pre-period by only two years leads to rejection of parallel pre-trends. \citet{kahnlang2019} emphasize that parallel pre-trends are neither necessary nor sufficient for post-treatment counterfactual parallel trends.

Typical applied DID analysis proceeds through a familiar sequence based on the widely understood approach for justifying the parallel trends assumption. Researchers first select treatment and comparison groups, estimate an event study specification, and examine the pre-treatment coefficients. When these coefficients are statistically insignificant, they are often interpreted as supporting the parallel trends assumption, with the belief that this assumption is likely to hold in the post-treatment period.  The researcher then proceeds to report a point-identified treatment effect. When the pre-treatment coefficients do not appear parallel, researchers consider various strategies, including changing the pre-treatment window, comparison group, covariates, functional form, or trend specification. Even when such changes are substantively defensible, choosing them after inspecting pre-treatment estimates makes the resulting evidence partly a product of specification search \citep{kahnlang2019, Roth2022}.

However, pre-treatment coefficients can have multiple (causal) interpretations. Before implementation, the observed treatment-control outcome gap combines selection bias and any behavioral response to future treatment. A nonparallel pre-trend may reflect changing selection bias, anticipation effect, or both. Conversely, a seemingly parallel pre-trend does not imply that either component is absent because changes in selection bias and anticipation effect may offset one another. In addition, rigorously speaking, the evidence of parallel pre-trends is different from the PT assumption needed in the post-treatment period. Given its inherently untestable nature, separate arguments need to be established for the validity of the PT assumption beyond pre-treatment period examination.\footnote{\cite{CallawaySantAnna2022} discuss this nuance in their vignette on pre-testing in a DID framework: ``importantly, this is just a pre-test; it is different from an actual test. Whether or not the parallel trends assumption holds in pre-treatment periods does not actually tell you if it holds in the current period (and this is when you need it to hold!). It is certainly possible for the identifying assumptions to hold in previous periods but not hold in current periods; it is also possible for identifying assumptions to be violated in previous periods but for them to hold in current periods.'' See also \cite{Freyaldenhoven_al2019}, \cite{kahnlang2019}, and \cite{Roth2022}.}

Motivated by these observations, this paper provides a roadmap for interpreting pre-treatment evidence in DID designs. Our objective is to develop identified sets that can be reported alongside conventional estimates, including TWFE estimates to provide a more complete view of policy effects under alternative interpretations of the pre-treatment evidence. As such, we do not propose that researchers simply abandon point identification. Specifically, we consider restrictions that allow PT to fail while maintaining No-AE, restrictions that allow (bounded) anticipation while maintaining PT, and restrictions that relax both assumptions. When researchers are unsure which source generated the pre-treatment trends, we also provide a misspecification-robust region across the competing models. These regions allow researchers to transparently report whether the conclusions remain supported under potential violations of their identifying assumptions.

The DID literature has developed extensive tools for addressing violations of parallel trends, while anticipation has received less attention. Yet anticipation can also be consequential for the validity of the DID design. We briefly highlight two applications illustrating the importance of this concern. \citet{alpert2016anticipatory} shows that reduced drug utilization ahead of Medicare Part D's implementation would substantially inflate the estimated treatment effect if unaccounted for. \citet{malani2015interpreting} finds an anticipatory increase in physician labor supply before tort reform, a policy not targeted at physicians but plausibly anticipated by them, which would underestimate or even lead to the wrong sign for the reform's effect if ignored. Institutional knowledge can guide responses such as excluding potentially contaminated pre-treatment periods or using announcement dates as treatment start points. These choices may still leave residual bias if anticipation begins before the selected cutoff.

This paper makes five salient contributions to the literature. First, we follow \cite{Heckman_al1998} to interpret the parallel trends assumption using a notion of \textit{selection bias} (also known as \textit{confounding bias} in statistics), which enables us to generalize the standard DID estimand by indexing it on a flexible baseline information set that can consist of multiple pre-treatment periods or baseline covariates.\footnote{Some recent papers also use similar interpretations of the PT assumption \citep{Henderson2023, Sofer2016, Parkal2023}.} 
Specifically, we introduce three new estimands: the \textit{parallel-trends-robust difference-in-differences (PT-RDID) estimand}, the \textit{anticipation-effects-robust difference-in-differences (AE-RDID) estimand}, and the \textit{jointly-robust difference-in-differences (RDID) estimand}.

Second, while maintaining the no-anticipation effect assumption, we replace standard parallel trends with \textit{bias set stability}, a strictly weaker identifying assumption when the selected baselines include period 0 and at least one earlier period. Specifically, we assume that the selection bias in the post-treatment period lies within the closed convex hull of selection biases in the selected baseline periods. We provide a sufficient condition for this assumption to hold. For example, we discuss and illustrate that this assumption may be plausible in economic settings where \citeauthor{Ashenfelter1978}'s (\citeyear{Ashenfelter1978}) dip is present. It is well documented that in such contexts, the PT assumption is usually not plausible (e.g., see \cite{AshenfelterCard1985}, \cite{HeckmanSmith1999}, \cite{Heckman_etal1999}). Under no anticipation and bias set stability, the resulting identified set contains the true ATT. When the baseline information set includes period 0, it also contains the standard DID estimand, recovering that estimand as a singleton when the pre-treatment outcome gaps are constant across the selected periods.

Our paper adds to a growing literature on relaxations of the PT assumption. Two closely related papers, \citet{ManskiPepper2018} and \citet{RambachanRoth2020}, develop \textit{trends-based} relaxations of the PT assumption. \citet{ManskiPepper2018} introduce bounded-variation assumptions that bound the absolute difference in untreated trends between treatment and control by a known sensitivity parameter, with PT recovered as a special case when the parameter is zero. \citet{RambachanRoth2020} generalize this idea by considering a large class of restrictions requiring that post-treatment violations of parallel trends not be ``too different'' from the pre-trends. Both approaches require the researcher to specify an explicit sensitivity parameter whose choice remains unclear in practice. Our approach differs along two dimensions. Substantively, we develop a \textit{selection-based} relaxation that can use information over time or across space, with the sensitivity parameter implicitly embedded in the baseline information set rather than chosen ex ante.\footnote{Our selection-based relaxation can use information over time (e.g., when the baseline information set is the set of pre-treatment periods) or across space (e.g., when the baseline information is the set of baseline covariates, which can include geographic region).} Formally, our bounding strategy falls within \citeauthor{RambachanRoth2020}'s broader class of pre-trend-based restrictions and can be viewed as a special case under a different lens, though the specific identifying assumptions we consider have not been examined in \citet{RambachanRoth2020}. \citet{ghanem2022selection} also frame the PT assumption in terms of selection into treatment and study alternative selection-based identifying conditions; our approach is complementary in developing partial identification bounds indexed on a baseline information set of pre-treatment periods or covariates. Our approach carries a corresponding limitation: when PT holds in the pre-treatment period but fails in the post-treatment period, our identified set coincides with the standard DID estimand, which does not identify the ATT.

Third, while maintaining parallel trends, we replace the no-anticipation-effect assumption with a strictly weaker identifying assumption, which we call \textit{bounded anticipation}. This addresses an empirically prevalent channel that has received less methodological attention than parallel-trends robustness. Specifically, we assume that the average anticipation effect is bounded in magnitude by the true ATT, and we provide a testable implication and a bound on the ATT. By incorporating information from multiple pre-treatment periods, we can tighten this bound under the assumption that the average anticipation effect is bounded and exponentially discounted. Here, the way to model the anticipation behavior is motivated by the exponential discounting looking-forward model, which is commonly used in finance and macroeconomics \citep{malani2015interpreting}. Combining these two key assumptions---bias set stability and bounded anticipation---yields a joint-relaxation identified set robust to simultaneous violations of PT and No-AE, which is a natural choice when the researcher is willing to impose both relaxations. With multiple pre-treatment periods, we also develop a misspecification-robust region following \citet{desire2024qe}. When the standard DID model is falsified, this region combines the identified sets from its minimum data-consistent relaxations, allowing researchers to assess what they can learn about the ATT without choosing between those relaxations in advance. By focusing on these minimum relaxations, it can yield a tighter region than the joint-relaxation set.

Our paper therefore contributes to a small theoretical literature on anticipation effects in DID designs. \citet{malani2015interpreting} and \citet{gong2021bounds} are the closest prior work. \citeauthor{malani2015interpreting} assumes a parametric outcome model with exponentially discounted expectations and uses instrumental variables to identify unobserved expectations. Our exponential $c$-bounded anticipation assumption relaxes this parametric requirement and does not rely on instruments. \citeauthor{gong2021bounds} bounds the ATT under a two-type model of anticipators and non-anticipators; our motivating model accommodates arbitrarily many heterogeneous types with varying discount factors, and our identification assumption is implied by his. Our bounds also do not require the researcher to know the sign of either the treatment or the anticipation effect in advance. This is particularly relevant in settings like Medicare Part D, where income and intertemporal-substitution channels can push anticipation in opposite directions \citep{alpert2016anticipatory}.

Fourth, we show how our framework extends to the DID with multiple treatment periods setting.
We first develop a testable implication for the extended PT and the extended No-AE and we derive bounds on each post-treatment period ATT when each of the assumptions is relaxed. 
This approach can help reveal the heterogeneity in the treatment effects over time. 
As before, our PT-robust bounds collapse to the usual DID estimand at each post-treatment date when the outcome gap between the two groups is constant across the selected baselines. We also extend the method to the identification of more causal parameters, including those considered in \cite{CallawaySantAnna2021}. The causal parameters we consider can also help reveal the dynamic effect of the treatment. This extension contributes to the causal interpretation of TWFE event-study coefficients under staggered adoption designs \citep{Borusyak_al2022, AtheyImbens2022, goodmanbacon2021ddtiming, CallawaySantAnna2021, deChaisemartinDHaultfoeuille2020, SunAbraham2021}, building on \citeauthor{Wooldridge2021}'s (\citeyear{Wooldridge2021}) TWFE regression idea to compute confidence sets when the baseline information set is the set of pre-treatment periods.

Fifth, we illustrate the empirical relevance of our methodology by revisiting Colombia's coca substitution program \citep{ladino2021one}, plastic-bag charges in Uruguay \citep{ccc}, and compulsory licensing of German patents after World War I \citep{baten}. On one hand, our 95\% ATT confidence sets based on the joint relaxation analysis supports the negative findings for coca cultivation and plastic-bag use. On the other hand, for the German field-level comparison, the estimated bounds remain positive, but the corresponding confidence sets include zero, weakening the statistical support for a positive effect. We believe that these applications are informative to applied researchers because they demonstrate how our methods can show robustness of their findings or show their dependence on their research design's identifying assumptions.

The rest of the paper is organized as follows. Section~\ref{sec:setting} introduces the econometric framework. Sections~\ref{sec:relax-pt}--\ref{sec:relax-both} develop identification results under relaxations of parallel trends, no anticipation, and both assumptions. Section~\ref{sec:inference} presents estimation and inference for the identified sets, and Section~\ref{sec:extension} considers multiple treatment periods. Section~\ref{sec:empirical} presents the empirical applications, and Section~\ref{sec:conclusion} concludes. The appendices provide proofs, various extensions, and supporting empirical analyses.

\section{Setting}\label{sec:setting}
\subsection{The baseline model}
Consider the following potential outcome model:
\begin{eqnarray}\label{seq1}
Y_t&=&Y_t(1)D+Y_t(0)(1-D)
\end{eqnarray}
where $(Y_t,D)$, $t\in \mathcal T_0 \cup \{1\}$ represents the observed data, while the vector $(Y_1(0), Y_1(1))$ is latent. $\mathcal T_0 \subseteq \{\ldots, -1, 0\}$ denotes the set of pre-treatment periods, with $0 \in \mathcal T_0$ being the last pre-treatment period. The variables $Y_{t_0}, Y_1 \in \mathcal Y$ are respectively the observed outcomes in the baseline period $t_0$ and the follow-up period 1, while $D\in \left\{0,1\right\}$ is the observed treatment that occurred between periods 0 and 1, $Y_1(0)$ and $Y_1(1)$ are the potential outcomes that would have been observed in period 1 had the treatment $D$ been externally set to 0 and~1, respectively. 

In this paper, we are interested in identifying the average treatment effect on the treated (ATT) defined as 
$$ATT\equiv \mathbb E[Y_1(1)-Y_1(0)\vert D=1].$$

Throughout, we assume that the first moments are well-defined for the observed and potential outcomes. We also assume that we observe individuals in both the treatment and control groups, i.e., $0<\mathbb P(D=1)<1$.\footnote{Our sharpness and data-consistency results allow unrestricted real-valued counterfactual outcomes. Known support restrictions can tighten the bounds and change which relaxations are data-consistent.}

Our framework starts from a decomposition of the standard ordinary least squares (OLS) estimand that reframes the identification of the ATT as the identification of a single unobserved quantity---the selection bias in period 1. We then show how the standard DID design is obtained as a special case of the more flexible framework we develop below. We can decompose the standard OLS estimand in period~1 and in any pre-treatment period $t_0 \in \mathcal T_0$ as
\begin{eqnarray}
\theta_1^{OLS}&\equiv& \frac{Cov(Y_1,D)}{Var(D)}=\mathbb{E}[Y_{1}\mid D=1]-\mathbb{E}[Y_{1}\mid D=0]  \nonumber\\
 & = & \underbrace{\mathbb{E}[Y_{1}(1)-Y_{1}(0)\mid D=1]}_{\text{ATT}}+\underbrace{\mathbb{E}[Y_{1}(0)\mid D=1]-\mathbb{E}[Y_{1}(0)\mid D=0]}_{\text{\ensuremath{SB_{1}} (Selection Bias)}}, \label{eq:ols1}\\
\theta_{t_0}^{OLS}&\equiv& \frac{Cov(Y_{t_0},D)}{Var(D)}= \mathbb{E}[Y_{t_0}\mid D=1]-\mathbb{E}[Y_{t_0}\mid D=0] \nonumber\\
 & = & \underbrace{\mathbb{E}[Y_{t_0}(1)-Y_{t_0}(0)\mid D=1]}_{AE_{t_0}}+\underbrace{\mathbb{E}[Y_{t_0}(0)\mid D=1]-\mathbb{E}[Y_{t_0}(0)\mid D=0]}_{SB_{t_0}}, \label{eq:ols0}
\end{eqnarray}
where $SB_t\equiv \mathbb E[Y_t(0) \vert D=1]-\mathbb E[Y_t(0) \vert D=0]$ is called the \textit{selection bias} in period $t$, using the terminology of \cite{heckman1998characterizing}, and $AE_{t}\equiv\mathbb{E}[Y_{t}(1)-Y_{t}(0)\mid D=1]$ is the anticipation effects in period $t$.
Note that we can rewrite \eqref{eq:ols1} as 
\begin{equation}
    \text{ATT} = \theta_1^{OLS} - SB_1, \label{eq:att}
\end{equation}
and the identification of the ATT can be re-framed as the identification of $SB_1$.

For the identification of $SB_1$ (and hence the ATT), the standard DID approach relies on two main assumptions: parallel trends (PT), or bias stability ($SB_1=SB_0$), and no anticipation effects (No-AE, $AE_0=0$). Together, these restrictions imply $SB_1=SB_0=\theta_0^{OLS}$.
Therefore, the standard DID estimand defined as
\begin{align}
    \theta^{DID} &\equiv \theta_1^{OLS} - \theta_0^{OLS} \nonumber \\
    &= \text{ATT} - AE_0 + (SB_1-SB_0), \label{eq:main}
\end{align}
identifies the ATT under the key two assumptions: $SB_1=SB_0$ and $AE_0=0$.\footnote{However, it is worth noting that none of these assumptions is necessary for the DID estimand to identify the ATT. From Equation \eqref{eq:main}, we can see that the DID estimand identifies the ATT if and only if the difference in the selection biases between periods 1 and 0 is equal to the anticipation effect in period 0 (i.e., $SB_1-SB_0=AE_0$).
}

However, when these above two assumptions are questionable, researchers doubt the validity of the DID design. This concern typically arises from evidence in the data such as detectable non-parallel pre-trends, which may either signal that selection bias may be unstable across periods or anticipatory behavior where treatment status may already be affecting outcomes before implementation. 

In this paper, we propose an identification approach that allows for violations of the PT and/or No-AE assumptions. Our framework exploits \eqref{eq:ols0} for arbitrary $t_0 \in \mathcal T_0$ by replacing the point identification of $SB_1$ with a set of candidate values determined by a chosen baseline information set $\mathcal I_0$.
In particular, equation \eqref{eq:att} inspires us to introduce what we call the \textit{robust difference-in-differences (RDID)} estimand with the baseline information set below. 
\begin{definition}[PT-RDID]\label{def:rdid}
Given a {nonempty} baseline information set $\mathcal I_0 \subseteq \mathcal T_0$, we define the \textit{parallel-trends-robust difference-in-differences (PT-RDID)} estimand as
\begin{eqnarray}\label{eq:RDID}
\Theta^{\text{PT-RDID}}\equiv \big\{ \theta_1^{OLS}- s: s \in \overline{\operatorname{conv}}\{\theta^{OLS}_{t_0}:t_0 \in \mathcal I_0\} \big\},
\end{eqnarray}
{where $\overline{\operatorname{conv}}\{\theta^{OLS}_{t_0}: t_0 \in \mathcal I_0\}$ is the closed convex hull of the OLS estimands for all pre-treatment periods in the information set $\mathcal I_0$. 
}
\end{definition}

Note that this is a type of RDID estimand we consider in this paper and thus named as PT-RDID because we show in Section~\ref{sec:relax-pt} that it partially identifies the ATT under No-AE allowing for violations of PT.
Based on this definition, it follows straightforwardly in the following that the standard DID design corresponds to the PT-RDID estimand with the smallest baseline information set $\mathcal I_0 = \{0\}$.

\begin{example}[Standard DID]
Consider $\mathcal I_0 = \{0\}$ where the information set contains only one pre-treatment period, the period 0.
We have
\begin{align*}
    \overline{\operatorname{conv}}\{\theta^{OLS}_{t_0}: t_0 \in \mathcal I_0\} 
        &= \big\{ \theta_0^{OLS} \big\},
\end{align*}
and $\Theta^{\text{PT-RDID}} = \{\theta^{DID}\}$, which point-identifies the ATT if $SB_1=SB_0$ (PT) and $AE_0=0$ (No-AE).
\end{example}

\section{Allowing for violation of parallel trends}\label{sec:relax-pt}\label{subsec:did-relax-pt} 

In this section, we assume that there is no anticipation effect of the treatment for the treated units, and we relax the parallel trends assumption. We first state the maintained no-anticipation assumption and introduce our new identifying assumption, \textit{bias set stability}, which restricts the (unobserved) period-1 selection bias to lie within the closed convex hull of the pre-treatment selection biases. We then derive a sharp identified set for the ATT (Proposition~\ref{prop1}) that contains the standard DID estimand when the baseline information set includes period 0 and recovers it as a singleton when the pre-treatment outcome gaps are constant across that set. Appendix~\ref{app:point-approaches} develops two additional approaches to determining $SB_1$: a policy-oriented approach based on a loss function (Appendix~\ref{subsec:po-rdid}) and projection-based forecasting for ordered baseline periods (Appendix~\ref{sec:orderedinfo}).

Formally, consider the following assumption first.
\begin{assumption}[No Anticipation Effects]\label{ass:NoAE}
$$
AE_{t_0}\equiv\mathbb{E}[Y_{t_0}(1)-Y_{t_0}(0)\mid D=1] = 0,
$$
for all $t_0\in \mathcal I_0$.
\end{assumption}

The baseline information set $\mathcal I_0$ need not contain every observed pre-treatment period. Institutional knowledge can guide the selection of periods in which no anticipation is credible, for example, periods preceding a policy announcement. Assumption~\ref{ass:NoAE} applies only to the selected periods, and bias set stability uses the range of their selection biases. The choice of $\mathcal I_0$ therefore determines both where no anticipation is required and the baseline bias range used to restrict the post-treatment bias.

Under this maintained no-anticipation assumption, each pre-treatment selection bias $SB_{t_0}$ becomes identified from the data as $\theta^{OLS}_{t_0}$ (see Remark~\ref{rm:sb-identified}). The following assumption then restricts the period-1 selection bias $SB_1$ to lie within the closed convex hull of these identified pre-treatment biases.

\begin{assumption}[Bias Set Stability]\label{sb:bounds}
\begin{eqnarray*}
{SB_1} \in \left[ \inf_{t_0\in \mathcal I_0} SB_{t_0}, \sup_{t_0 \in \mathcal I_0} SB_{t_0}\right] \equiv \Delta_{SB(\mathcal I_0)}.\end{eqnarray*}
\end{assumption} 
\begin{remark}
    Assumption \ref{sb:bounds} is weaker than the standard ``parallel/common trends'' assumption when $\mathcal I_0$ includes period 0 and at least one earlier period. 
    Consider $\mathcal I_0 = \mathcal T_0 = \{0\}$ again.
    In this case, Assumption \ref{sb:bounds} implies ${SB_1}= SB_0,$ which is equivalent to the parallel trends assumption.
\end{remark}
\begin{remark}\label{rm:sb-identified}
    When there is no anticipation effect of the treatment ($AE_{t_0} = 0$ for all $t_0 \in \mathcal I_0$), $SB_{t_0}$ is identified as
    \begin{align*}
        SB_{t_0}&=\mathbb E[Y_{t_0} \vert D=1]-\mathbb E[Y_{t_0} \vert D=0]\\
        &\equiv \theta^{OLS}_{t_0}.
    \end{align*}
\end{remark}
Note that $\overline{\operatorname{conv}}\{\theta^{OLS}_{t_0}: t_0 \in \mathcal I_0\} = \Delta_{SB(\mathcal I_0)}$ with the pre-treatment biases identified and Assumption~\ref{sb:bounds}, and
the following proposition holds.
\begin{proposition}\label{prop1}
Suppose that model \eqref{seq1} along with Assumptions \ref{ass:NoAE} and \ref{sb:bounds} holds, {and suppose $\inf_{t_0\in\mathcal I_0}\theta_{t_0}^{OLS}$ and $\sup_{t_0\in\mathcal I_0}\theta_{t_0}^{OLS}$ are finite}. Then, the following bounds hold for the ATT: 
\begin{eqnarray*}
\mathrm{ATT} \in \left[ \theta^{OLS}_1- \sup_{t_0 \in \mathcal I_0} \theta^{OLS}_{t_0}, \theta^{OLS}_1 - \inf_{t_0 \in \mathcal I_0} \theta^{OLS}_{t_0} \right] = \Theta^{\text{PT-RDID}}.
\end{eqnarray*} 
These bounds are sharp, and $\Theta^{\text{PT-RDID}}$ is the identified set for the ATT.
\end{proposition}
{ A proof is relegated to Appendix \ref{sec:prop1}.}
Intuitively, the ATT bounds have the form of ``the OLS estimand in period 1 minus a candidate for $SB_1$.'' Because $SB_1$ is only known up to the closed convex hull of the pre-treatment biases, plugging in the largest candidate ($\sup_{t_0} \theta^{OLS}_{t_0}$) yields the lower ATT bound, and the smallest candidate ($\inf_{t_0} \theta^{OLS}_{t_0}$) yields the upper ATT bound. When only period 0 is in the information set, both extremes coincide with $\theta^{OLS}_0$ and the bound collapses to the standard DID estimand.

\begin{remark}\label{rm.rdid1.1}
    The identified set $\Theta^{\text{PT-RDID}}$ is never empty. If $0\in\mathcal I_0$, it contains the standard DID estimand $\theta^{DID}=\theta^{OLS}_1-\theta^{OLS}_0$. It collapses to this singleton when the pre-treatment selection biases are constant across $\mathcal I_0$, including the special case $\mathcal I_0=\{0\}$. In this case, the candidate set $\overline{\operatorname{conv}}\{\theta^{OLS}_{t_0}: t_0 \in \mathcal I_0\} $ reduces to $\{\theta_0^{OLS}\}$ and the robust framework recovers exactly the standard DID estimand.
\end{remark}

Beyond the operational form of Assumption~\ref{sb:bounds}, one may ask under what data generating processes it holds. Appendix~\ref{sec:sufficient-conditions} provides two sufficient conditions for Assumption~\ref{sb:bounds}: a linear interactive-fixed-effects structure with time-symmetric factor loadings in expectation, and a nonlinear variant that requires pointwise matching across periods. Both formalize the Ashenfelter-dip intuition above by exhibiting a class of DGPs under which Assumption~\ref{sb:bounds} holds while PT fails.
 
\section{Allowing for anticipation effects}\label{sec:relax-noae} 
In this section, we maintain the parallel trends assumption in its strong form and relax the no-anticipation assumption. Like Section~\ref{subsec:did-relax-pt}, the section proceeds in three steps. We first state the maintained strong parallel trends assumption and introduce the our new identifying assumption on anticipation effects, which we call \textit{exponential $c$-bounded anticipation}: the magnitude of the anticipation effect at each pre-treatment period is bounded by an exponentially discounted fraction of the true ATT. We then establish a testable implication (Lemma~\ref{lem:samesign}) that the DID estimand at each pre-treatment period must share the sign of the ATT, and derive a sharp identified set for the ATT (Proposition~\ref{bound_multiple_preperiods}). Additional justification and sufficient conditions for the exponential discounting model are provided in Appendix~\ref{ap:motivating}. Throughout this section and the next, $\mathcal I_0$ is finite.\footnote{Finiteness ensures that the displayed maxima and minima are attained and permits the sharpness constructions to select baseline periods satisfying the relevant extremal inequalities.}

Recall from Section~\ref{sec:setting} that for $t_0\in\mathcal{I}_{0}$, we define
\begin{align*}
    \theta_{t_0}^{OLS} & \equiv\mathbb{E}[Y_{t_0}\mid D=1]-\mathbb{E}[Y_{t_0}\mid D=0]\\
SB_{t_0}&\equiv\mathbb{E}[Y_{t_0}(0)\mid D=1]-\mathbb{E}[Y_{t_0}(0)\mid D=0]\\
AE_{t_0}&\equiv\mathbb{E}[Y_{t_0}(1)-Y_{t_0}(0)\mid D=1],
\end{align*}
and, we have $\theta_{t_0}^{OLS}=AE_{t_0}+SB_{t_0}$.

\begin{assumption}[Strong Parallel Trends]\label{ass:strongPT}
$$
SB_{t_0}=SB_{1}\text{ for all }t_0\in\mathcal{I}_{0}
$$

\end{assumption}

Maintaining strong parallel trends alone does not identify the ATT because $AE_{t_0}$ remains unrestricted.\footnote{We use \textit{PT} or \textit{standard PT} for the two-period version of parallel trends ($SB_1 = SB_0$) and use \textit{Strong PT} for the multi-period version just stated in Assumption~\ref{ass:strongPT}. Section~\ref{sec:extension} further introduces \textit{Extended PT} for the setting with multiple treatment periods.} Our new identifying assumption helps us to make progress by bounding the magnitude of anticipation effects.

\begin{assumption}[Exponential c-bounded Anticipation Effects]\label{ass:expoAE} For some known $c\in[0,1)$,
\[
\mid AE_{t_0}\mid\leq c^{|t_0|+1}\cdot\mid \mathrm{ATT}\mid\text{ for all }t_0\in\mathcal{I}_{0}
\]

\end{assumption}

Assumption \ref{ass:expoAE} states that individuals are forward looking regarding the true ATT, and the anticipation effects on average are discounted from ATT by a parameter c. This assumption can be justified by the observation that at some point (unknown to econometricians) before the treatment, only a small number of people are aware of upcoming intervention or policy, they might respond to that, suggesting that on average the anticipation effects should not be significantly large initially. However, as information spreads exponentially, more and more people are informed, and these effects should grow substantially as well. Therefore, when viewed in reverse, the average anticipation effects are expected to diminish exponentially as one moves further away from the treatment period. Alternative justification of assumption \ref{ass:expoAE} is discussed in Appendix \ref{ap:motivating} and a sufficient condition is provided.
\begin{remark}
Assumption \ref{ass:expoAE} is similar to the exponential discounting model in \citet{malani2015interpreting} where they assume the anticipation effect is equal to discounted ATT if there is only one post-treatment period, while our assumption only states that the magnitude of AE is bounded by the magnitude of discounted ATT.
\end{remark}
{
Beyond identifying the ATT, the pair of assumptions above delivers a testable implication that we exploit both here and in the multi-period extension of Section~\ref{sec:extension}.

\begin{lemma}\label{lem:samesign}
    Let $\theta^{DID}_{t_0}$ denote the DID estimand using $t_0 \in \mathcal I_0$:
    $$
    \theta^{DID}_{t_0} \equiv \theta_1^{OLS} - \theta_{t_0}^{OLS}.
    $$
    Under Assumptions \ref{ass:strongPT} and \ref{ass:expoAE}, we have
    $$
    \operatorname{sgn}(\theta^{DID}_{t_0}) = \operatorname{sgn}(\mathrm{ATT}),
    $$
    for all $t_0 \in \mathcal I_0$.
\end{lemma}
A proof is relegated to Appendix \ref{sec:lem:samesign}.

\begin{remark}
    Note that Lemma \ref{lem:samesign} provides a testable implication for the model \eqref{seq1} with Assumptions \ref{ass:strongPT} and \ref{ass:expoAE} when there are more than one pre-treatment periods in $\mathcal I_0$:
    $$
    \operatorname{sgn}(\theta^{DID}_{t_0}) = \operatorname{sgn}(\theta^{DID}_{t_0'}),
    $$
    for all distinct $t_0, t_0' \in \mathcal I_0$.
\end{remark}

To characterize the identified set under Assumptions \ref{ass:strongPT} and \ref{ass:expoAE}, consider a candidate value $\theta$ for the ATT. Strong PT gives $SB_{t_0}=SB_1$, and hence
\[
\theta_{t_0}^{DID}=\theta-AE_{t_0}.
\]
Therefore, the bounded-anticipation restriction is equivalent to
\begin{equation}\label{eq:rdid2-feasibility}
|\theta-\theta_{t_0}^{DID}|
\leq c^{|t_0|+1}|\theta|
\qquad\text{for every }t_0\in\mathcal I_0.
\end{equation}
Solving \eqref{eq:rdid2-feasibility} separately on $\theta\geq0$ and $\theta\leq0$ yields the two intersection intervals in Proposition \ref{bound_multiple_preperiods}, whose union we define as \textit{the anticipation-effects-robust difference-in-differences (AE-RDID) estimand}. This characterization also makes clear that the model is falsified whenever both intervals are empty.

\begin{proposition}\label{bound_multiple_preperiods}
Under Assumptions \ref{ass:strongPT} and \ref{ass:expoAE}, bound on ATT is given by
\begin{align}
\label{equation_bound_multiple_preperiods}
ATT\in & \left[\max_{t_0\in\mathcal{I}_{0}}\left\{ \frac{\theta_{t_0}^{DID}}{1+c^{|t_0|+1}}\right\} ,\min_{t_0\in\mathcal{I}_{0}}\left\{ \frac{\theta_{t_0}^{DID}}{1-c^{|t_0|+1}}\right\} \right]\bigcup\nonumber \\
 & \qquad\qquad\qquad\qquad\left[\max_{t_0\in\mathcal{I}_{0}}\left\{ \frac{\theta_{t_0}^{DID}}{1-c^{|t_0|+1}}\right\} ,\min_{t_0\in\mathcal{I}_{0}}\left\{ \frac{\theta_{t_0}^{DID}}{1+c^{|t_0|+1}}\right\} \right] \\
 &\equiv \Theta^{\text{AE-RDID}} \nonumber
\end{align}
where $\theta_{t_0}^{DID}\equiv\theta_{1}^{OLS}-\theta_{t_0}^{OLS}$.
These bounds are sharp.
\end{proposition}
A proof is relegated to Appendix \ref{sec:bound_multiple_preperiods}. Intuitively, each pre-treatment period $t_0$ contributes an interval of ATT candidates that satisfy the bounded-anticipation restriction at $t_0$, and the identified set is the intersection of these intervals, further partitioned by the sign of the ATT (positive-ATT and negative-ATT branches). The union structure in \eqref{equation_bound_multiple_preperiods} reflects the ambiguity about the sign of the ATT under \eqref{eq:rdid2-feasibility}. Remark~\ref{rm.rdid2.convex} below shows that once the sign is determined by Lemma~\ref{lem:samesign}, the feasible values form a single interval.
\begin{remark}\label{rm.rdid2.convex}
    Lemma \ref{lem:samesign} implies that $\Theta^{\text{AE-RDID}}$ is always convex under Assumptions \ref{ass:strongPT} and \ref{ass:expoAE}:
    $$
    \Theta^{\text{AE-RDID}} = \left\{\begin{array}{lcl}
        \big[\max_{t_0\in\mathcal{I}_{0}}\big\{ \frac{\theta_{t_0}^{DID}}{1+c^{|t_0|+1}}\big\} ,\min_{t_0\in\mathcal{I}_{0}}\big\{ \frac{\theta_{t_0}^{DID}}{1-c^{|t_0|+1}}\big\} \big] & \mathrm{if} & \mathrm{ATT} \geq 0 \\
        \big[\max_{t_0\in\mathcal{I}_{0}}\big\{ \frac{\theta_{t_0}^{DID}}{1-c^{|t_0|+1}}\big\} ,\min_{t_0\in\mathcal{I}_{0}}\big\{ \frac{\theta_{t_0}^{DID}}{1+c^{|t_0|+1}}\big\} \big] & \mathrm{if} & \mathrm{ATT} < 0
    \end{array}\right. .
    $$
\end{remark}

\begin{remark}
    If $\mathcal I_0 = \mathcal T_0 = \{0\}$, then
    \begin{align*}
        \Theta^{\text{AE-RDID}} 
                &= \left[ \min\left\{\frac{\theta_{0}^{DID}}{1+c}, \frac{\theta_{0}^{DID}}{1-c} \right\}, \max\left\{ \frac{\theta_{0}^{DID}}{1-c}, \frac{\theta_{0}^{DID}}{1+c} \right\} \right],
    \end{align*}
    and $\theta_{0}^{DID} \in \Theta^{\text{AE-RDID}}$.
        
\end{remark}
}

The bounds in \eqref{equation_bound_multiple_preperiods} take the form of a union of two intersection bounds. They are tighter than the bounds derived from just two pre-treatment periods, and plotting them as a function of $c$ quantifies how failures of the no-anticipation assumption translate into non-identification of the ATT. The next section relaxes both parallel trends and no-anticipation simultaneously.

\section{Relaxing parallel trends and no anticipation assumptions}\label{sec:relax-both}

This section combines the two new identifying assumptions introduced in Sections~\ref{subsec:did-relax-pt} and~\ref{sec:relax-noae} to allow the ATT to be identified when both PT and No-AE may be violated. Section~\ref{subsec:joint-relaxation} imposes bias set stability and exponential $c$-bounded anticipation simultaneously, delivering an identified set $\Theta^{\text{RDID}}$ that is robust to violations of both DID assumptions. Section~\ref{subsec:misspec-robust} takes a complementary perspective following \citet{desire2024qe}. When the standard DID model is falsified, it characterizes the minimum relaxations that restore compatibility with the observed data and reports the union of their identified sets.

{
\subsection{Joint Relaxation}\label{subsec:joint-relaxation}

When Assumption~\ref{sb:bounds} (bias set stability) and Assumption~\ref{ass:expoAE} (exponential $c$-bounded anticipation) are imposed together, we define the following \textit{jointly-robust difference-in-differences (RDID) estimand} which provides a joint-relaxation identified set for the ATT.

\begin{proposition}\label{bound_relaxtwo}
Under Assumption \ref{ass:expoAE} and \ref{sb:bounds}, bound on the ATT is given by
\begin{align}\label{relaxboth}
    ATT&\in \Bigg[ 
    \min_{t_0\in\mathcal{I}_{0}}\left\{ \min\Big\{ \frac{\theta_{t_0}^{DID}}{1+c^{|t_0|+1}}, \frac{\theta_{t_0}^{DID}}{1-c^{|t_0|+1}}\Big\} \right\}, \nonumber\\
    & \qquad \qquad \qquad \max_{t_0\in\mathcal{I}_{0}}\left\{ \max\Big\{ \frac{\theta_{t_0}^{DID}}{1+c^{|t_0|+1}}, \frac{\theta_{t_0}^{DID}}{1-c^{|t_0|+1}}\Big\} \right\}
    \Bigg],\\
 &\equiv \Theta^{\text{RDID}}, \nonumber
\end{align}
where $\theta_{t_0}^{DID}\equiv\theta_{1}^{OLS}-\theta_{t_0}^{OLS}$.
These bounds are sharp.\end{proposition}
A proof is relegated to Appendix \ref{sec:bound_relaxtwo}, where the sharpness of the bounds is also established. Intuitively, each pre-treatment period $t_0$ contributes two candidate values for the ATT, one under each sign branch of the bounded-anticipation restriction, and the outer interval takes the extreme of these across all $t_0 \in \mathcal I_0$. This bound is generically wider than either $\Theta^{\text{PT-RDID}}$ or $\Theta^{\text{AE-RDID}}$; Appendix~\ref{sec:inc} formalizes the containments $\Theta(A_{\text{PT-R}}) \subseteq \Theta(A_{\text{R}})$ and $\Theta(A_{\text{AE-R}}) \subseteq \Theta(A_{\text{R}})$.

\begin{remark}
    Note that $\Theta^{\text{RDID}}$ takes the form of a union bound. It is wider than the bounds before, but it is robust to the violation of two identifying assumptions at the same time. Also, note that since $1\pm c^{|t_0|+1}>0$, the joint bound contains zero exactly when the PT-RDID bound contains zero.
\end{remark}

The joint-relaxation analysis allows both original assumptions to fail while maintaining bias set stability and bounded anticipation. A complementary question arises when researchers wish to retain as much of the standard model as the observed data can permit. We therefore characterize its minimum data-consistent relaxations and combine their identified sets without the need to select a single relaxation in advance.
}

{

\subsection{Misspecification Robust Bounds}\label{subsec:misspec-robust}
\setlength{\emergencystretch}{2em}

\leavevmode{Following the discordant-relaxations framework of \citet{desire2024qe}, we construct the region $\Theta^{*}(A_0)$.} When the standard DID model is falsified, the construction retains its minimum data-consistent relaxations within a specified model family and reports the union of their identified sets.

{
{Let $Q$ be a joint law of the observed variables and the potential outcomes $(Y_t(0),Y_t(1))$ for $t\in\mathcal T_0\cup\{1\}$.} For a restriction bundle $B$, let $\mathcal M(B)$ denote the class of full-data laws satisfying $B$. For two bundles $B$ and $C$, write
\[
B\succeq C
\quad\Longleftrightarrow\quad
\mathcal M(B)\subseteq\mathcal M(C),
\]
so that $B\succeq C$ means that $B$ is logically at least as strong as $C$. Write $B\sim C$ when both $B\succeq C$ and $C\succeq B$, and write $B\succ C$ when $B\succeq C$ but $C\not\succeq B$. {The menu below uses equivalence classes under $\sim$, represented by the same bundle symbols.}

Fix an observed-data distribution $P$. For a full-data law $Q$, let $Q_O$ denote its observed-data marginal, and define
\[
\mathcal M_P(B)
\equiv
\{Q\in\mathcal M(B):Q_O=P\},
\qquad
\Theta_P(B)
\equiv
\{ATT(Q):Q\in\mathcal M_P(B)\}.
\]
We call $B$ data-consistent at $P$ when $\mathcal M_P(B)\neq\varnothing$. We suppress the subscript $P$ when no confusion can arise. {Thus $B\succeq C$ implies $\Theta(B)\subseteq\Theta(C)$.}

Define the standard DID benchmark and its three pre-specified relaxations by
\begin{align*}
A_0&\equiv\{\text{No anticipation, Strong parallel trends}\},\\
A_{\text{PT-R}}&\equiv\{\text{No anticipation, Bias set stability}\},\\
A_{\text{AE-R}}&\equiv\{\text{Strong parallel trends, Exponential $c$-bounded anticipation}\},\\
A_{\text{R}}&\equiv\{\text{Bias set stability, Exponential $c$-bounded anticipation}\}.
\end{align*}
Thus, $A_{\text{PT-R}}$ relaxes only strong PT, $A_{\text{AE-R}}$ relaxes only no anticipation, and $A_{\text{R}}$ relaxes both. Because no anticipation implies exponential $c$-bounded anticipation and strong PT implies bias-set stability,
\[
A_0\succeq A_{\text{PT-R}}\succeq A_{\text{R}},
\qquad
A_0\succeq A_{\text{AE-R}}\succeq A_{\text{R}}.
\]
For the fixed observed-data distribution,
\[
\Theta(A_{\text{PT-R}})=\Theta^{\text{PT-RDID}},
\qquad
\Theta(A_{\text{AE-R}})=\Theta^{\text{AE-RDID}},
\qquad
\Theta(A_{\text{R}})=\Theta^{\text{RDID}},
\]
and $\Theta(A_0)$ is the standard-DID singleton whenever $A_0$ is data-consistent. {The corresponding set inclusions are shown in Appendix~\ref{sec:inc}.}

{Let $\mathfrak A=\{A_0,A_{\text{PT-R}},A_{\text{AE-R}},A_{\text{R}}\}$ be this finite menu.}

\begin{definition}[Minimum data-consistent relaxation; MR bound]\label{def:mrbound}
A bundle $B\in\mathfrak A$ is a \emph{data-consistent relaxation} of $A_0$ within $\mathfrak A$ if $A_0\succeq B$ and $\mathcal M_P(B)\neq\varnothing$. It is a \emph{minimum data-consistent relaxation} if no data-consistent relaxation $C\in\mathfrak A$ is strictly stronger than $B$. Let $\mathcal R_P^{*}(A_0;\mathfrak A)$ denote the set of such bundles. The \emph{misspecification-robust (MR) bound} for the ATT is
\[
\Theta_P^{*}(A_0;\mathfrak A)
\equiv
\bigcup_{B\in\mathcal R_P^{*}(A_0;\mathfrak A)}\Theta_P(B).
\]
We write $\mathcal R^{*}(A_0)$ and $\Theta^{*}(A_0)$ after suppressing $P$ and the fixed menu $\mathfrak A$.
\end{definition}

{We use logical strength rather than the subset inclusion in \citet{desire2024qe} because each relaxation replaces a restriction with a weaker one. The minimum relaxations and MR bound depend on the chosen menu.}

{Data consistency depends on whether the pre-period OLS estimands are constant and whether $\Theta(A_{\text{AE-R}})$ is empty.} If {$\theta_{t_0}^{OLS}$ is constant over $t_0\in\mathcal I_0$, with common value $\bar s$}, define the common DID value
\[
\theta_{\mathrm{com}}^{DID}\equiv\theta_1^{OLS}-\bar s.
\]
When $0\in\mathcal I_0$, this equals the standard DID estimand $\theta^{DID}$. We restrict attention to $c\in(0,1)$ in Assumption \ref{ass:expoAE}; Remark \ref{rm:mr-degenerate} records the degenerate cases.

\begin{proposition}\label{prop:mrbound}
Suppose that model \eqref{seq1} holds and that $c\in(0,1)$ in Assumption \ref{ass:expoAE}. Then:
\begin{enumerate}[(i)]
\item $A_{\text{PT-R}}$ and $A_{\text{R}}$ are data-consistent at every $P$. The benchmark $A_0$ is data-consistent at $P$ if and only if $\theta_{t_0}^{OLS}$ is constant over $t_0\in\mathcal I_0$, in which case $\Theta(A_0)=\{\theta_{\mathrm{com}}^{DID}\}$. The bundle $A_{\text{AE-R}}$ is data-consistent at $P$ if and only if $\Theta^{\text{AE-RDID}}\neq\varnothing$, where $\Theta^{\text{AE-RDID}}$ is computed from $P$ by Proposition \ref{bound_multiple_preperiods}.
\item {In the three cases of \eqref{eq:mrbound-cases}, $\mathcal R^{*}(A_0)$ equals $\{A_0\}$, $\{A_{\text{PT-R}}\}$, and $\{A_{\text{PT-R}},A_{\text{AE-R}}\}$, respectively.}
\item Consequently,
\begin{equation}\label{eq:mrbound-cases}
\Theta^{*}(A_0)=
\begin{cases}
\Theta(A_0)=\{\theta_{\mathrm{com}}^{DID}\},
&\text{if $\theta_{t_0}^{OLS}$ is constant over $\mathcal I_0$},\\
\Theta(A_{\text{PT-R}}),
&\text{if $\theta_{t_0}^{OLS}$ is nonconstant and $\Theta(A_{\text{AE-R}})=\varnothing$},\\
\Theta(A_{\text{PT-R}})\cup\Theta(A_{\text{AE-R}}),
&\text{if $\theta_{t_0}^{OLS}$ is nonconstant and $\Theta(A_{\text{AE-R}})\neq\varnothing$}.
\end{cases}
\end{equation}
In every case, $\Theta^{*}(A_0)\neq\varnothing$ and
\[
\Theta^{*}(A_0)
\subseteq
\Theta(A_{\text{PT-R}})\cup\Theta(A_{\text{AE-R}})
\subseteq
\Theta(A_{\text{R}}).
\]
\end{enumerate}
\end{proposition}
A proof is relegated to Appendix \ref{sec:prop:mrbound}.

\begin{remark}\label{rm:mr-signdiag}
{By Lemma~\ref{lem:samesign}, conflicting signs of $\theta^{DID}_{t_0}$, $t_0\in\mathcal I_0$, imply $\Theta(A_{\text{AE-R}})=\varnothing$. Same-signed comparisons can also give an empty set under Proposition~\ref{bound_multiple_preperiods}. The regimes in \eqref{eq:mrbound-cases} must therefore be distinguished by emptiness of $\Theta(A_{\text{AE-R}})$, not by sign agreement alone.}
\end{remark}

\begin{remark}\label{rm:mr-degenerate}
If $c=0$, exponential $c$-bounded anticipation is equivalent to no anticipation, so $A_{\text{AE-R}}\sim A_0$ and $A_{\text{R}}\sim A_{\text{PT-R}}$; after identifying logically equivalent bundles, \eqref{eq:mrbound-cases} continues to apply. If $\mathcal I_0$ is a singleton, bias-set stability and strong PT coincide on $\mathcal I_0$, so $A_{\text{PT-R}}\sim A_0$ and the benchmark is data-consistent at every $P$; hence $\Theta^{*}(A_0)=\{\theta_{\mathrm{com}}^{DID}\}$.
\end{remark}
}

}

The three cases have direct interpretations. When $A_0$ is data-consistent (the standard DID model is not falsified), the misspecification-robust region collapses to the standard DID estimand. When only the PT-relaxation regime $A_{\text{PT-R}}$ is data-consistent, meaning that no candidate ATT under the No-AE-relaxation regime $A_{\text{AE-R}}$ is compatible with the observed pre-treatment DIDs, the region reduces to the PT-relaxation identified set alone. When both single-assumption regimes are data-consistent, the region reports the union of the two, reflecting the ambiguity between the two data-consistent relaxations.

The next section develops estimation and inference procedures for each of the three identified sets $\Theta^{\text{PT-RDID}}$, $\Theta^{\text{AE-RDID}}$, and $\Theta^{\text{RDID}}$.

\section{Estimation and Inference}\label{sec:inference}

We estimate the identified sets by replacing the population DID estimands with their sample counterparts. For a finite baseline set \(\mathcal I_0\), collect the baseline-specific DID estimands in \(\bm{\theta}^{{DID}}\equiv(\theta_{t_0}^{{DID}})_{t_0\in\mathcal I_0}\). We estimate these estimands jointly from a saturated treatment-by-period regression, with standard errors clustered at the independent sampling unit. The joint covariance accounts for the common post-treatment outcome gap. This construction concerns one focal post-treatment period without covariates. Covariate adjustment and pooling post-treatment periods require the corresponding identification and first-stage arguments. Appendix~\ref{app:inference-details} gives the full procedures and regularity conditions.

We distinguish confidence sets for the scalar ATT, denoted by $C^P_j$, from confidence envelopes for the entire identified set, denoted by $C^S_j$. An envelope must cover all admissible ATT values on the same sampling event. A scalar confidence set covers each admissible ATT at the stated level under the corresponding identifying restrictions. An envelope therefore also provides conservative ATT inference. Coverage is asymptotic as the number of independent clusters grows. We keep $|\mathcal I_0|$ fixed and restrict $c$ to $[0,\bar c]$, where $\bar c<1$.

For \(A_{\text{PT-R}}\), let \(g_r=\theta_{t_0}^{{DID}}\) for \(r=t_0\in\mathcal I_0\). For \(A_{\text{R}}\), let \(g_r=\theta_{t_0}^{{DID}}/(1+\sigma c_{t_0})\) for \(r=(t_0,\sigma)\in\mathcal I_0\times\{-1,+1\}\), where \(c_{t_0}=c^{|t_0|+1}\). In both cases, the identified set is \([\min_r g_r,\max_r g_r]\). If $s_r$ is the standard error of $\widehat g_r$, a level-$(1-\alpha)$ envelope is
\begin{equation}
C^S_j=\left[\min_r\{\widehat g_r-z_{1-\alpha/2}s_r\},\,
                 \max_r\{\widehat g_r+z_{1-\alpha/2}s_r\}\right],
\qquad j\in\{\text{PT-R},\text{R}\},
\label{eq:concise-union-envelope}
\end{equation}
where $z_p$ is the standard normal quantile. We use the same interval for $C^P_j$. Appendix~\ref{sec:union-inference} gives the endpoint coverage argument and the associated $p$-values.

For $A_{\text{AE-R}}$, an ATT value $\vartheta$ is admissible if $|\theta_{t_0}^{{DID}}-\vartheta|\leq c_{t_0}|\vartheta|$ for every $t_0\in\mathcal I_0$. We obtain $C^P_{\text{AE-R}}(c)$ by inverting tests of these inequalities using generalized moment selection \citep{AndrewsSoares2010}. For the entire identified set, we use the intersection-bound correction of \citet{chernozhukov2013intersection}, as implemented by \citet{CKLR2015}. Appendix~\ref{sec:inf2} gives both procedures and their conditions. An empty estimated identified set alone does not reject the model.

For the MR target in Proposition~\ref{prop:mrbound}, we use $C^P_*(c)=C^P_{\text{PT-R}}\cup C^P_{\text{AE-R}}(c)$ when $|\mathcal I_0|>1$. Each component is calibrated at level $1-\alpha$, with no further adjustment for this scalar target. When $|\mathcal I_0|=1$, the MR set is the single DID and we use $C^P_*=C^P_{\text{PT-R}}$. We do not select an MR regime by a preliminary test. An envelope for the entire MR set instead requires joint protection of its components, as described in Appendix~\ref{sec:mrinference}.

For simultaneous sensitivity analysis, we construct one joint confidence rectangle for $\bm \theta^{{DID}}$ using common normal multiplier draws. Projecting this rectangle through the identified-set maps gives bands that cover the identified sets simultaneously for all $c\in[0,\bar c]$ and all models. For MR, projecting the union of the two single-assumption sets gives a conservative confidence envelope. Appendix~\ref{sec:simultaneous-c-bands} gives the formulas. These bands provide simultaneous coverage beyond the intervals calibrated at a fixed $c$.

\section{Extension to Multiple Treatment Periods}\label{sec:extension}

{Throughout this section, $\mathcal T_0$ is nonempty and contains the reference period $0$. We allow $\mathcal T_0$ to be infinite. Proposition \ref{prop1ex} requires only that its displayed baseline extrema be finite, whereas Proposition \ref{bound_multiple} additionally requires every cohort-specific comparison set $\mathcal C_h$ to be finite.}

{

{

We generalize our analysis to a setting where treatment may be received in multiple periods. Let
\[
\mathcal D\equiv\{0\}\times\{0,1\}^T
\]
denote the treatment-path space, and consider the potential-outcome model
\begin{equation}\label{eq:ext}
Y_t
=
\sum_{\bm d\in\mathcal D}
Y_t(\bm d)\mathbbm{1}\{\bm D=\bm d\},
\end{equation}
for $t\in\mathcal T_0\cup\{1,\ldots,T\}$, where $\bm D\equiv(0,D_1,\ldots,D_T)$ is the observed treatment path and $Y_t(\bm d)\equiv Y_t(0,d_1,\ldots,d_T)$ is the potential outcome under $\bm d\equiv(0,d_1,\ldots,d_T)\in\mathcal D$.\footnote{See \cite{Robins1986, Robins1987}, and \cite{Han2021} for a similar definition of the potential outcome model.} We use $\bm d$ for an intervention path and $\bm d'\in\operatorname{supp}(\bm D)$ for the realized path defining a conditioning group. The fixed reference and target paths are $\bm d^0$ and $\bm d^1$.
}

In the above framework, the parameter of interest is the average treatment effect in period $t$ on the treated group following the path $\bm d^0 \equiv (0,d_1^0,\ldots, d_T^0)$ to $\bm d^1 \equiv (0,d_1^1,\ldots, d_T^1)$, which is defined as:$$
ATT_t[\bm d^0 \rightarrow \bm d^1] \equiv \mathbb E\left[Y_t(\bm d^1)-Y_t(\bm d^0) \vert \bm D=\bm d^1\right].
$$

Similarly to what we have in the one post-treatment setting, we can write the difference-in-means estimand $(\theta^{DIM}_t)$ between the two groups $\bm d^0$ and $\bm d^1$ in period $t$ as: 
\begin{align*}
\theta^{DIM}_t[\bm d^0 \rightarrow \bm d^1]&\equiv \mathbb E\left[Y_t \vert \bm D=\bm d^1\right]-\mathbb E\left[Y_t \vert \bm D=\bm d^0\right]\\
& =ATT_t[\bm d^0\rightarrow \bm d^1]+SB_t[\bm d^0\rightarrow \bm d^1], 
\end{align*}
where $SB_t[\bm d^0\rightarrow \bm d^1]\equiv \mathbb E\left[Y_t(\bm d^0) \vert \bm D=\bm d^1\right]-\mathbb E\left[Y_t(\bm d^0) \vert \bm D=\bm d^0\right]$.

\subsection{Testable Implication for Extended PT and No-AE}

{

\begin{assumption}[Extended Strong Parallel Trends]\label{ass:extpt}
Let $\bm d_{\infty}\equiv(0,\ldots,0)$. For every realized treatment path $\bm d'\in\operatorname{supp}(\bm D)$,
\[
\mathbb{E}[Y_t(\bm d_{\infty})-Y_0(\bm d_{\infty})\mid\bm D=\bm d']
=
\mathbb{E}[Y_t(\bm d_{\infty})-Y_0(\bm d_{\infty})]
\]
for all $t\in\mathcal T_0\cup\{1,\ldots,T\}$.
\end{assumption}

For $t\in\mathcal T_0\cup\{1,\ldots,T\}$, define the set of paths with no treatment through period $t$ by
\[
\mathcal D_t^0
\equiv
\left\{\bm d=(0,d_1,\ldots,d_T):d_s=0
\text{ for every treatment period }s\leq t\right\}.
\]
When $t\leq0$, the condition on $d_s$ is vacuous and $\mathcal D_t^0=\mathcal D$.

\begin{assumption}[Extended No Anticipation]\label{ass:extnoae}
For every $t\in\mathcal T_0\cup\{1,\ldots,T\}$, every $\bm d\in\mathcal D_t^0$, and every conditioning path $\bm d'\in\operatorname{supp}(\bm D)$,
\[
\mathbb{E}[Y_t(\bm d)-Y_t(\bm d_{\infty})\mid\bm D=\bm d']=0.
\]
\end{assumption}
}

Assumption \ref{ass:extpt} states that the average outcomes of all groups at any time would have followed the same trend in the absence of treatment. 
Assumption \ref{ass:extnoae} states that for any group of people, if they have not been treated by period $t$, no matter what happens later, the average potential outcome at $t$ is the same as the average potential outcome of never being treated. 
In other words, if individuals have not received treatment before period $t$, they will not expect any treatment in the future to have an impact on the current situation. 
In this case, the future treatment does not affect the current outcome. 
However, if individuals have been treated before period $t$, they might have anticipation effects if this happens again in the future, then the future treatment will potentially affect the current outcome.

{

\begin{proposition}\label{multiTI}
Suppose Assumptions \ref{ass:extpt} and \ref{ass:extnoae} hold. For every $t\in\mathcal T_0\cup\{1,\ldots,T\}$, every $\bm d'\in\mathcal D_t^0\cap\operatorname{supp}(\bm D)$, and $\bm d_{\infty}\in\operatorname{supp}(\bm D)$,
\begin{equation}
\label{implication}
\mathbb{E}[Y_t-Y_0\mid\bm D=\bm d']
=
\mathbb{E}[Y_t-Y_0\mid\bm D=\bm d_{\infty}].
\end{equation}
\end{proposition}
}

{ A proof is relegated to Appendix \ref{sec:multiTI}.}

Proposition \ref{multiTI} implies that all ``not-yet-treated'' groups at period $t$, including the never-treated, have equal (linear) trends between periods 0 and $t$. The implication (\ref{implication}) is testable since all variables and groups involved are observed.

\begin{remark}
In the staggered adoption design, we propose the following test procedure. Suppose we observe $T$ treatment periods. Define\\ \vspace{0.1cm}
$D^{g} \equiv\mathbbm{1}\{\bm D =(0,0,...,d_{g}=1,...,1)\}$ \\ 
\vspace{0.1cm}
$D^{0} \equiv\mathbbm{1}\{\bm D =(0,...,0)\}$ \\ 
\vspace{0.1cm}
$\theta_{s}^{g} \equiv\left(\mathbb{E}[Y_{it}\mid D_{i}^{g}=1,t=s]-\mathbb{E}[Y_{it}\mid D_{i}^{0}=1,t=s]\right)-\left(\mathbb{E}[Y_{it}\mid D_{i}^{g}=1,t=0]-\mathbb{E}[Y_{it}\mid D_{i}^{0}=1,t=0]\right)$
    where $g$ refers to the first time being treated, $i$ refers to individual $i$. $D_i^g=1$ if individual $i$ is treated at period $g$ for the first time, and we say individual $i$ is in group $g$. Then, we can run the following regression (TWFE approach) to obtain the parameter $\theta_{s}^{g}$ (see \citealp{wooldridge2021two,ban2022generalized}); for $t\in\{0,1,...,T\}$,
\begin{equation}
\label{Robust_TWFE}
Y_{it}=\beta+\sum_{g=1}^{T}\gamma^{g}D_{i}^{g}+\sum_{s=1}^{T}\delta_{s}\mathds{1}\{t=s\}+\sum_{s=1}^{T}\sum_{g=1}^{T} { \theta_{s}^{g}}  
   D_{i}^{g}\mathds{1}\{t=s\}+\varepsilon_{it}
\end{equation} By Proposition \ref{multiTI}, if Assumptions \ref{ass:extpt} and \ref{ass:extnoae} hold, $\theta_{s}^{g}=0$
when $g>s$. 
A joint Wald test of $\theta_s^g=0$ for all observed cohorts and periods with $1\leq s<g\leq T$ tests a necessary implication of Assumptions~\ref{ass:extpt} and~\ref{ass:extnoae}. Rejection is evidence that at least one assumption fails.
\end{remark}

\subsection{Multi-period Extension of PT-RDID}\label{sec:rdid1multi}

Consider a weaker version of Assumption \ref{ass:extnoae} where individuals are now allowed to anticipate the effects of the treatment for the rest of the period after 0.\footnote{Note that we do not require Assumption \ref{ass:extnoae} to hold for all time periods but only for $t_0 \in \mathcal T_0$. That is, we allow the individuals to anticipate the effects of the treatment for the rest of the period. This is because Assumption \ref{sb:boundsex} restricts each post-treatment bias using only the pre-treatment biases. This assumption is less restrictive than the commonly used no-anticipatory effects assumption.}

{

\begin{assumption}[Extended No Anticipation in Pre-treatment Periods]\label{ass:extnoae0}
For every $t_0\in\mathcal T_0$, every treatment path $\bm d\in\mathcal D$, and every conditioning path $\bm d'\in\operatorname{supp}(\bm D)$,
\[
\mathbb{E}[Y_{t_0}(\bm d)-Y_{t_0}(\bm d_{\infty})\mid\bm D=\bm d']=0,
\]
where $\bm d_{\infty}\equiv(0,\ldots,0)$.
\end{assumption}
}

The following Lemma \ref{lem:bsmulti} shows that for any paths of interest $\bm d^0$ and $\bm d^1$, the pre-treatment period selection bias, $SB_{t_0}[\bm d^0\rightarrow \bm d^1]$, is identified as $\theta^{DIM}_{t_0}[\bm d^0 \rightarrow \bm d^1]$.

\begin{lemma}\label{lem:bsmulti}
    For any paths of interest $\bm d^0 \equiv (0,d_1^0,\ldots, d_T^0)$ and $\bm d^1 \equiv (0,d_1^1,\ldots, d_T^1)$ satisfying $\mathbb P(\bm D=\bm d^0)>0$ and $\mathbb P(\bm D=\bm d^1)>0$, Assumption \ref{ass:extnoae0} implies
    $$
    SB_{t_0}[\bm d^0\rightarrow \bm d^1] = \theta^{DIM}_{t_0}[\bm d^0 \rightarrow \bm d^1],
    $$
    for all $t_0 \in \mathcal T_0$.
\end{lemma}
{ A proof is relegated to Appendix \ref{sec:lem:bsmulti}.}
Accordingly, we extend Assumption \ref{sb:bounds}. 

\begin{assumption}[Extended bias set stability]\label{sb:boundsex}
For each $t \in \{1, \ldots, T\}$, 
\begin{align*}
SB_t[\bm d^0\rightarrow \bm d^1] &\in \left[ \inf_{t_0\in \mathcal T_0} SB_{t_0}[\bm d^0\rightarrow \bm d^1], \sup_{t_0 \in \mathcal T_0} SB_{t_0}[\bm d^0\rightarrow \bm d^1]\right] \\
&\equiv \Delta_{SB(\mathcal T_0)}[\bm d^0\rightarrow \bm d^1],
\end{align*}
where $SB_{t_0}[\bm d^0\rightarrow \bm d^1]$ is the baseline selection bias with respect to the treatment status in period $t_0 \in \mathcal T_0$ when the information set $\mathcal I_0$ is equal to $\mathcal T_0$. 
\end{assumption} 
Assumption \ref{sb:boundsex} is a generalization of Assumption \ref{sb:bounds}, and we have the following identification results for $ATT_t[\bm d^0 \rightarrow \bm d^1]$ with different treatment paths.

\begin{proposition}\label{prop1ex}
Suppose model \eqref{eq:ext} and Assumptions \ref{ass:extnoae0} and \ref{sb:boundsex} hold, $\mathbb P(\bm D=\bm d^0)>0$, $\mathbb P(\bm D=\bm d^1)>0$, and the displayed baseline extrema are finite. Then the following bounds are valid for $ATT_t$: 
\begin{align*}
&ATT_t[\bm d^0 \rightarrow \bm d^1] \\
&\qquad \in \left[ \theta^{DIM}_t[\bm d^0\rightarrow \bm d^1] - \sup_{t_0 \in \mathcal T_0} \theta^{DIM}_{t_0}[\bm d^0 \rightarrow \bm d^1], \theta^{DIM}_t[\bm d^0\rightarrow \bm d^1]- \inf_{t_0 \in \mathcal T_0} \theta^{DIM}_{t_0}[\bm d^0 \rightarrow \bm d^1] \right], \\
& \qquad \equiv \Theta_t^{\text{PT-R}}[\bm d^0 \rightarrow \bm d^1].
\end{align*}
These bounds are sharp, and $\Theta_t^{\text{PT-R}}[\bm d^0 \rightarrow \bm d^1]$ is the identified set for $ATT_t[\bm d^0 \rightarrow \bm d^1]$.\end{proposition}
{ A proof is relegated to Appendix \ref{sec:prop1ex}.}
In a staggered design setting, the average treatment effect in period $t$ on units who are treated for the first time in period $g$ could be an interesting parameter, as considered in \cite{CallawaySantAnna2021}:
 $$ATT_t^g \equiv ATT_t[(0,\ldots, 0)\rightarrow (0,\ldots,0,d_g=1,1,\ldots, 1)].$$  

By defining $G$ as
$$
G = \left\{\begin{array}{ccl}
g & \mathrm{if} & \bm D = (0, \ldots, 0, d_g = 1, 1, \ldots, 1) \\
\infty & \mathrm{if} & \bm D = (0, \ldots, 0)
\end{array}\right. ,
$$
where $G = \infty$ denotes the never-treated cohort as our control group, we can write
\begin{align*}
\theta_t^{DIM, g} &\equiv  \theta_t^{DIM}[(0, \ldots, 0)\rightarrow (0, \ldots, 0, d_g = 1, 1, \ldots, 1)] \\
&= \mathbb{E}[Y_t \vert G = g] - \mathbb{E}[Y_t \vert G = \infty] \\
SB_t^g &\equiv SB_t[(0, \ldots, 0)\rightarrow (0, \ldots, 0, d_g = 1, 1, \ldots, 1)] \\
	&= \mathbb{E}[Y_t(\infty) \vert G = g] - \mathbb{E}[Y_t(\infty) \vert G = \infty],
\end{align*}
for $t \in \mathcal T_0 \cup \{1, \ldots, T \}$, by abusing the notations of $g$ and $\infty$ for paths of potential outcomes.

Hence, Assumption \ref{sb:boundsex} can be re-written as
\begin{align*}
SB_t^g &\in \left[ \inf_{t_0\in \mathcal T_0} SB_{t_0}^g, \sup_{t_0 \in \mathcal T_0} SB_{t_0}^g\right] \\
&\equiv \Delta_{SB(\mathcal T_0, g)},
\end{align*}
for each $t \in \{1, \ldots, T \}$, and $SB_{t_0}^g$ is identified as $\theta_{t_0}^{DIM, g}$ for $t_0 \in \mathcal T_0$.
Thus, we have 
\begin{align*}
ATT_t^g &\in \left[ \theta^{DIM, g}_t - \sup_{t_0 \in \mathcal T_0} \theta_{t_0}^{DIM, g}, \theta^{DIM, g}_t - \inf_{t_0 \in \mathcal T_0} \theta_{t_0}^{DIM, g} \right].\\
&\equiv \Theta_t^g
\end{align*}

}

{ 
\subsection{Multi-period Extension of AE-RDID with Staggered Adoption Design}

In the setting of multiple post-treatment periods in Section \ref{sec:rdid1multi}, we considered scenarios where individuals are allowed to anticipate after the first time the initial treated group receives treatment (Assumption \ref{ass:extnoae0}).
However, we can relax further, allowing people to anticipate even earlier,\footnote{It is plausible that people begin to anticipate after observing other groups being treated. However, it is also possible that some people change their behavior even before the first treatment of other groups. This could be because of having inside information, being influenced by the congressional negotiation process, or exposed to some rumors.} while maintaining Assumption \ref{ass:extpt} (Extended Strong Parallel Trends).

In the staggered adoption design, recall that
\begin{align*}
ATT_t^g &\equiv ATT_t[\bm d_{\infty} \rightarrow \bm d^1], \\
\theta_t^{DIM, g} &\equiv  \theta_t^{DIM}[\bm d_{\infty} \rightarrow \bm d^1] \\
SB_t^g &\equiv SB_t[\bm d_{\infty} \rightarrow \bm d^1]\textbf{}
\end{align*}
where $\bm d_{\infty} \equiv (0, \ldots, 0)$ and $\bm d^1 = (0,\ldots,0,d_g=1,1,\ldots, 1)$, and we define 
\begin{align*}
    AE_t^g &\equiv AE_{t}[\bm d_{\infty} \rightarrow \bm d^1], \\
    &= AE_{t}[(0, \ldots, 0) \rightarrow (0,\ldots,0,d_g=1,1,\ldots, 1)],
\end{align*}
accordingly. 

\begin{assumption}[Extended bounded anticipation effects]
\label{extendedBAE} For some known $c\in[0,1)$, given any cohort $g\in \{1, \ldots, T\}$,
$$
\big| AE_t^g\big| \leq  c^{g-t}\big| ATT_g^g\big|,$$ 
for all $ t < g$.
\end{assumption}

In a forward-looking model, the effect at adoption can reflect both the contemporaneous effect and expected future treatment effects. This motivates using $|ATT_g^g|$ as the scale for the bound on average anticipation before adoption.

On the other hand, we have the following Lemma \ref{lem2} suggesting that the extended parallel trends assumption is equivalent to the stability of selection bias, a perspective consistently used throughout our paper.

\begin{lemma}\label{lem2}
Suppose $\operatorname{supp}(G)\subseteq\{1,\ldots,T,\infty\}$ and $\mathbb P(G=\infty)>0$. Then Assumption \ref{ass:extpt} is equivalent to
\[
SB_t^g=SB_0^g
\]
for every finite $g\in\operatorname{supp}(G)$ and every $t\in\mathcal T_0\cup\{1,\ldots,T\}$.\end{lemma}

{ A proof is relegated to Appendix \ref{sec:lem2}.}

{

Since we are interested in effects after adoption, we use the preceding assumptions to identify $ATT_t^g$ for $t\geq g$. For $t,t'\in\mathcal T_0\cup\{1,\ldots,T\}$, define
\[
\theta_{t,t'}^{DID,g}
\equiv
\theta_t^{DIM,g}-\theta_{t'}^{DIM,g}.
\]
For each finite $g\in\operatorname{supp}(G)$, define the set of observed comparison periods by
\[
\mathcal C_g
\equiv
\{t_0\in\mathcal T_0\cup\{1,\ldots,T\}:t_0<g\}.
\]
When $\mathcal C_g$ is nonempty and finite, define the cohort-specific feasibility set
{\begin{equation}\label{eq:cohort-feasibility-definition}
\mathcal F_g(c)
\equiv
\left\{
\vartheta\in\mathbb R:
\left|\vartheta-\theta_{g,t_0}^{DID,g}\right|
\leq c^{g-t_0}|\vartheta|
\text{ for every }t_0\in\mathcal C_g
\right\}.
\end{equation}
Thus $\mathcal F_g(c)$ is exactly the set of candidate values for $ATT_g^g$ that satisfy every cohort-$g$ bounded-anticipation restriction together with extended PT. Solving \eqref{eq:cohort-feasibility-definition} separately by the sign of $\vartheta$ gives}
\begin{align}
\mathcal F_g(c)
={}&
\left[
\max_{t_0\in\mathcal C_g}
\left\{\frac{\theta_{g,t_0}^{DID,g}}{1+c^{g-t_0}}\right\},
\min_{t_0\in\mathcal C_g}
\left\{\frac{\theta_{g,t_0}^{DID,g}}{1-c^{g-t_0}}\right\}
\right]
\nonumber\\
&\quad\cup
\left[
\max_{t_0\in\mathcal C_g}
\left\{\frac{\theta_{g,t_0}^{DID,g}}{1-c^{g-t_0}}\right\},
\min_{t_0\in\mathcal C_g}
\left\{\frac{\theta_{g,t_0}^{DID,g}}{1+c^{g-t_0}}\right\}
\right].
\label{eq:cohort-feasibility-explicit}
\end{align}

\begin{proposition}\label{bound_multiple}
Suppose $\operatorname{supp}(G)\subseteq\{1,\ldots,T,\infty\}$ and $\mathbb P(G=\infty)>0$, and suppose $\mathcal C_h$ is nonempty and finite for every finite $h\in\operatorname{supp}(G)$. Under Assumptions \ref{ass:extpt} and \ref{extendedBAE}, for every finite $g\in\operatorname{supp}(G)$ and $t\geq g$,
\[
ATT_t^g
\in
\theta_{t,g}^{DID,g}+\mathcal F_g(c).
\]
Equivalently,
{\small
\begin{align}
ATT_t^g\in{}&
\left[
\theta_{t,g}^{DID,g}
+\max_{t_0\in\mathcal C_g}
\left\{\frac{\theta_{g,t_0}^{DID,g}}{1+c^{g-t_0}}\right\},
\theta_{t,g}^{DID,g}
+\min_{t_0\in\mathcal C_g}
\left\{\frac{\theta_{g,t_0}^{DID,g}}{1-c^{g-t_0}}\right\}
\right]
\nonumber\\
&\quad\cup
\left[
\theta_{t,g}^{DID,g}
+\max_{t_0\in\mathcal C_g}
\left\{\frac{\theta_{g,t_0}^{DID,g}}{1-c^{g-t_0}}\right\},
\theta_{t,g}^{DID,g}
+\min_{t_0\in\mathcal C_g}
\left\{\frac{\theta_{g,t_0}^{DID,g}}{1+c^{g-t_0}}\right\}
\right].
\label{eq:staggered-att-set}
\end{align}}
In particular, $ATT_g^g\in\mathcal F_g(c)$, whose explicit form is given in \eqref{eq:cohort-feasibility-explicit}. If $\mathcal F_h(c)=\varnothing$ for any finite $h\in\operatorname{supp}(G)$, the joint model is falsified and every identified set under the maintained model is empty. If $\mathcal F_h(c)\neq\varnothing$ for every such cohort, the displayed bounds are sharp.\end{proposition}
A proof is relegated to Appendix \ref{sec:bound_multiple}.
}
\begin{remark}
    Note that the same argument as in Lemma \ref{lem:samesign} can be applied in this staggered setting. 
    Specifically, for any group $g\in \{1, \ldots, T\}$ under Assumptions \ref{extendedBAE} and \ref{ass:extpt}, we have
    $$
    \operatorname{sgn}(\theta_{g,t}^{DID,g}) = \operatorname{sgn}(ATT_g^g),
    $$
    for all $t < g$.
    This further implies that the bounds in Proposition \ref{bound_multiple} are always convex:
    \begin{align*}
        ATT_t^g \in \left\{\begin{array}{lcl}
            \Big[ \theta_{t,g}^{DID,g}+\max_{t_{0}<g}\Big\{ \frac{\theta_{g,t_{0}}^{DID,g}}{1+c^{g-t_{0}}}\Big\} , \theta_{t,g}^{DID,g}+\min_{t_{0}<g}\Big\{ \frac{\theta_{g,t_{0}}^{DID,g}}{1-c^{g-t_{0}}}\Big\} \Big] & \mathrm{if} & ATT_g^g \geq 0 \\
            \Big[\theta_{t,g}^{DID,g}+\max_{t_{0}<g}\Big\{ \frac{\theta_{g,t_{0}}^{DID,g}}{1-c^{g-t_{0}}}\Big\}  ,\theta_{t,g}^{DID,g} +\min_{t_{0}<g}\Big\{ \frac{\theta_{g,t_{0}}^{DID,g}}{1+c^{g-t_{0}}}\Big\}  \Big]  & \mathrm{if} & ATT_g^g < 0 
        \end{array}\right.
    \end{align*}
\end{remark}

}

\section{Empirical illustrations}\label{sec:empirical}

In this section, we illustrate our theoretical results using three empirical applications. Our objective is to demonstrate the usefulness of the proposed methods and provide guidance on how researchers may use them. We revisit Colombia's coca substitution program \citep{ladino2021one}, the introduction of charges for plastic bags in Uruguay \citep{ccc}, and compulsory licensing of German patents after World War I \citep{baten}. These applications demonstrate how researchers can use our proposed methods to assess the robustness of their conclusions to potential violations of their identifying assumptions.  

For each application, we first replicate the key published findings based on our analysis sample. We then report estimated ATT sets under bias set stability with no anticipation ($A_{\text{PT-R}}$), strong parallel trends with bounded anticipation ($A_{\text{AE-R}}$), and both replacement restrictions ($A_{\text{R}}$). We also report the misspecification-robust region developed in Section~\ref{subsec:misspec-robust}. For the $A_{\text{PT-R}}$ comparisons using smaller baseline information sets, we retain the regression fit from the full analysis sample and vary only the baseline periods used to construct the bounds.

For Uruguay and Germany, the regressions estimate a common post-treatment coefficient across the implementation window. A pooled ATT interpretation requires the adjusted post-treatment gap to equal the window's ATT plus its untreated selection bias. We impose the bias restrictions on this pooled bias and bound pre-treatment anticipation relative to the absolute pooled ATT.

We compare our results in the main text based on $c=0.5$, with additional $A_{\text{AE-R}}$ sensitivity results reported in Appendix~\ref{app:empirical-diagnostics}. For each application, let \(h=1-t_0\) denote the number of observation periods from baseline \(t_0\leq0\) to the first post-treatment date, indexed by \(1\). Thus \(h=1\) for the last baseline. We count years for coca cultivation and patents and months for plastic bags. Under this $c=0.5$ benchmark, the magnitude of anticipation at baseline \(t_0\) is bounded by \(0.5^h\) times the absolute ATT. Appendix~\ref{app:empirical-diagnostics} varies \(c\) under \(A_{\text{AE-R}}\), maintaining strong parallel trends, to assess sensitivity to varying degrees of allowed anticipation effects. Appendix~\ref{app:empirical-diagnostics} also reports observed group means, full pre-treatment diagnostics, and confidence envelopes for the entire identified set.

Figure~\ref{fig:application-baselines} reports DID estimates that compare a common post-treatment treated--control outcome gap with the gap in each pre-treatment baseline period. Note that typical event-study pre-treatment coefficients compare each pre-treatment gap with a fixed pre-treatment reference gap.\footnote{Figure~\ref{fig:application-eventstudy} in that appendix reports the corresponding pre-treatment event-study style estimates.} Under no anticipation in the selected baseline periods, the DID estimates in Figure~\ref{fig:application-baselines} satisfy
\[
\theta_1^{OLS}-\theta_{t_0}^{OLS}=(ATT+SB_1)-SB_{t_0}.
\]
As such, the variation in the pre-treatment estimates in Figure~\ref{fig:application-baselines} traces the pre-treatment selection biases after a common shift and reversal of sign. Their smallest and largest values over $\mathcal I_0$ give the endpoints of the $A_{\text{PT-R}}$ identified set. These estimates reflect the variation in the historical bias set stability and its implications for the ATT. Together with institutional knowledge, the historical pattern helps assess whether the post-treatment bias could reasonably lie within the selected baseline range.

\subsection{Effect of Colombia's coca substitution program}

\citet{ladino2021one} examine how the announcement and implementation of Colombia's National Comprehensive Program for the Substitution of Illicit Crops (PNIS) affected coca cultivation. The program offered payments and support for alternative livelihoods to farmers that agreed to eradicate coca. The PNIS program was announced in 2014 and was implemented in 2017. The eligibility for the program  depended on how much coca the farmers had cultivated \textit{after} the announcement (late 2016 and early 2017) but \textit{before} the implementation in 2017. These features created an incentive to cultivate coca in anticipation of future payments. The authors find that the announcement increased cultivation in areas with FARC presence, while implementation reduced cultivation among program recipients. This setting is therefore useful for examining how an implementation result changes when we relax both no-anticipation effects and parallel trends assumptions.

Our illustration focuses on the 2018 implementation effect using annual observations of coca hectares in one-square-kilometer grids from 2010 to 2018. The analysis compares grids first paid in 2018 with grids not paid through 2018. We exclude 148 grids already treated in 2017 to focus on the cohort first treated in 2018, ensuring that its baseline observations precede implementation.\footnote{\citet{ladino2021one} note that 2017 was the first year in which PNIS was ``effectively implemented'' and ``8.8\% of the grids received PNIS payments." As such, 2017 outcomes for already-paid grids can contain actual implementation effects, and we use 2017 purely as a pre-implementation baseline for the 2018 cohort.} The benchmark finding for our comparison is the simple implementation specification in Table 3, column 1 of the \citet{ladino2021one}, which reports a reduction of 0.880 hectares per grid.\footnote{The original article's preferred specification also includes an interaction for grids neighboring recipients and reports an implementation estimate of -1.04 hectares. Our illustration concerns the selected implementation comparison; the original announcement analysis uses a different treatment definition.} 
Table~\ref{tab:ladino-results} reports our standard DID replication estimate of a -0.827 hectares on our analysis sample, which is statistically significant and consistent with the original article.

\begin{figure}[p]
\centering
\captionsetup{font={small,stretch=1},skip=3pt}
\captionsetup[subfigure]{labelfont=normalfont,font={small,stretch=1}}
\begin{subfigure}{.76\textwidth}
\centering
\begin{tikzpicture}
\node[anchor=south west,inner sep=0] (ladinoplot) at (0,0) {\includegraphics[width=\linewidth,height=.23\textheight,keepaspectratio,trim=0 66bp 0 6bp,clip]{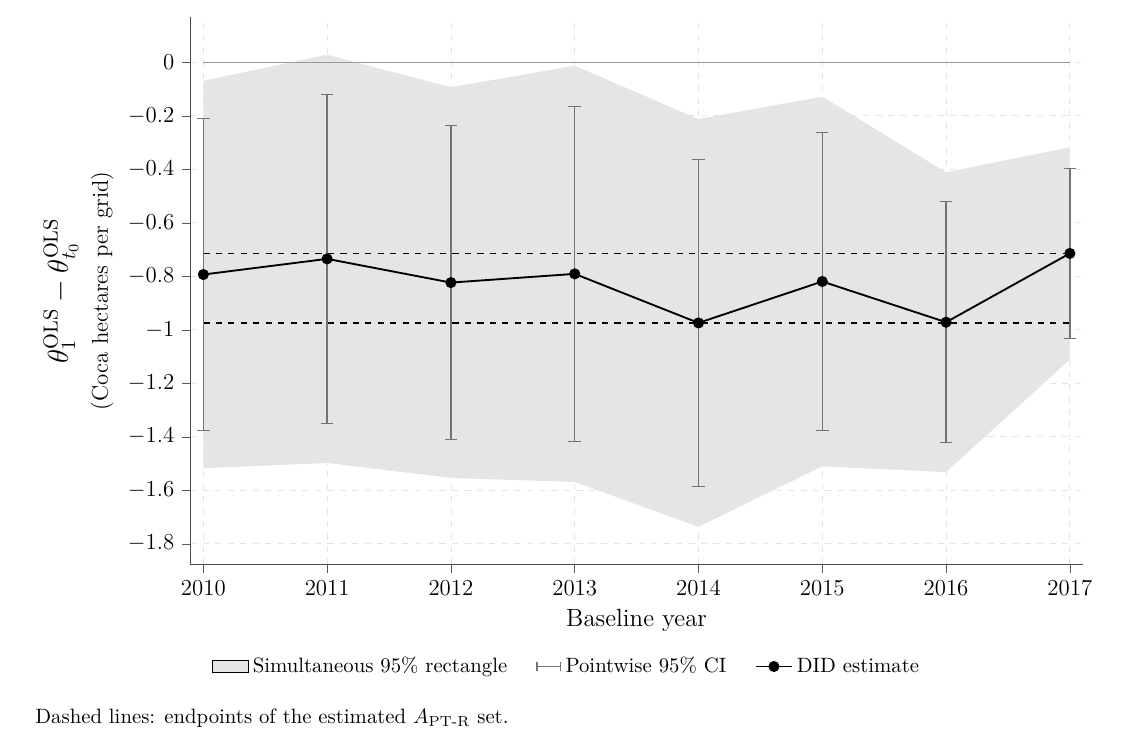}};
\begin{scope}[x={(ladinoplot.south east)},y={(ladinoplot.north west)}]
\draw[black,densely dashed,thin] (.614,.085) -- (.614,.98);
\node[black,anchor=north east,font=\scriptsize,fill=white,inner sep=1pt] at (.945,.97) {2014 announcement};
\end{scope}
\end{tikzpicture}
\caption{Colombia's coca substitution program (baseline year)}
\end{subfigure}
\par
\begin{subfigure}{.76\textwidth}
\centering
\includegraphics[width=\linewidth,height=.23\textheight,keepaspectratio,trim=0 66bp 0 6bp,clip]{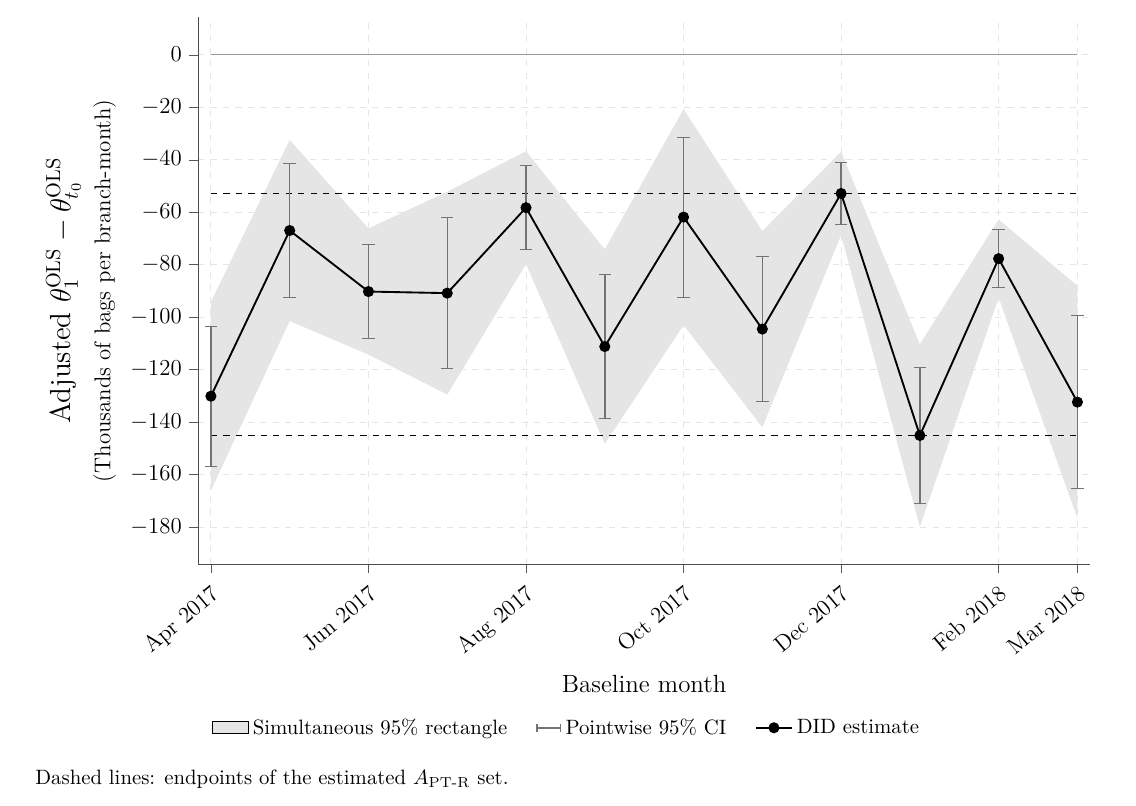}
\caption{Plastic-bag charges in Uruguay (baseline month)}
\end{subfigure}
\par
\begin{subfigure}{.76\textwidth}
\centering
\includegraphics[width=\linewidth,height=.23\textheight,keepaspectratio,trim=0 66bp 0 6bp,clip]{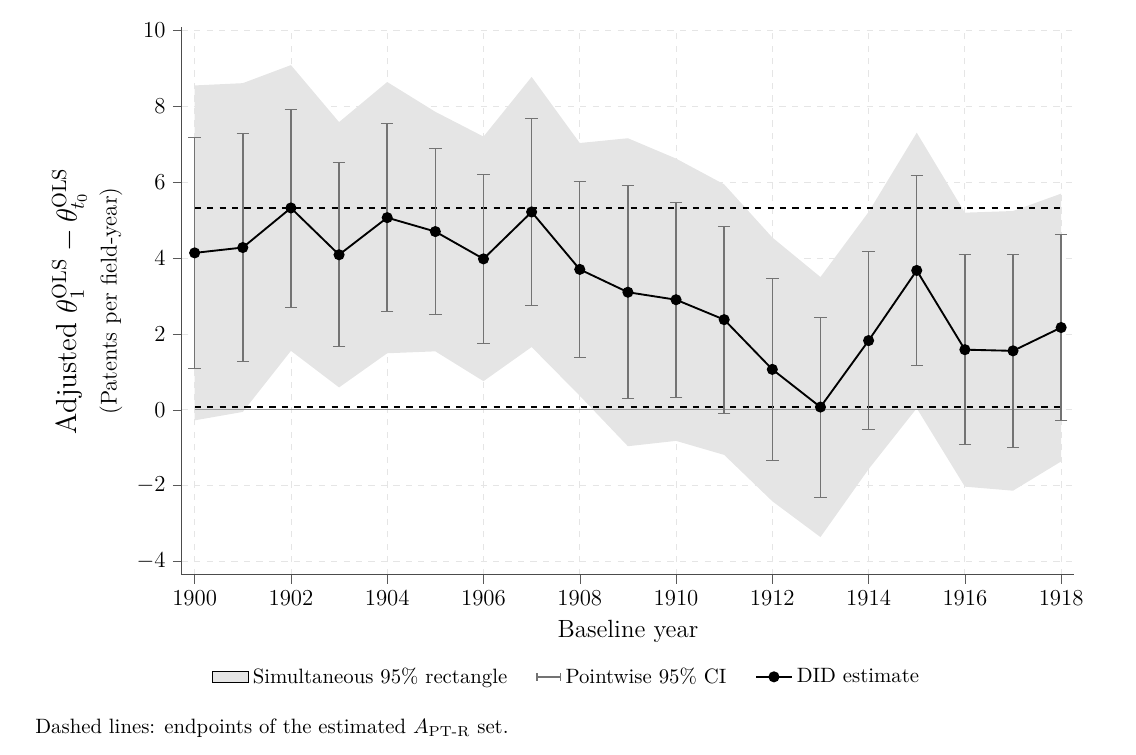}
\caption{Compulsory licensing in Germany (baseline year)}
\end{subfigure}
\caption{Figure reports baseline-specific, pre-treatment DID estimates and ATT bounds under bias set stability. Each point uses the same post-treatment period and a different baseline and are different from typical event-study pre-treatment coefficients. Horizontal dashed lines reproduce the estimated $A_{\text{PT-R}}$ bounds using all displayed baselines in Tables~\ref{tab:ladino-results}--\ref{tab:baten0923-results}. Vertical bars show pointwise 95\% confidence intervals; gray shading shows simultaneous 95\% bounds for the DID estimates. The vertical dashed line in (a) marks the 2014 announcement. Units: coca hectares per grid (a), thousands of bags per branch-month (b), and patents per field-year (c).}
\label{fig:application-baselines}
\end{figure}

\FloatBarrier

\begin{table}[H]
\centering\small
\caption{Effect of PNIS on coca cultivation (2018 treatment design)}
\label{tab:ladino-results}
\begin{threeparttable}
\begin{tabular}{lccc}
\toprule
Source or model & \shortstack{Estimate (SE)\\or identified set} & \shortstack{95\% ATT\\confidence set} & \(p\)-value \\
\midrule
Original article &  -0.880 (  0.230) & Not reported & \(<0.01\) \\
Same-sample replication &  -0.827 (  0.241) & [ -1.301,  -0.354] & 0.0007 \\
RDID common \(A_0\) &  -0.828 (  0.241) & [ -1.299,  -0.356] & 0.0006 \\
\addlinespace
\(A_{\text{PT-R}}\), baseline 2010--2017 & \([-0.974, -0.714]\) & \([-1.587, -0.121]\) & 0.0190 \\
\(A_{\text{PT-R}}\), baseline 2010--2015 & \([-0.974, -0.735]\) & \([-1.587, -0.121]\) & 0.0190 \\
\(A_{\text{PT-R}}\), baseline 2010--2013 & \([-0.824, -0.735]\) & \([-1.417, -0.121]\) & 0.0190 \\
\(A_{\text{AE-R}}\) & \(\varnothing\) & \([-1.470, -0.346]\) & \(\leq 0.0001\) \\
\(A_{\text{R}}\) & \([-1.429, -0.476]\) & \([-2.067, -0.120]\) & 0.0190 \\
MR & \([-0.974, -0.714]\) & \([-1.587, -0.121]\) & 0.0190 \\
\bottomrule
\end{tabular}

\begin{tablenotes}[flushleft]
\footnotesize\setstretch{1}\setlength{\baselineskip}{11pt}
\item \textbf{Notes:} The outcome is coca hectares per $1$ km$^2$ grid. The first row reports Table 3, column 1 of \citet{ladino2021one}. Other rows use 2010--2018 observations, comparing grids first paid in 2018 with grids unpaid through 2018. Baseline years specify $\mathcal I_0$. The anticipation restrictions use $c=0.5$. Parentheses contain standard errors; brackets report estimated ATT sets in column 2 and 95\% ATT confidence sets in column 3. The $p$-values test $ATT=0$; $\varnothing$ denotes an empty estimated set.
\end{tablenotes}
\end{threeparttable}
\end{table}

In their original event study analysis, the authors write that ``no coefficient before PNIS implementation is statistically significant, supporting the parallel trends assumption'' \citep[p.~7]{ladino2021one}. Our proposed bounds provide complementary evidence by examining the robustness of the the program's effects under alternative restrictions on anticipation and selection bias.

Figure \ref{fig:application-baselines}(a) reports DID estimates using 2018 as the post-treatment year and each year from 2010 to 2017 as the baseline. Differences across these estimates reflect variation in the treated–control outcome gap during the pre-treatment period. All eight baseline years yield negative estimates and a joint test rejects equality across these eight estimates at the 5\% level ($p=0.0141$; Appendix~\ref{app:ladino-diagnostics}, Table~\ref{tab:ladino-pretests}). The pre-treatment gaps do not show a sustained upward or downward trend. This relative stability lends support to our bias set stability assumption. That is, the historical range of selection biases provides a reasonable basis for bounding the post-treatment bias and, consequently, the ATT.

Table~\ref{tab:ladino-results} reports our bounds under the alternative identifying restrictions. When we replace parallel trends with bias set stability while maintaining no anticipation ($A_{\text{PT-R}}$), the estimated ATT is between -0.974 and -0.714 hectares per grid. Allowing departures from both parallel trends and no anticipation under the joint replacement restrictions ($A_{\text{R}}$) widens the estimated interval to [-1.429,-0.476] hectares per grid. The corresponding 95\% ATT confidence set is [-2.067,-0.120], which remains entirely negative. Thus, the finding that implementation reduced coca cultivation remains statistically significant when both assumptions are relaxed.

We also use contextual information from the orginal article to vary the baseline information set under $A_{\text{PT-R}}$. We consider 2010--2015, which excludes the years surrounding eligibility determination and enrollment, and 2010--2013, which excludes all years from the 2014 announcement onward. As such, the latter reasonably places the no-anticipation effects assumption entirely on periods preceding the public announcement. The estimated ATT sets are [-0.974,-0.735] and [-0.824,-0.735] hectares per grid, respectively, and continue to remain entirely negative.

The estimated misspecification-robust region (MR) also equals the $A_{\text{PT-R}}$ interval in this application. It reduces to this interval because the estimated set under strong parallel trends and bounded anticipation ($A_{\text{AE-R}}$) is empty at $c=0.5$. The reported $A_{\text{R}}$ and MR results continue to use the full 2010--2017 baseline information set. Taken together, the results provide a useful robustness finding the range of possible effect magnitudes for the program's effect is wider, albeit the conclusion that implementation reduced coca cultivation remains unchanged.

\subsection{Pricing plastic bags in Uruguay}

\citet{ccc} examine whether modest charges for disposable plastic bags reduce their use when competing stores may continue providing bags free of charge. We focus on the first implementation wave in Salto, where three supermarket branches began charging two Uruguayan pesos per bag in April 2018. The initiative was announced in December 2017, which gave customers time to stock up on free bags before the charge began. The authors find increased  use of bags during the announcement period, followed by a substantial reduction after implementation. This setting is useful for assessing the sensitivity of these findings to potential violations of the no-anticipation effects assumption.

In their discussion of non-parallel pre-treatment event study coefficients immediately before implementation, the authors note that: ``[t]hese could be the consequence of anticipation effects'' \citep[p.~11]{ccc}. They subsequently discuss the rise in bag use before pricing and conclude: ``[w]e cannot rule out this possibility in our case study'' \citep[p.~16]{ccc}. To account for anticipation, their equation~(6) uses a four-month announcement period and estimates separate announcement and implementation coefficients.\footnote{Equation~(6) in \citet{ccc} uses a pre-announcement reference period to estimate separate announcement and implementation coefficients. Interpreting these coefficients as causal effects requires no anticipation in the reference period and parallel trends in the post-period. However, thr addition of announcement interaction does not itself separate anticipation from violations of parallel trends.} In this regard, our analysis complements their point-identified results by allowing departures from parallel trends. The December 2017 announcement also motivates our additional $A_{\text{PT-R}}$ specification using April--November 2017, which excludes the four-month announcement period.

Our illustration focuses on the first 12 months after implementation, from April 2018 to March 2019. We use monthly observations from April 2017 to March 2019 and compare the Salto branches with branches that did not charge for bags during this period. The outcome is disposable bags (in thousands) used per branch-month. The benchmark specification includes branch and month fixed effects, with standard errors clustered by branch. Table 5, column A of \citet{ccc} reports a reduction of 93.520 thousand bags per branch-month, approximately 75\% of the treated branches' pre-treatment mean of 124.9 thousand bags. Table~\ref{tab:ccc0923key} reports our standard DID replication estimate of -93.516 thousand bags. 

\begin{table}[H]
\centering\small
\caption{Effect of plastic bag charges on disposable-bag use in Salto}
\label{tab:ccc0923key}
\begin{threeparttable}
\begin{tabular}{lccc}
\toprule
Source or model & \shortstack{Estimate (SE)\\or identified set} & \shortstack{95\% ATT\\confidence set} & \(p\)-value \\
\midrule
Original article & -93.520 (  7.440) & Not reported & \(<0.01\) \\
Same-sample replication & -93.516 (  7.442) & [-108.392, -78.640] & \(\leq 0.0001\) \\
RDID common \(A_0\) & -93.516 (  7.442) & [-108.102, -78.930] & \(\leq 0.0001\) \\
\addlinespace
\(A_{\text{PT-R}}\), baseline Apr 2017--Mar 2018 & \([-145.115, -52.870]\) & \([-170.923, -31.320]\) & \(\leq 0.0001\) \\
\(A_{\text{PT-R}}\), baseline Apr--Nov 2017 & \([-130.130, -58.285]\) & \([-156.827, -31.320]\) & \(\leq 0.0001\) \\
\(A_{\text{AE-R}}\) & \(\varnothing\) & \(\varnothing\) & \(\leq 0.0001\) \\
\(A_{\text{R}}\) & \([-264.762, -49.760]\) & \([-330.498, -30.838]\) & \(\leq 0.0001\) \\
MR & \([-145.115, -52.870]\) & \([-170.923, -31.320]\) & \(\leq 0.0001\) \\
\bottomrule
\end{tabular}

\begin{tablenotes}[flushleft]
\footnotesize\setstretch{1}\setlength{\baselineskip}{11pt}
\item \textbf{Notes:} The outcome is disposable bags (in thousands) used per branch-month. The first row reports Table 5, column A of \citet{ccc}. Other rows use April 2017--March 2019 observations, comparing Salto branches with branches not charging for bags during this period. Baseline periods specify $\mathcal I_0$. The anticipation restrictions use $c=0.5$. Parentheses contain standard errors; brackets report estimated ATT sets in column 2 and 95\% ATT confidence sets in column 3. The $p$-values test $ATT=0$; $\varnothing$ denotes an empty set.
\end{tablenotes}
\end{threeparttable}
\end{table}

Again, Figure~\ref{fig:application-baselines}(b) reports DID estimates using the pooled April 2018--March 2019 implementation period as the post-treatment period and each month from April 2017 to March 2018 as the baseline. All the estimates are negative and  vary substantially from month to month, reflecting variation in the pre-treatment treated--control outcome gap. Moreover, a joint test rejects equality across these estimates ($p\leq0.0001$; Appendix~\ref{app:ccc-diagnostics}, Table~\ref{tab:ccc0923pre}). Our bias set stability assumption uses the range of baseline biases to restrict the post-treatment bias, partially identifying the ATT. 

Table~\ref{tab:ccc0923key} reports our bounds under the alternative identifying restrictions. The $A_{\text{PT-R}}$ estimates maintain no-anticipation and replace parallel trends with bias set stability, leading to the estimated ATT between -145.115 and -52.870 thousand bags per branch-month. Allowing departures from both parallel trends and no anticipation under the joint replacement restrictions ($A_{\text{R}}$) widens the estimated interval to [-264.762,-49.760]. The corresponding 95\% ATT confidence sets are [-170.923,-31.320] and [-330.498,-30.838], respectively, and both remain entirely negative. As such, despite the significant variation in the pre-treatment treated--control outcome gap in Figure~\ref{fig:application-baselines}(b), the finding that the initiative reduced disposable-bag use remains statistically significant when both assumptions are relaxed.\footnote{Only three branches were treated, all in Salto, so the branch-clustered confidence sets and $p$-values warrant caution.}

Using the pre-announcement months under $A_{\text{PT-R}}$ yields an estimated ATT set of [-130.130,-58.285] thousand bags per branch-month and a 95\% ATT confidence set of [-156.827,-31.320]. However, the qualitative conclusions remain unchanged when the baseline information excludes the announcement period.

The estimated misspecification-robust region (MR) also equals the $A_{\text{PT-R}}$ interval in this application because the estimated set under strong parallel trends and bounded anticipation ($A_{\text{AE-R}}$) is empty at $c=0.5$. The reported $A_{\text{R}}$ and MR results use the full April 2017--March 2018 baseline information set. The 95\% ATT confidence set under $A_{\text{AE-R}}$ is also empty.\footnote{This result concerns the combination of strong parallel trends and the specified anticipation restriction, and we cannot isolate which assumption accounts for the incompatibility.} As in the coca illustration, our analysis produced a wider range of effect magnitudes, but preserves the overall conclusion that the initiative reduced bag use.

\subsection{Compulsory licensing and German invention}

Finally, we revisit \citet{baten}, which examines whether compulsory licensing discourages or encourages invention. Under the US Trading with the Enemy Act of 1917, German-owned US patents became available for compulsory licensing, with most licenses issued between 1919 and 1922. The authors compare German patenting in chemical research fields with and without licensed patents after 1918. Licensing disproportionately affected fields related to wartime products such as dyes and explosives, whose demand changed after the war. We examine the baseline field-level DID specification to assess how the results change when we allow for some violations of the parallel trends assumption.

The authors observe that ``patenting may have begun to increase at a slightly higher rate for fields with licensing before 1918'' \citep[p.~234]{baten}. They address this concern using alternative controls for pre-treatment trends and find their positive result robust to those specifications. Our exercise complements their analysis by considering a different restriction, which is bias set stability, permitting the selection bias in the post-period to vary within the range of selected pre-treatment biases.

Our illustration uses annual observations for 212 chemical research fields from 1900 to 1930, comprising 6,572 field-year observations. The treated group consists of 101 fields with at least one licensed German-owned US patent and the remaining 111 fields serve as the comparison group. The outcome is successful German patent applications by German inventors per field-year. We use 1900--1918 as the pre-treatment period and pool 1919--1930 for the post-treatment period. Our benchmark specification includes field and year fixed effects and a control for German patents by non-US foreign inventors, with standard errors clustered by field.\footnote{In the replication data, the foreign-patent control is set to zero in field-years with zero domestic patents. The bounds therefore assess sensitivity of the published adjusted specification. Interpreting the adjustment causally requires a separate justification for conditioning on this outcome-dependent control.} Table 2, column 1 of \citet{baten} reports an increase of 3.194 patents per field-year, approximately 30\% of the treated fields' pre-treatment mean of 10.77. Our own DID estimates replicate this finding (Table~\ref{tab:baten0923-results}).

\begin{table}[htbp]
\centering\small
\caption{Effect of compulsory licensing on German patenting}
\label{tab:baten0923-results}
\begin{threeparttable}
\begin{tabular}{lccc}
\toprule
Source or model & \shortstack{Estimate (SE)\\or identified set} & \shortstack{95\% ATT\\confidence set} & \(p\)-value \\
\midrule
Original article &   3.194 (  0.953) & Not reported & \(<0.01\) \\
Same-sample replication &   3.194 (  0.953) & [  1.315,   5.073] & 0.0010 \\
RDID common \(A_0\) &   3.194 (  0.969) & [  1.295,   5.093] & 0.0010 \\
\addlinespace
\(A_{\text{PT-R}}\), baseline 1900--1918 & \([0.073, 5.325]\) & \([-2.302, 7.934]\) & 0.9519 \\
\(A_{\text{PT-R}}\), baseline 1900--1914 & \([0.073, 5.325]\) & \([-2.302, 7.934]\) & 0.9519 \\
\(A_{\text{AE-R}}\) & \(\varnothing\) & \([1.683, 3.521]\) & 0.0005 \\
\(A_{\text{R}}\) & \([0.072, 5.325]\) & \([-2.339, 9.234]\) & 0.9519 \\
MR & \([0.073, 5.325]\) & \([-2.302, 7.934]\) & 0.9519 \\
\bottomrule
\end{tabular}

\begin{tablenotes}[flushleft]
\footnotesize\setstretch{1}\setlength{\baselineskip}{11pt}
\item \textbf{Notes:} The outcome is successful German patent applications by German inventors per field-year. The first row reports Table 2, column 1 of \citet{baten}. Other rows use 212 fields over 1900--1930, with 1919--1930 as the post-treatment period. Baseline periods specify $\mathcal I_0$. The anticipation restrictions use $c=0.5$. Parentheses contain standard errors; brackets report estimated ATT sets in column 2 and 95\% ATT confidence sets in column 3. The $p$-values test $ATT=0$; $\varnothing$ denotes an empty estimated set.
\end{tablenotes}
\end{threeparttable}
\end{table}

Figure~\ref{fig:application-baselines}(c) again reports DID estimates using the pooled 1919--1930 post-treatment period and each year from 1900 to 1918 as the baseline. The estimates are generally smaller for later baseline years, reflecting variation in the treated--control outcome gap before treatment. All 19 estimates are positive, and a joint test rejects equality across these comparisons at the 5\% level ($p=0.0292$; Appendix~\ref{app:baten-diagnostics}, Table~\ref{tab:baten0923-pretests}). This changing historical pattern is relevant when assessing whether the post-treatment bias could plausibly remain within the selected baseline range. Our bound estimates based on bias set stability allow this range of pre-treatment selection biases to occur in the post-period.  

Table~\ref{tab:baten0923-results} reports our bounds under the alternative identifying restrictions. We first maintain no anticipation and we replace parallel trends with bias set stability ($A_{\text{PT-R}}$), and find that the estimated ATT is between 0.073 and 5.325 patents per field-year. Allowing departures from both parallel trends and no anticipation under the joint replacement restrictions ($A_{\text{R}}$) yields a similar estimated interval, [0.072,5.325]. The corresponding 95\% ATT confidence sets are [-2.302,7.934] and [-2.339,9.234], respectively. Although the estimated bounds remain positive, the confidence sets include zero, so the positive effect is no longer statistically significant under these relaxations.

We also consider 1900--1914 as the baseline information set under $A_{\text{PT-R}}$, motivated by the original article's exclusion of 1915--1918 in its wartime robustness analysis \citep[p.~235]{baten}. The estimated ATT set is [0.073,5.325] patents per field-year, and the 95\% ATT confidence set is [-2.302,7.934]. These results are identical to those using 1900--1918: the estimated bounds remain positive, while the confidence set includes zero.

The estimated misspecification-robust region (MR) also equals the $A_{\text{PT-R}}$ interval because the estimated set under strong parallel trends and bounded anticipation ($A_{\text{AE-R}}$) is empty at $c=0.5$.\footnote{The 95\% ATT confidence set under $A_{\text{AE-R}}$ is nevertheless nonempty and positive, [1.683,3.521].} The reported $A_{\text{R}}$ and MR results use the full 1900--1918 baseline information set.

The three applications show how to use our proposed methods to investigate the robustness of DID estimates when we relax the standard identifying assumptions. In the coca and plastic-bag applications, the finding of negative treatment effects is preserved under the joint replacement restrictions. In the patent application, the estimated bounds remain positive, but are no longer statistically significant as reported in the original article. As such, researchers do not need to abandon point identification per se; they can report our proposed bounds alongside their preferred point estimates to allow reader's to assess the robustness of their findings under clearly stated relaxations of the stronger identifying restrictions.

\section{Conclusion}\label{sec:conclusion}

This paper develops a unified framework for assessing DID conclusions when parallel trends, no anticipation, or both assumptions may fail. We use selection bias to formulate bias set stability and combine it with bounded anticipation to derive identified sets for the ATT. The framework provides separate analyses under each replacement restriction, a joint-relaxation analysis, and a minimum-relaxation region for comparing data-consistent alternatives to the standard model. These results give researchers a way to interpret pre-treatment evidence and report how conclusions depend on the restrictions they are willing to maintain.

The three empirical applications illustrate the value of this comparison. Under the joint replacement restrictions at $c=0.5$, the 95\% ATT confidence sets preserve the negative findings for coca cultivation in Colombia and plastic-bag use in Uruguay. In the selected German patent comparison, estimated bounds remain positive, but the confidence sets include zero, weakening support for a positive effect once sampling uncertainty is considered. The applications therefore show that the proposed analysis can document robustness of a qualitative conclusion or reveal that its statistical support depends on stronger identifying restrictions.

Researchers can report their preferred point estimate together with estimated identified sets under the alternative restrictions and the corresponding ATT confidence sets. This presentation distinguishes the range of admissible effect magnitudes from sampling uncertainty and gives a clearer account of what the data and identifying assumptions support.

\bibliographystyle{jpe}
\bibliography{mybib}

\clearpage

\appendix

\section{Proofs}

\subsection{A proof of Proposition \ref{prop1}}\label{sec:prop1}

\subsubsection*{Validity of the bounds}
By \eqref{eq:att}, $ATT=\theta_1^{OLS}-SB_1$. Under Assumption \ref{ass:NoAE}, $SB_{t_0}=\theta_{t_0}^{OLS}$ for every $t_0\in\mathcal I_0$. Assumption \ref{sb:bounds} therefore gives
\[
SB_1\in\left[\inf_{t_0\in\mathcal I_0}\theta_{t_0}^{OLS},
\sup_{t_0\in\mathcal I_0}\theta_{t_0}^{OLS}\right],
\]
which yields the displayed bounds.

{

\subsubsection*{Sharpness of the bounds}
\begin{proof}
Fix an arbitrary candidate $\theta\in\Theta^{\text{PT-RDID}}$ and define the period-1 selection bias required to attain that candidate by
\begin{equation}\label{eq:prop1-sharp-s}
s\equiv\theta_1^{OLS}-\theta.
\end{equation}
By the definition of $\Theta^{\text{PT-RDID}}$, candidate membership is equivalent to
\begin{equation}\label{eq:prop1-sharp-feasibility}
s\in
\left[
\inf_{t_0\in\mathcal I_0}\theta_{t_0}^{OLS},
\sup_{t_0\in\mathcal I_0}\theta_{t_0}^{OLS}
\right].
\end{equation}

We now construct a full vector of potential outcomes on the original observed-data probability space. For every $t\in\mathcal I_0\cup\{1\}$, retain each observed potential-outcome cell:
\begin{equation}\label{eq:prop1-sharp-observed}
\widetilde Y_t(1)=Y_t\quad\text{on }\{D=1\},
\qquad
\widetilde Y_t(0)=Y_t\quad\text{on }\{D=0\}.
\end{equation}
For each $t_0\in\mathcal I_0$, complete the treated units' missing untreated baseline outcome by
\begin{equation}\label{eq:prop1-sharp-baseline-completion}
\widetilde Y_{t_0}(0)=Y_{t_0}
\quad\text{on }\{D=1\}.
\end{equation}
Complete their missing untreated period-1 outcome by
\begin{equation}\label{eq:prop1-sharp-followup-completion}
\widetilde Y_1(0)
=\mathbb E[Y_1\mid D=0]+s
\quad\text{on }\{D=1\}.
\end{equation}
Finally, set the remaining missing cells, $\widetilde Y_t(1)$ on $\{D=0\}$, equal to zero. The variables in \eqref{eq:prop1-sharp-baseline-completion}--\eqref{eq:prop1-sharp-followup-completion} are real-valued and integrable under the maintained regularity conditions. Moreover, \eqref{eq:prop1-sharp-observed} implies
\begin{equation}\label{eq:prop1-sharp-consistency}
\widetilde Y_t(1)D+\widetilde Y_t(0)(1-D)=Y_t
\qquad\text{almost surely},
\end{equation}
so the completion satisfies consistency and leaves the joint distribution of the observed data unchanged.

We next verify each identifying restriction. By \eqref{eq:prop1-sharp-observed} and \eqref{eq:prop1-sharp-baseline-completion}, for every $t_0\in\mathcal I_0$,
\begin{equation}\label{eq:prop1-sharp-noae}
AE_{t_0}
=\mathbb E[\widetilde Y_{t_0}(1)-\widetilde Y_{t_0}(0)\mid D=1]
=\mathbb E[Y_{t_0}-Y_{t_0}\mid D=1]
=0.
\end{equation}
Thus Assumption \ref{ass:NoAE} holds. The same construction gives
\begin{align}
SB_{t_0}
&=\mathbb E[\widetilde Y_{t_0}(0)\mid D=1]
 -\mathbb E[\widetilde Y_{t_0}(0)\mid D=0]\nonumber\\
&=\mathbb E[Y_{t_0}\mid D=1]-\mathbb E[Y_{t_0}\mid D=0]
=\theta_{t_0}^{OLS}
\qquad(t_0\in\mathcal I_0).
\label{eq:prop1-sharp-baseline-bias}
\end{align}
In period 1, \eqref{eq:prop1-sharp-observed} and \eqref{eq:prop1-sharp-followup-completion} imply
\begin{align}
SB_1
&=\mathbb E[\widetilde Y_1(0)\mid D=1]
 -\mathbb E[\widetilde Y_1(0)\mid D=0]\nonumber\\
&=\{\mathbb E[Y_1\mid D=0]+s\}-\mathbb E[Y_1\mid D=0]
=s.
\label{eq:prop1-sharp-post-bias}
\end{align}
Equation \eqref{eq:prop1-sharp-baseline-bias} identifies the two endpoints in \eqref{eq:prop1-sharp-feasibility} with the infimum and supremum of the constructed baseline selection biases, while \eqref{eq:prop1-sharp-post-bias} gives $SB_1=s$. Hence \eqref{eq:prop1-sharp-feasibility} verifies Assumption \ref{sb:bounds} for the constructed DGP.

Finally, the ATT generated by the completion is
\begin{align}
ATT
&=\mathbb E[\widetilde Y_1(1)-\widetilde Y_1(0)\mid D=1]\nonumber\\
&=\mathbb E[Y_1\mid D=1]
 -\{\mathbb E[Y_1\mid D=0]+s\}\nonumber\\
&=\theta_1^{OLS}-s
=\theta,
\label{eq:prop1-sharp-att}
\end{align}
where the last equality follows from \eqref{eq:prop1-sharp-s}. Thus, for every $\theta\in\Theta^{\text{PT-RDID}}$, the construction preserves the observed-data distribution, satisfies all maintained assumptions, and attains $ATT=\theta$. This proves sharpness.
\end{proof}
}

\subsection{A proof of Lemma \ref{lem:samesign}}\label{sec:lem:samesign}
\begin{proof}
Under Assumption \ref{ass:strongPT}, we have
\begin{equation}
    \theta^{DID}_{t_0} = \mathrm{ATT} - AE_{t_0}, \label{eq:SPT_DID}
\end{equation}
for any given $t_0 \in \mathcal I_0$.
\begin{enumerate}
    \item $\theta^{DID}_{t_0} =0 \Rightarrow \mathrm{ATT} = 0$.

    Suppose not, $\mathrm{ATT} \neq 0$.
    If $\theta^{DID}_{t_0} =0$, \eqref{eq:SPT_DID} implies $\mathrm{ATT} = AE_{t_0} \neq 0$, which contradicts Assumption \ref{ass:expoAE}.

    \item $\mathrm{ATT} = 0 \Rightarrow \theta^{DID}_{t_0} =0$.
    
    Note that $AE_{t_0} = 0$ if $\mathrm{ATT} = 0$ under Assumption \ref{ass:expoAE}.
    Thus, $\theta^{DID}_{t_0}=0$ by \eqref{eq:SPT_DID}.

    \item $\theta^{DID}_{t_0} >0 \Rightarrow \mathrm{ATT} > 0$
    Suppose not, $\mathrm{ATT} \leq 0$.
    From \eqref{eq:SPT_DID}, we have $\mathrm{ATT} = \theta^{DID}_{t_0} + AE_{t_0} \leq 0$.
    This further implies $\vert \mathrm{ATT} \vert = - \theta^{DID}_{t_0} - AE_{t_0}$, and $\vert \mathrm{ATT} \vert < - AE_{t_0}$ as $\theta^{DID}_{t_0}>0$. Thus, we have $\vert \mathrm{ATT} \vert < \vert AE_{t_0} \vert$, which contradicts Assumption \ref{ass:expoAE}.

    \item $\mathrm{ATT} > 0 \Rightarrow \theta^{DID}_{t_0} >0$.

    Suppose not, $\theta^{DID}_{t_0} \leq 0$.
    From \eqref{eq:SPT_DID}, we have $\vert \mathrm{ATT} \vert = \theta^{DID}_{t_0} + AE_{t_0}$ as $\mathrm{ATT} > 0$, which implies $\vert \mathrm{ATT} \vert \leq AE_{t_0}$ as $\theta^{DID}_{t_0} \leq 0$. Thus, we have $\vert \mathrm{ATT} \vert \leq \vert AE_{t_0} \vert$, which contradicts Assumption \ref{ass:expoAE}.

    \item $\theta^{DID}_{t_0} <0 \Rightarrow \mathrm{ATT} < 0$

    Suppose not, $\mathrm{ATT} \geq 0$.
    From \eqref{eq:SPT_DID}, we have $\vert \mathrm{ATT} \vert = \theta^{DID}_{t_0} + AE_{t_0}$, and $\vert \mathrm{ATT} \vert < AE_{t_0}$ as $\theta^{DID}_{t_0} < 0$. 
    Thus, we have $\vert \mathrm{ATT} \vert < \vert AE_{t_0} \vert$, which contradicts Assumption \ref{ass:expoAE}.

    \item $\mathrm{ATT} < 0 \Rightarrow \theta^{DID}_{t_0} <0$.

    Suppose not, $\theta^{DID}_{t_0} \geq 0$.
    From \eqref{eq:SPT_DID}, we have $\vert \mathrm{ATT} \vert = - \theta^{DID}_{t_0} - AE_{t_0}$ as $\mathrm{ATT} < 0$, which implies $\vert \mathrm{ATT} \vert \leq - AE_{t_0}$ as $\theta^{DID}_{t_0} \geq 0$. 
    Thus, we have $\vert \mathrm{ATT} \vert \leq \vert AE_{t_0} \vert$, which contradicts Assumption \ref{ass:expoAE}.
\end{enumerate}
\end{proof}

{\subsection{A proof of Proposition \ref{bound_multiple_preperiods}}\label{sec:bound_multiple_preperiods}

\subsubsection*{Validity of the bounds}
\begin{proof}
For notational convenience, retain only
\[
c_{t_0}\equiv c^{|t_0|+1}.
\]
Under Assumption \ref{ass:strongPT}, $SB_{t_0}=SB_1$. Therefore, for every $t_0\in\mathcal I_0$,
\begin{align}
\theta_{t_0}^{DID}
&=\theta_1^{OLS}-\theta_{t_0}^{OLS}\nonumber\\
&=(ATT+SB_1)-(AE_{t_0}+SB_{t_0})\nonumber\\
&=ATT-AE_{t_0}.
\label{eq:rdid2-proof-decomposition}
\end{align}
Consequently, a candidate value $\theta$ for the ATT is compatible with the two assumptions only if its implied anticipation effect $\theta-\theta_{t_0}^{DID}$ satisfies
\begin{equation}\label{eq:rdid2-proof-feasibility}
|\theta-\theta_{t_0}^{DID}|
\leq c_{t_0}|\theta|
\qquad\text{for every }t_0\in\mathcal I_0.
\end{equation}
If $\theta\geq0$, condition \eqref{eq:rdid2-proof-feasibility} is equivalent to
\[
\frac{\theta_{t_0}^{DID}}{1+c_{t_0}}
\leq\theta\leq
\frac{\theta_{t_0}^{DID}}{1-c_{t_0}}
\qquad\text{for every }t_0\in\mathcal I_0,
\]
which gives the first intersection interval in \eqref{equation_bound_multiple_preperiods}. If $\theta\leq0$, it is equivalent to
\[
\frac{\theta_{t_0}^{DID}}{1-c_{t_0}}
\leq\theta\leq
\frac{\theta_{t_0}^{DID}}{1+c_{t_0}}
\qquad\text{for every }t_0\in\mathcal I_0,
\]
which gives the second intersection interval. For $0<c<1$, the first interval can be nonempty only when every $\theta_{t_0}^{DID}\geq0$, in which case it lies in $[0,\infty)$; the second can be nonempty only when every $\theta_{t_0}^{DID}\leq0$, in which case it lies in $(-\infty,0]$. At $c=0$ the two intervals coincide. They equal the common DID singleton if all baseline contrasts agree and are empty otherwise. Thus the union in \eqref{equation_bound_multiple_preperiods} is exactly the set of candidate values satisfying \eqref{eq:rdid2-proof-feasibility}, proving validity.
\end{proof}

\subsubsection*{Sharpness of the bounds}
\begin{proof}
Fix an arbitrary $\theta\in\Theta^{\text{AE-RDID}}$. For each $t_0\in\mathcal I_0$, let $c_{t_0}\equiv c^{|t_0|+1}$ and define
\begin{equation}\label{eq:rdid2-sharp-targets}
s\equiv\theta_1^{OLS}-\theta,
\qquad
a_{t_0}\equiv\theta-\theta_{t_0}^{DID}.
\end{equation}
By the characterization in \eqref{eq:rdid2-proof-feasibility},
\begin{equation}\label{eq:rdid2-sharp-budget}
|a_{t_0}|\leq c_{t_0}|\theta|
\qquad\text{for every }t_0\in\mathcal I_0.
\end{equation}
The scalar $s$ is the period-1 selection bias that the completion must generate, while $a_{t_0}$ is the required anticipation effect in baseline period $t_0$.

Construct the full potential-outcome vector on the original observed-data probability space. Retain all observed cells:
\begin{equation}\label{eq:rdid2-sharp-observed}
\widetilde Y_u(1)=Y_u\quad\text{on }\{D=1\},
\qquad
\widetilde Y_u(0)=Y_u\quad\text{on }\{D=0\},
\quad u\in\mathcal I_0\cup\{1\}.
\end{equation}
For every $u\in\mathcal I_0\cup\{1\}$, complete the treated units' missing untreated outcome by
\begin{equation}\label{eq:rdid2-sharp-completion}
\widetilde Y_u(0)=\mathbb E[Y_u\mid D=0]+s
\quad\text{on }\{D=1\}.
\end{equation}
Set every remaining missing cell, $\widetilde Y_u(1)$ on $\{D=0\}$, equal to zero. These assignments are real-valued and integrable. Moreover, \eqref{eq:rdid2-sharp-observed} implies
\begin{equation}\label{eq:rdid2-sharp-consistency}
\widetilde Y_u(1)D+\widetilde Y_u(0)(1-D)=Y_u
\qquad\text{almost surely},
\end{equation}
so consistency holds and the observed-data distribution is unchanged.

The selection bias induced by \eqref{eq:rdid2-sharp-completion} is, for every $u\in\mathcal I_0\cup\{1\}$,
\begin{align}
SB_u
&=\mathbb E[\widetilde Y_u(0)\mid D=1]
 -\mathbb E[\widetilde Y_u(0)\mid D=0]\nonumber\\
&=\{\mathbb E[Y_u\mid D=0]+s\}-\mathbb E[Y_u\mid D=0]
=s.
\label{eq:rdid2-sharp-bias}
\end{align}
Thus $SB_{t_0}=SB_1=s$ for every $t_0\in\mathcal I_0$, which verifies Assumption \ref{ass:strongPT}. Next, for each baseline period,
\begin{align}
AE_{t_0}
&=\mathbb E[\widetilde Y_{t_0}(1)-\widetilde Y_{t_0}(0)\mid D=1]\nonumber\\
&=\mathbb E[Y_{t_0}\mid D=1]
 -\{\mathbb E[Y_{t_0}\mid D=0]+s\}\nonumber\\
&=\theta_{t_0}^{OLS}-s\nonumber\\
&=\theta-\big(\theta_1^{OLS}-\theta_{t_0}^{OLS}\big)\nonumber\\
&=\theta-\theta_{t_0}^{DID}
=a_{t_0},
\label{eq:rdid2-sharp-ae}
\end{align}
where the fourth equality uses $s=\theta_1^{OLS}-\theta$ from \eqref{eq:rdid2-sharp-targets}. The constructed ATT is
\begin{align}
ATT
&=\mathbb E[\widetilde Y_1(1)-\widetilde Y_1(0)\mid D=1]\nonumber\\
&=\mathbb E[Y_1\mid D=1]
 -\{\mathbb E[Y_1\mid D=0]+s\}\nonumber\\
&=\theta_1^{OLS}-s
=\theta.
\label{eq:rdid2-sharp-att}
\end{align}
Combining \eqref{eq:rdid2-sharp-budget}, \eqref{eq:rdid2-sharp-ae}, and \eqref{eq:rdid2-sharp-att} gives
$|AE_{t_0}|\leq c_{t_0}|ATT|$ for every $t_0$, so Assumption \ref{ass:expoAE} also holds. Hence every $\theta\in\Theta^{\text{AE-RDID}}$ is attained by a DGP that preserves the observed-data distribution and satisfies both maintained assumptions. This proves sharpness.
\end{proof}

}

{

\subsection{A proof of Proposition \ref{bound_relaxtwo}}\label{sec:bound_relaxtwo}

\subsubsection*{Validity of the bounds}
\begin{proof}
{For each $t_0\in\mathcal I_0$, write
\[
c_{t_0}\equiv c^{|t_0|+1}.
\]
Consider any DGP satisfying Assumptions \ref{ass:expoAE} and \ref{sb:bounds}. By \eqref{eq:att}, its period-1 selection bias is
\begin{equation}\label{eq:rdid3-proof-post-bias}
s\equiv\theta_1^{OLS}-ATT=SB_1.
\end{equation}
For each $t_0\in\mathcal I_0$, bounded anticipation and the decomposition $\theta_{t_0}^{OLS}=AE_{t_0}+SB_{t_0}$ imply
\begin{equation}\label{eq:rdid3-proof-baseline-range}
SB_{t_0}
\in
\left[
\theta_{t_0}^{OLS}-c_{t_0}|ATT|,
\theta_{t_0}^{OLS}+c_{t_0}|ATT|
\right].
\end{equation}
Because $\mathcal I_0$ is finite, Assumption \ref{sb:bounds} is equivalent to the existence of at least one baseline bias weakly below $s$ and at least one weakly above $s$. For any $t_0\in\mathcal I_0$, equation \eqref{eq:rdid3-proof-post-bias} and $\theta_{t_0}^{DID}=\theta_1^{OLS}-\theta_{t_0}^{OLS}$ give
\begin{align}
SB_{t_0}\leq s
&\quad\Longleftrightarrow\quad
AE_{t_0}\geq ATT-\theta_{t_0}^{DID},\nonumber\\
SB_{t_0}\geq s
&\quad\Longleftrightarrow\quad
AE_{t_0}\leq ATT-\theta_{t_0}^{DID}.
\label{eq:rdid3-proof-ordering}
\end{align}
Since Assumption \ref{ass:expoAE} restricts $AE_{t_0}$ to $[-c_{t_0}|ATT|,c_{t_0}|ATT|]$, the first ordering in \eqref{eq:rdid3-proof-ordering} can hold for some baseline period only if
\[
ATT-\theta_{t_0}^{DID}\leq c_{t_0}|ATT|
\quad\text{for some }t_0\in\mathcal I_0,
\]
and the second can hold for some baseline period only if
\[
ATT-\theta_{t_0}^{DID}\geq-c_{t_0}|ATT|
\quad\text{for some }t_0\in\mathcal I_0.
\]
Thus every admissible ATT satisfies
\begin{equation}\label{eq:rdid3-feasibility}
\begin{aligned}
ATT-\theta_{t_0}^{DID}
&\leq c_{t_0}|ATT|
&&\text{for some }t_0\in\mathcal I_0,\\
ATT-\theta_{t_0}^{DID}
&\geq-c_{t_0}|ATT|
&&\text{for some }t_0\in\mathcal I_0.
\end{aligned}
\end{equation}
The sharpness construction below shows that these conditions are also sufficient.

Define
\begin{align*}
L^+&\equiv\min_{t_0\in\mathcal I_0}\frac{\theta_{t_0}^{DID}}{1+c_{t_0}},
&U^+&\equiv\max_{t_0\in\mathcal I_0}\frac{\theta_{t_0}^{DID}}{1-c_{t_0}},\\
L^-&\equiv\min_{t_0\in\mathcal I_0}\frac{\theta_{t_0}^{DID}}{1-c_{t_0}},
&U^-&\equiv\max_{t_0\in\mathcal I_0}\frac{\theta_{t_0}^{DID}}{1+c_{t_0}}.
\end{align*}
For $ATT\geq0$, conditions \eqref{eq:rdid3-feasibility} are equivalent to $L^+\leq ATT\leq U^+$; for $ATT\leq0$, they are equivalent to $L^-\leq ATT\leq U^-$. Hence every admissible ATT belongs to
\begin{equation}\label{eq:rdid3-signsets}
\big([L^+,U^+]\cap[0,\infty)\big)
\cup
\big([L^-,U^-]\cap(-\infty,0]\big).
\end{equation}
This union equals $[\min\{L^+,L^-\},\max\{U^+,U^-\}]$. Indeed, if every $\theta_{t_0}^{DID}\geq0$, the nonpositive component is empty except possibly at zero, and \eqref{eq:rdid3-signsets} equals $[L^+,U^+]$. If every $\theta_{t_0}^{DID}\leq0$, the nonnegative component is empty except possibly at zero, and the union equals $[L^-,U^-]$. If the $\theta_{t_0}^{DID}$ values include both signs, then $L^-\leq L^+<0<U^-\leq U^+$, so the two sign-restricted components meet at zero and their union is $[L^-,U^+]$. Substituting the four endpoint definitions gives exactly the interval in Proposition \ref{bound_relaxtwo}, proving validity.
}
\end{proof}

\subsubsection*{Sharpness of the bounds}
\begin{proof}
{Fix an arbitrary candidate $\theta\in\Theta^{\text{RDID}}$ and define, for every $t_0\in\mathcal I_0$,
\begin{equation}\label{eq:rdid3-sharp-basic-targets}
c_{t_0}\equiv c^{|t_0|+1},
\qquad
s\equiv\theta_1^{OLS}-\theta.
\end{equation}
Because $\theta\in\Theta^{\text{RDID}}$, the same sign-specific algebra used to obtain \eqref{eq:rdid3-signsets} implies that there exist periods $t_0^-,t_0^+\in\mathcal I_0$ satisfying
\begin{equation}\label{eq:rdid3-sharp-bracketing-periods}
\theta-\theta_{t_0^-}^{DID}\leq c_{t_0^-}|\theta|,
\qquad
\theta-\theta_{t_0^+}^{DID}\geq-c_{t_0^+}|\theta|.
\end{equation}
The superscripts are chosen according to the bias relations established below: the first selected period yields $b_{t_0^-}\leq s$, whereas the second yields $b_{t_0^+}\geq s$.

For every $t_0\in\mathcal I_0$, define the target anticipation effect and the associated target baseline selection bias by
\begin{align}
a_{t_0}
&\equiv
\min\left\{
 c_{t_0}|\theta|,
 \max\{-c_{t_0}|\theta|,\theta-\theta_{t_0}^{DID}\}
\right\},\nonumber\\
b_{t_0}
&\equiv\theta_{t_0}^{OLS}-a_{t_0}.
\label{eq:rdid3-sharp-targets}
\end{align}
By construction,
\begin{equation}\label{eq:rdid3-sharp-ae-budget}
|a_{t_0}|\leq c_{t_0}|\theta|
\qquad\text{for every }t_0\in\mathcal I_0.
\end{equation}
Moreover, \eqref{eq:rdid3-sharp-basic-targets} and $\theta_{t_0}^{DID}=\theta_1^{OLS}-\theta_{t_0}^{OLS}$ imply
\begin{equation}\label{eq:rdid3-sharp-bias-gap}
b_{t_0}-s
=\theta-\theta_{t_0}^{DID}-a_{t_0}.
\end{equation}

For $t_0^-$, the first inequality in \eqref{eq:rdid3-sharp-bracketing-periods} gives
\[
\theta-\theta_{t_0^-}^{DID}\leq c_{t_0^-}|\theta|,
\]
while
\[
\max\{-c_{t_0^-}|\theta|,\theta-\theta_{t_0^-}^{DID}\}
\geq\theta-\theta_{t_0^-}^{DID}.
\]
Thus both arguments of the outer minimum in \eqref{eq:rdid3-sharp-targets} are at least $\theta-\theta_{t_0^-}^{DID}$, and therefore
\begin{equation}\label{eq:rdid3-sharp-minus-truncation}
a_{t_0^-}\geq\theta-\theta_{t_0^-}^{DID}.
\end{equation}
Substituting \eqref{eq:rdid3-sharp-minus-truncation} into \eqref{eq:rdid3-sharp-bias-gap} yields
\begin{equation}\label{eq:rdid3-sharp-minus-bias}
b_{t_0^-}\leq s.
\end{equation}

For $t_0^+$, the second inequality in \eqref{eq:rdid3-sharp-bracketing-periods} gives $\theta-\theta_{t_0^+}^{DID}\geq-c_{t_0^+}|\theta|$, so the inner maximum in \eqref{eq:rdid3-sharp-targets} equals $\theta-\theta_{t_0^+}^{DID}$. Hence
\begin{equation}\label{eq:rdid3-sharp-plus-truncation}
a_{t_0^+}
=\min\{c_{t_0^+}|\theta|,\theta-\theta_{t_0^+}^{DID}\}
\leq\theta-\theta_{t_0^+}^{DID}.
\end{equation}
Substituting \eqref{eq:rdid3-sharp-plus-truncation} into \eqref{eq:rdid3-sharp-bias-gap} yields
\begin{equation}\label{eq:rdid3-sharp-plus-bias}
b_{t_0^+}\geq s.
\end{equation}
Equations \eqref{eq:rdid3-sharp-minus-bias} and \eqref{eq:rdid3-sharp-plus-bias} imply
\begin{equation}\label{eq:rdid3-sharp-bracketing}
s\in
\left[
\min_{t_0\in\mathcal I_0}b_{t_0},
\max_{t_0\in\mathcal I_0}b_{t_0}
\right].
\end{equation}
}

We now construct the potential outcomes on the original observed-data probability space. Retain every observed cell:
\begin{equation}\label{eq:rdid3-sharp-observed}
\widetilde Y_u(1)=Y_u\quad\text{on }\{D=1\},
\qquad
\widetilde Y_u(0)=Y_u\quad\text{on }\{D=0\},
\quad u\in\mathcal I_0\cup\{1\}.
\end{equation}
For each $t_0\in\mathcal I_0$, complete the treated units' missing untreated baseline outcome by
\begin{equation}\label{eq:rdid3-sharp-baseline-completion}
\widetilde Y_{t_0}(0)
=\mathbb E[Y_{t_0}\mid D=0]+b_{t_0}
\quad\text{on }\{D=1\},
\end{equation}
and complete their period-1 untreated outcome by
\begin{equation}\label{eq:rdid3-sharp-post-completion}
\widetilde Y_1(0)
=\mathbb E[Y_1\mid D=0]+s
\quad\text{on }\{D=1\}.
\end{equation}
Set every remaining missing cell, $\widetilde Y_u(1)$ on $\{D=0\}$, equal to zero. All assignments are real-valued and integrable. Equation \eqref{eq:rdid3-sharp-observed} implies
\begin{equation}\label{eq:rdid3-sharp-consistency}
\widetilde Y_u(1)D+\widetilde Y_u(0)(1-D)=Y_u
\qquad\text{almost surely},
\end{equation}
so consistency holds and the observed-data distribution is preserved.

The baseline selection biases induced by \eqref{eq:rdid3-sharp-baseline-completion} satisfy
\begin{align}
SB_{t_0}
&=\mathbb E[\widetilde Y_{t_0}(0)\mid D=1]
 -\mathbb E[\widetilde Y_{t_0}(0)\mid D=0]\nonumber\\
&=\{\mathbb E[Y_{t_0}\mid D=0]+b_{t_0}\}
 -\mathbb E[Y_{t_0}\mid D=0]
=b_{t_0}.
\label{eq:rdid3-sharp-baseline-bias}
\end{align}
Similarly, \eqref{eq:rdid3-sharp-post-completion} gives
\begin{align}
SB_1
&=\mathbb E[\widetilde Y_1(0)\mid D=1]
 -\mathbb E[\widetilde Y_1(0)\mid D=0]\nonumber\\
&=\{\mathbb E[Y_1\mid D=0]+s\}-\mathbb E[Y_1\mid D=0]
=s.
\label{eq:rdid3-sharp-post-bias}
\end{align}
Equations \eqref{eq:rdid3-sharp-bracketing}, \eqref{eq:rdid3-sharp-baseline-bias}, and \eqref{eq:rdid3-sharp-post-bias} therefore verify Assumption \ref{sb:bounds}.

For each baseline period, the resulting anticipation effect is
\begin{align}
AE_{t_0}
&=\mathbb E[\widetilde Y_{t_0}(1)-\widetilde Y_{t_0}(0)\mid D=1]\nonumber\\
&=\theta_{t_0}^{OLS}-b_{t_0}
=a_{t_0},
\label{eq:rdid3-sharp-ae}
\end{align}
where the last equality follows from \eqref{eq:rdid3-sharp-targets}. The constructed ATT is
\begin{align}
ATT
&=\mathbb E[\widetilde Y_1(1)-\widetilde Y_1(0)\mid D=1]\nonumber\\
&=\theta_1^{OLS}-s
=\theta.
\label{eq:rdid3-sharp-att}
\end{align}
Combining \eqref{eq:rdid3-sharp-ae-budget}, \eqref{eq:rdid3-sharp-ae}, and \eqref{eq:rdid3-sharp-att} verifies Assumption \ref{ass:expoAE}. Thus every candidate in $\Theta^{\text{RDID}}$ is attained by a DGP that preserves the observed distribution and satisfies both maintained assumptions. This proves sharpness.
\end{proof}
}

\subsection{A proof of Proposition \ref{multiTI}}\label{sec:multiTI}

{

\begin{proof}
Fix $t\in\mathcal T_0\cup\{1,\ldots,T\}$ and $\bm d'\in\mathcal D_t^0\cap\operatorname{supp}(\bm D)$. Extended PT applied to $\bm d'$ and to $\bm d_{\infty}$ gives
\begin{align*}
\mathbb E[Y_t(\bm d_{\infty})-Y_0(\bm d_{\infty})\mid\bm D=\bm d']
&=\mathbb E[Y_t(\bm d_{\infty})-Y_0(\bm d_{\infty})\mid\bm D=\bm d_{\infty}]\\
&=\mathbb E[Y_t-Y_0\mid\bm D=\bm d_{\infty}],
\end{align*}
where the second equality uses consistency. Since $\bm d'\in\mathcal D_t^0$ and also $\bm d'\in\mathcal D_0^0$, extended no anticipation gives
\[
\mathbb E[Y_t(\bm d')-Y_t(\bm d_{\infty})\mid\bm D=\bm d']=0,
\qquad
\mathbb E[Y_0(\bm d')-Y_0(\bm d_{\infty})\mid\bm D=\bm d']=0.
\]
Subtracting the two equalities and using consistency on the group $\bm D=\bm d'$ yields
\[
\mathbb E[Y_t-Y_0\mid\bm D=\bm d']
=
\mathbb E[Y_t-Y_0\mid\bm D=\bm d_{\infty}],
\]
as claimed.
\end{proof}
}

\subsection{A proof of Lemma \ref{lem:bsmulti}}\label{sec:lem:bsmulti}
{

\begin{proof}
Consider any paths of interest $\bm d^0 \equiv (0,d_1^0,\ldots, d_T^0)$ and $\bm d^1 \equiv (0,d_1^1,\ldots, d_T^1)$.
For a given $t_0 \in \mathcal T_0$, Assumption \ref{ass:extnoae0}, applied with the conditioning path $\bm d' = \bm d^1$, implies
\begin{align*}
    \mathbb{E}[Y_{t_0}(\bm d^0)-Y_{t_0}(\bm d_{\infty})\mid \bm D=\bm d^{1}] = 0,\\
    \mathbb{E}[Y_{t_0}(\bm d^1)-Y_{t_0}(\bm d_{\infty})\mid \bm D=\bm d^{1}] = 0,
\end{align*}
where $\bm d_{\infty} \equiv (0, \ldots, 0)$.
Note that these together imply 
$$
\mathbb{E}[Y_{t_0}(\bm d^1)-Y_{t_0}(\bm d^0)\mid \bm D=\bm d^{1}] \equiv AE_{t_0}[\bm d^0 \rightarrow \bm d^1] = 0.
$$
Hence, for any $t_0 \in \mathcal T_0$,
\begin{align*}
    \theta^{DIM}_{t_0}[\bm d^0 \rightarrow \bm d^1] &=AE_{t_0}[\bm d^0\rightarrow \bm d^1]+SB_{t_0}[\bm d^0\rightarrow \bm d^1] \\
    &= SB_{t_0}[\bm d^0\rightarrow \bm d^1].
\end{align*}
\end{proof}
}

\subsection{A proof of Proposition \ref{prop1ex}}\label{sec:prop1ex}

\subsubsection*{Validity of the bounds}
By Lemma \ref{lem:bsmulti}, extended no anticipation identifies
\[
SB_{t_0}[\bm d^0\to\bm d^1]
=\theta_{t_0}^{DIM}[\bm d^0\to\bm d^1]
\]
for every $t_0\in\mathcal T_0$. Since
\[
ATT_t[\bm d^0\to\bm d^1]
=\theta_t^{DIM}[\bm d^0\to\bm d^1]
-SB_t[\bm d^0\to\bm d^1],
\]
the bounds follow from Assumption \ref{sb:boundsex}.

{

\subsubsection*{Sharpness of the bounds}
\begin{proof}
If $\bm d^0=\bm d^1$, then $ATT_t[\bm d^0\to\bm d^1]=0$ and the displayed interval is the singleton $\{0\}$, so sharpness is immediate. Suppose henceforth that $\bm d^0\neq\bm d^1$.

Fix an arbitrary post-treatment period $t$ and candidate $\theta^t\in\Theta_t^{\text{PT-R}}[\bm d^0\to\bm d^1]$. Define the period-$t$ reference-path selection bias required to attain $\theta^t$ by
\begin{equation}\label{eq:multi-rdid1-sharp-kappa}
\kappa
\equiv
\theta_t^{DIM}[\bm d^0\to\bm d^1]-\theta^t.
\end{equation}
By candidate membership,
\begin{equation}\label{eq:multi-rdid1-sharp-feasibility}
\kappa\in
\left[
\inf_{t_0\in\mathcal T_0}
\theta_{t_0}^{DIM}[\bm d^0\to\bm d^1],
\sup_{t_0\in\mathcal T_0}
\theta_{t_0}^{DIM}[\bm d^0\to\bm d^1]
\right].
\end{equation}
For every supported realized path $\bm d'\in\operatorname{supp}(\bm D)$ and every period $s\in\mathcal T_0\cup\{1,\ldots,T\}$, write
\begin{equation}\label{eq:multi-rdid1-sharp-means}
m_s(\bm d')\equiv\mathbb E[Y_s\mid\bm D=\bm d'].
\end{equation}

We construct all path-specific potential outcomes jointly on the original observed-data probability space. First, retain the observed path cell in every supported group:
\begin{equation}\label{eq:multi-rdid1-sharp-observed}
\widetilde Y_s(\bm d')=Y_s
\quad\text{on }\{\bm D=\bm d'\},
\qquad
\bm d'\in\operatorname{supp}(\bm D).
\end{equation}
For every baseline period $t_0\in\mathcal T_0$, every supported conditioning path $\bm d'$, and every intervention path $\bm d\neq\bm d'$, define the missing baseline cell by
\begin{equation}\label{eq:multi-rdid1-sharp-baseline-completion}
\widetilde Y_{t_0}(\bm d)=m_{t_0}(\bm d')
\quad\text{on }\{\bm D=\bm d'\}.
\end{equation}
Equations \eqref{eq:multi-rdid1-sharp-observed} and \eqref{eq:multi-rdid1-sharp-baseline-completion} imply that, for every baseline period, supported conditioning path, and intervention path,
\begin{equation}\label{eq:multi-rdid1-sharp-baseline-means}
\mathbb E[\widetilde Y_{t_0}(\bm d)\mid\bm D=\bm d']
=m_{t_0}(\bm d').
\end{equation}
In particular, applying \eqref{eq:multi-rdid1-sharp-baseline-means} to $\bm d$ and to $\bm d_\infty$ gives
\begin{equation}\label{eq:multi-rdid1-sharp-noae}
\mathbb E[\widetilde Y_{t_0}(\bm d)-\widetilde Y_{t_0}(\bm d_\infty)
\mid\bm D=\bm d']=0
\end{equation}
for every $t_0\in\mathcal T_0$, every intervention path $\bm d$, and every supported $\bm d'$. Thus Assumption \ref{ass:extnoae0} holds jointly for all conditioning paths.

Because $\bm d^0\neq\bm d^1$, equation \eqref{eq:multi-rdid1-sharp-baseline-means} also yields, for every $t_0\in\mathcal T_0$,
\begin{align}
SB_{t_0}[\bm d^0\to\bm d^1]
&=\mathbb E[\widetilde Y_{t_0}(\bm d^0)\mid\bm D=\bm d^1]
 -\mathbb E[\widetilde Y_{t_0}(\bm d^0)\mid\bm D=\bm d^0]\nonumber\\
&=m_{t_0}(\bm d^1)-m_{t_0}(\bm d^0)\nonumber\\
&=\theta_{t_0}^{DIM}[\bm d^0\to\bm d^1].
\label{eq:multi-rdid1-sharp-baseline-bias}
\end{align}

For every post-treatment period $s\in\{1,\ldots,T\}$, complete the target group's missing reference-path outcome by
\begin{equation}\label{eq:multi-rdid1-sharp-post-completion}
\widetilde Y_s(\bm d^0)
=m_s(\bm d^0)+\kappa
\quad\text{on }\{\bm D=\bm d^1\}.
\end{equation}
Set every remaining missing post-treatment path cell equal to zero. All assignments are real-valued and integrable. Since every observed path cell was retained in \eqref{eq:multi-rdid1-sharp-observed},
\begin{equation}\label{eq:multi-rdid1-sharp-consistency}
\widetilde Y_s(\bm D)=Y_s
\qquad\text{almost surely},
\end{equation}
so consistency holds and the observed-data distribution is unchanged.

For every post-treatment period $s$, equations \eqref{eq:multi-rdid1-sharp-observed} and \eqref{eq:multi-rdid1-sharp-post-completion} give
\begin{align}
SB_s[\bm d^0\to\bm d^1]
&=\mathbb E[\widetilde Y_s(\bm d^0)\mid\bm D=\bm d^1]
 -\mathbb E[\widetilde Y_s(\bm d^0)\mid\bm D=\bm d^0]\nonumber\\
&=\{m_s(\bm d^0)+\kappa\}-m_s(\bm d^0)
=\kappa.
\label{eq:multi-rdid1-sharp-post-bias}
\end{align}
Equation \eqref{eq:multi-rdid1-sharp-baseline-bias} identifies the endpoints in \eqref{eq:multi-rdid1-sharp-feasibility} with the infimum and supremum of the constructed baseline selection biases, while \eqref{eq:multi-rdid1-sharp-post-bias} gives every post-treatment selection bias as $\kappa$. Therefore \eqref{eq:multi-rdid1-sharp-feasibility} verifies Assumption \ref{sb:boundsex} for every post-treatment period.

Finally, the target treatment effect generated in period $t$ is
\begin{align}
ATT_t[\bm d^0\to\bm d^1]
&=\mathbb E[\widetilde Y_t(\bm d^1)-\widetilde Y_t(\bm d^0)
\mid\bm D=\bm d^1]\nonumber\\
&=m_t(\bm d^1)-\{m_t(\bm d^0)+\kappa\}\nonumber\\
&=\theta_t^{DIM}[\bm d^0\to\bm d^1]-\kappa
=\theta^t,
\label{eq:multi-rdid1-sharp-att}
\end{align}
where the last equality follows from \eqref{eq:multi-rdid1-sharp-kappa}. Thus every point of $\Theta_t^{\text{PT-R}}[\bm d^0\to\bm d^1]$ is attained by a joint completion that preserves the observed distribution and satisfies Assumptions \ref{ass:extnoae0} and \ref{sb:boundsex}. This proves sharpness for an arbitrary reference-target path pair.
\end{proof}
}

\subsection{A proof of Lemma \ref{lem2}}\label{sec:lem2}

\begin{proof}
First suppose Assumption \ref{ass:extpt} holds. Applying it to cohort $g$ and to the never-treated cohort gives
\[
\mathbb E[Y_t(\infty)-Y_0(\infty)\mid G=g]
=
\mathbb E[Y_t(\infty)-Y_0(\infty)\mid G=\infty].
\]
Rearranging yields $SB_t^g=SB_0^g$.

Conversely, suppose $SB_t^g=SB_0^g$ for every $g$ and $t$. Rearrangement gives
\[
\mathbb E[Y_t(\infty)-Y_0(\infty)\mid G=g]
=
\mathbb E[Y_t(\infty)-Y_0(\infty)\mid G=\infty]
\]
for every finite $g\in\operatorname{supp}(G)$. Thus the conditional never-treated trend has the same value for every element of $\operatorname{supp}(G)$. By the law of iterated expectations, this common value equals the unconditional trend
\[
\mathbb E[Y_t(\infty)-Y_0(\infty)].
\]
This is Assumption \ref{ass:extpt} for every supported treatment path.
\end{proof}

{

{\subsection{A proof of Proposition \ref{bound_multiple}}\label{sec:bound_multiple}

\subsubsection*{Validity of the bounds}
\begin{proof}
Fix a finite supported cohort $g$. By Lemma \ref{lem2}, extended PT is equivalent to
\[
SB_\tau^g=SB_g^g
\qquad\text{for every period }\tau.
\]
For each $t_0\in\mathcal C_g$,
\begin{align}
AE_{t_0}^g
&=\theta_{t_0}^{DIM,g}-SB_{t_0}^g\nonumber\\
&=\theta_{t_0}^{DIM,g}-SB_g^g\nonumber\\
&=\theta_{t_0}^{DIM,g}-\big(\theta_g^{DIM,g}-ATT_g^g\big)\nonumber\\
&=ATT_g^g-\theta_{g,t_0}^{DID,g}.
\label{eq:staggered-validity-ae}
\end{align}
Consequently, Assumption \ref{extendedBAE} holds for cohort $g$ if and only if
\begin{equation}\label{eq:staggered-validity-feasibility}
\left|ATT_g^g-\theta_{g,t_0}^{DID,g}\right|
\leq c^{g-t_0}|ATT_g^g|
\qquad\text{for every }t_0\in\mathcal C_g,
\end{equation}
which is exactly the statement $ATT_g^g\in\mathcal F_g(c)$.

For any $t\geq g$, the same bias stability gives
\begin{align}
ATT_t^g
&=\theta_t^{DIM,g}-SB_t^g\nonumber\\
&=\theta_t^{DIM,g}-SB_g^g\nonumber\\
&=\theta_t^{DIM,g}-\big(\theta_g^{DIM,g}-ATT_g^g\big)\nonumber\\
&=\theta_{t,g}^{DID,g}+ATT_g^g.
\label{eq:staggered-validity-post}
\end{align}
Combining $ATT_g^g\in\mathcal F_g(c)$ with \eqref{eq:staggered-validity-post} yields
$ATT_t^g\in\theta_{t,g}^{DID,g}+\mathcal F_g(c)$, and substituting the explicit form of $\mathcal F_g(c)$ gives \eqref{eq:staggered-att-set}. If $\mathcal F_h(c)$ is empty for any finite supported cohort $h$, \eqref{eq:staggered-validity-feasibility} cannot hold for that cohort, so no DGP can satisfy the joint restrictions.
\end{proof}

\subsubsection*{Sharpness of the bounds}
\begin{proof}
Assume $\mathcal F_h(c)\neq\varnothing$ for every finite $h\in\operatorname{supp}(G)$. Fix an arbitrary target cohort $g$, period $t\geq g$, and candidate
\begin{equation}\label{eq:staggered-sharp-candidate}
\theta\in\theta_{t,g}^{DID,g}+\mathcal F_g(c).
\end{equation}
Define the target adoption-date effect for cohort $g$ by
\begin{equation}\label{eq:staggered-sharp-target-cohort}
\vartheta_g
\equiv
\theta-\theta_{t,g}^{DID,g}.
\end{equation}
Equation \eqref{eq:staggered-sharp-candidate} implies $\vartheta_g\in\mathcal F_g(c)$. For every other finite supported cohort $h$, select an arbitrary $\vartheta_h\in\mathcal F_h(c)$. Thus, for every finite supported $h$,
\begin{equation}\label{eq:staggered-sharp-feasibility}
|\vartheta_h-\theta_{h,t_0}^{DID,h}|
\leq c^{h-t_0}|\vartheta_h|
\qquad\text{for every }t_0\in\mathcal C_h.
\end{equation}
Define the target time-invariant selection bias for cohort $h$ and the never-treated observed mean by
\begin{equation}\label{eq:staggered-sharp-targets}
s^h\equiv\theta_h^{DIM,h}-\vartheta_h,
\qquad
m_{\infty,\tau}\equiv\mathbb E[Y_\tau\mid G=\infty].
\end{equation}

For each finite cohort $h$, let $\bm d^h$ denote its realized staggered-adoption path, and let $\bm d_\infty$ denote the never-treated path. Construct the full potential-outcome vector on the original observed-data probability space by retaining every observed cohort-path cell:
\begin{equation}\label{eq:staggered-sharp-observed}
\widetilde Y_\tau(\bm d^h)=Y_\tau
\quad\text{on }\{G=h\},
\qquad
\widetilde Y_\tau(\bm d_\infty)=Y_\tau
\quad\text{on }\{G=\infty\}.
\end{equation}
For every finite supported cohort $h$ and every period $\tau\in\mathcal T_0\cup\{1,\ldots,T\}$, define its missing never-treated outcome by
\begin{equation}\label{eq:staggered-sharp-completion}
\widetilde Y_\tau(\bm d_\infty)
=m_{\infty,\tau}+s^h
\quad\text{on }\{G=h\}.
\end{equation}
Set all remaining missing treatment-path potential outcomes equal to zero. The assignments are jointly defined, real-valued, and integrable. Since \eqref{eq:staggered-sharp-observed} retains the potential outcome along every group's realized path,
\begin{equation}\label{eq:staggered-sharp-consistency}
\widetilde Y_\tau(\bm D)=Y_\tau
\qquad\text{almost surely},
\end{equation}
so the construction satisfies consistency and preserves the complete observed-data distribution.

We first verify extended PT. For every finite supported cohort $h$ and every period $\tau$, equations \eqref{eq:staggered-sharp-observed} and \eqref{eq:staggered-sharp-completion} imply
\begin{align}
SB_\tau^h
&=\mathbb E[\widetilde Y_\tau(\bm d_\infty)\mid G=h]
 -\mathbb E[\widetilde Y_\tau(\bm d_\infty)\mid G=\infty]\nonumber\\
&=\{m_{\infty,\tau}+s^h\}-m_{\infty,\tau}
=s^h.
\label{eq:staggered-sharp-bias}
\end{align}
Thus $SB_\tau^h=SB_0^h$ for every $h$ and $\tau$, so Lemma \ref{lem2} implies Assumption \ref{ass:extpt}. Equivalently, the cohort shift cancels from the never-treated trend:
\begin{align}
\mathbb E[\widetilde Y_\tau(\bm d_\infty)-\widetilde Y_0(\bm d_\infty)\mid G=h]
&=(m_{\infty,\tau}+s^h)-(m_{\infty,0}+s^h)\nonumber\\
&=m_{\infty,\tau}-m_{\infty,0}\nonumber\\
&=\mathbb E[Y_\tau-Y_0\mid G=\infty].
\label{eq:staggered-sharp-trends}
\end{align}
The same equality holds for $G=\infty$; hence all supported cohorts have the same never-treated trend. By iterated expectations, this common conditional trend equals the unconditional never-treated trend required by Assumption \ref{ass:extpt}.

Next, for every finite supported cohort $h$ and $t_0\in\mathcal C_h$, the anticipation effect generated by the completion is
\begin{align}
AE_{t_0}^h
&=\mathbb E[\widetilde Y_{t_0}(\bm d^h)-\widetilde Y_{t_0}(\bm d_\infty)\mid G=h]\nonumber\\
&=\mathbb E[Y_{t_0}\mid G=h]-\{m_{\infty,t_0}+s^h\}\nonumber\\
&=\theta_{t_0}^{DIM,h}-s^h\nonumber\\
&=\vartheta_h-\theta_{h,t_0}^{DID,h},
\label{eq:staggered-sharp-ae}
\end{align}
where the last equality follows from \eqref{eq:staggered-sharp-targets} and $\theta_{h,t_0}^{DID,h}=\theta_h^{DIM,h}-\theta_{t_0}^{DIM,h}$. The cohort-$h$ treatment effect at its adoption date is
\begin{align}
ATT_h^h
&=\mathbb E[\widetilde Y_h(\bm d^h)-\widetilde Y_h(\bm d_\infty)\mid G=h]\nonumber\\
&=\theta_h^{DIM,h}-s^h
=\vartheta_h.
\label{eq:staggered-sharp-adoption-att}
\end{align}
Combining \eqref{eq:staggered-sharp-feasibility}, \eqref{eq:staggered-sharp-ae}, and \eqref{eq:staggered-sharp-adoption-att} gives
\[
|AE_{t_0}^h|\leq c^{h-t_0}|ATT_h^h|
\qquad(t_0\in\mathcal C_h),
\]
so Assumption \ref{extendedBAE} holds simultaneously for every positive-probability cohort.

Finally, for the target cohort $g$ and target period $t$,
\begin{align}
ATT_t^g
&=\mathbb E[\widetilde Y_t(\bm d^g)-\widetilde Y_t(\bm d_\infty)\mid G=g]\nonumber\\
&=\theta_t^{DIM,g}-s^g\nonumber\\
&=\theta_t^{DIM,g}-\theta_g^{DIM,g}+\vartheta_g\nonumber\\
&=\theta_{t,g}^{DID,g}+\vartheta_g
=\theta,
\label{eq:staggered-sharp-target-att}
\end{align}
where the last equality follows from \eqref{eq:staggered-sharp-target-cohort}. Thus every candidate in $\theta_{t,g}^{DID,g}+\mathcal F_g(c)$ is attained by a single joint DGP that preserves the observed distribution and satisfies both maintained assumptions for all supported cohorts. This proves sharpness whenever every cohort-specific feasibility set is nonempty.
\end{proof}

}

\subsection{$\Theta(A_{\text{PT-R}}) \subseteq \Theta(A_{\text{R}})$ and $\Theta(A_{\text{AE-R}}) \subseteq \Theta(A_{\text{R}})$}\label{sec:inc}
First, for any $x_0 \in \Theta(A_{\text{PT-R}}) = \Theta^{\text{PT-RDID}}$, we have
$$
\min_{t_0 \in \mathcal I_0}\{ \theta_{t_0}^{DID} \} \leq x_0 \leq \max_{t_0 \in \mathcal I_0}\{ \theta_{t_0}^{DID} \}.
$$
Also, note that
$$
\min \bigg\{\frac{\theta_{t_0}^{DID}}{1+c^{|t_0|+1}}, \frac{\theta_{t_0}^{DID}}{1-c^{|t_0|+1}} \bigg\} \leq \theta_{t_0}^{DID} \leq \max \bigg\{\frac{\theta_{t_0}^{DID}}{1+c^{|t_0|+1}}, \frac{\theta_{t_0}^{DID}}{1-c^{|t_0|+1}} \bigg\},
$$
for any $t_0 \in \mathcal I_0$.
Thus, we have
$$
\min_{t_0 \in \mathcal I_0} \Bigg\{ \min\bigg\{\frac{\theta_{t_0}^{DID}}{1+c^{|t_0|+1}}, \frac{\theta_{t_0}^{DID}}{1-c^{|t_0|+1}} \bigg\} \Bigg\} \leq \min_{t_0 \in \mathcal I_0}\{ \theta_{t_0}^{DID} \},
$$
and
$$
\max_{t_0 \in \mathcal I_0}\{ \theta_{t_0}^{DID} \} \leq \max_{t_0 \in \mathcal I_0} \Bigg\{ \max\bigg\{\frac{\theta_{t_0}^{DID}}{1+c^{|t_0|+1}}, \frac{\theta_{t_0}^{DID}}{1-c^{|t_0|+1}} \bigg\} \Bigg\}.
$$
Hence, we have $x_0 \in \Theta(A_{\text{R}})$, which implies $\Theta(A_{\text{PT-R}}) \subseteq \Theta(A_{\text{R}})$.

On the other hand, consider an arbitrary $x_0 \in \Theta(A_{\text{AE-R}}) = \Theta^{\text{AE-RDID}}$. If $x_0$ belongs to the first intersection interval of $\Theta^{\text{AE-RDID}}$ (see Remark \ref{rm.rdid2.convex}), we have 
$$
\min_{t_0 \in \mathcal I_0} \bigg\{ \frac{\theta_{t_0}^{DID}}{1+c^{|t_0|+1}} \bigg\} \leq \max_{t_0 \in \mathcal I_0} \bigg\{ \frac{\theta_{t_0}^{DID}}{1+c^{|t_0|+1}} \bigg\} \leq x_0 \leq \min_{t_0 \in \mathcal I_0} \bigg\{ \frac{\theta_{t_0}^{DID}}{1-c^{|t_0|+1}} \bigg\} \leq \max_{t_0 \in \mathcal I_0} \bigg\{ \frac{\theta_{t_0}^{DID}}{1-c^{|t_0|+1}} \bigg\}.
$$
Similarly, if $x_0$ belongs to the second intersection interval of $\Theta^{\text{AE-RDID}}$, we have 
$$
\min_{t_0 \in \mathcal I_0} \bigg\{ \frac{\theta_{t_0}^{DID}}{1-c^{|t_0|+1}} \bigg\} \leq \max_{t_0 \in \mathcal I_0} \bigg\{ \frac{\theta_{t_0}^{DID}}{1-c^{|t_0|+1}} \bigg\} \leq x_0 \leq \min_{t_0 \in \mathcal I_0} \bigg\{ \frac{\theta_{t_0}^{DID}}{1+c^{|t_0|+1}} \bigg\} \leq \max_{t_0 \in \mathcal I_0} \bigg\{ \frac{\theta_{t_0}^{DID}}{1+c^{|t_0|+1}} \bigg\}.
$$
Therefore, we have $x_0 \in \Theta(A_{\text{R}})$, which implies $\Theta(A_{\text{AE-R}}) \subseteq \Theta(A_{\text{R}})$.

\subsection{Proof of validity of the inference method}\label{sec:inferenceproof}

{

For each $t_0\in\mathcal I_0$, write
\[
\vartheta(t_0)
\equiv
\theta_{t_0}^{DID},
\]
and let $[L_{t_0},U_{t_0}]$ satisfy
\[
\mathbb P\{\vartheta(t_0)<L_{t_0}\}\leq\alpha/2,
\qquad
\mathbb P\{\vartheta(t_0)>U_{t_0}\}\leq\alpha/2.
\]
Set
\[
L\equiv\min_{t_0\in\mathcal I_0}L_{t_0},
\qquad
U\equiv\max_{t_0\in\mathcal I_0}U_{t_0}.
\]
Because $\mathcal I_0$ is finite, choose
\[
t_0^{\min}\in\arg\min_{t_0\in\mathcal I_0}\vartheta(t_0),
\qquad
t_0^{\max}\in\arg\max_{t_0\in\mathcal I_0}\vartheta(t_0).
\]
Then
\[
\Theta^{\text{PT-RDID}}
=
[\vartheta(t_0^{\min}),\vartheta(t_0^{\max})].
\]
The event that the confidence envelope fails to contain the entire identified set is
\[
\{\Theta^{\text{PT-RDID}}\nsubseteq[L,U]\}
=
\{\vartheta(t_0^{\min})<L\}
\cup
\{\vartheta(t_0^{\max})>U\}.
\]
Since $L\leq L_{t_0^{\min}}$ and $U\geq U_{t_0^{\max}}$,
\begin{align*}
\mathbb P\{\Theta^{\text{PT-RDID}}\nsubseteq[L,U]\}
&\leq
\mathbb P\{\vartheta(t_0^{\min})<L_{t_0^{\min}}\}
+
\mathbb P\{\vartheta(t_0^{\max})>U_{t_0^{\max}}\}\\
&\leq\alpha.
\end{align*}
Therefore
\[
\mathbb P\{\Theta^{\text{PT-RDID}}\subseteq[L,U]\}\geq1-\alpha.
\]
This directly proves simultaneous containment of the entire identified set, and hence coverage of the true ATT whenever it belongs to that set.
}

}

\subsection{A proof of Proposition \ref{prop:mrbound}}\label{sec:prop:mrbound}
Fix the observed-data distribution $P$. We write $\Theta(B)$ for $\Theta_P(B)$, suppressing $P$ as in the main text. We retain $\mathcal M_P(B)$ for laws with observed marginal $P$ and $\mathcal M(B)$ for all laws satisfying $B$.
By the validity and sharpness parts of Propositions \ref{prop1}, \ref{bound_multiple_preperiods}, and \ref{bound_relaxtwo},
\[
\Theta(A_{\text{PT-R}})=\Theta^{\text{PT-RDID}},
\qquad
\Theta(A_{\text{AE-R}})=\Theta^{\text{AE-RDID}},
\qquad
\Theta(A_{\text{R}})=\Theta^{\text{RDID}}.
\]
A bundle $B$ is data-consistent at $P$ if and only if $\Theta(B)\neq\varnothing$, since the maintained integrability conditions make $ATT(Q)$ well defined. The sharpness constructions cited above retain every observed outcome. For every $t\in\mathcal T_0\setminus\mathcal I_0$, retain $Y_t(D)=Y_t$ and set $Y_t(1-D)=0$. The resulting full-data laws preserve the entire observed-data distribution $P$. The baseline restrictions are checked only for periods in $\mathcal I_0$. The proof proceeds in three steps.

\subsubsection*{Step 1: Data consistency}

Under Assumption \ref{ass:NoAE}, $SB_{t_0}=\theta_{t_0}^{OLS}$ for every $t_0\in\mathcal I_0$, so
\[
\Theta^{\text{PT-RDID}}
=
\left[
\theta_1^{OLS}-\max_{t_0\in\mathcal I_0}\theta_{t_0}^{OLS},
\;
\theta_1^{OLS}-\min_{t_0\in\mathcal I_0}\theta_{t_0}^{OLS}
\right].
\]
Because $\mathcal I_0$ is finite and the maintained regularity conditions make every displayed mean finite, this interval is nonempty. By sharpness, $A_{\text{PT-R}}$ is data-consistent at every $P$. Appendix \ref{sec:inc} gives $\Theta(A_{\text{R}})\supseteq\Theta(A_{\text{PT-R}})\neq\varnothing$, so $A_{\text{R}}$ is also data-consistent at every $P$. The characterization of $A_{\text{AE-R}}$ follows immediately from $\Theta(A_{\text{AE-R}})=\Theta^{\text{AE-RDID}}$.

Now suppose $\mathcal M_P(A_0)\neq\varnothing$ and take any $Q\in\mathcal M_P(A_0)$. No anticipation and strong PT imply, for every $t_0\in\mathcal I_0$,
\[
\theta_{t_0}^{OLS}=AE_{t_0}+SB_{t_0}=SB_1(Q).
\]
The right-hand side is common to all selected baseline periods, so their OLS estimands must have one value $\bar s$. This value is determined by $P$. Every such $Q$ therefore satisfies
\[
ATT(Q)=\theta_1^{OLS}-SB_1(Q)=\theta_1^{OLS}-\bar s
=\theta_{\mathrm{com}}^{DID}.
\]

Conversely, suppose $\theta_{t_0}^{OLS}=\bar s$ for all $t_0\in\mathcal I_0$. The displayed $A_{\text{PT-R}}$ interval is then $\{\theta_{\mathrm{com}}^{DID}\}$, and sharpness ensures that $\mathcal M_P(A_{\text{PT-R}})\neq\varnothing$. For any $Q\in\mathcal M_P(A_{\text{PT-R}})$, no anticipation gives $SB_{t_0}=\bar s$ on $\mathcal I_0$. Bias set stability forces $SB_1\in[\bar s,\bar s]$, so strong PT also holds. Thus $\mathcal M_P(A_{\text{PT-R}})\subseteq\mathcal M_P(A_0)$. The reverse inclusion follows from $A_0\succeq A_{\text{PT-R}}$. At this $P$, these two classes are therefore equal and nonempty, and $\Theta(A_0)=\{\theta_{\mathrm{com}}^{DID}\}$. This proves part (i).

\subsubsection*{Step 2: The order structure and the minimum relaxations}

The implications displayed in the text give
\[
A_0\succeq A_{\text{PT-R}}\succeq A_{\text{R}},
\qquad
A_0\succeq A_{\text{AE-R}}\succeq A_{\text{R}}.
\]
If $A_0$ is data-consistent at $P$, its equivalence class is the greatest element of the menu. Hence $\mathcal R^*(A_0)=\{A_0\}$. This also covers a singleton $\mathcal I_0$.

Now suppose $A_0$ is not data-consistent. Step 1 implies that the baseline OLS estimands are nonconstant, so $\mathcal I_0$ contains at least two periods. Choose distinct $t',t''\in\mathcal I_0$. The next two laws establish the required comparisons between the unrestricted classes $\mathcal M(B)$. They need not have observed marginal $P$.

First, let $Q_a$ satisfy $D\sim\operatorname{Bernoulli}(1/2)$, with all potential outcomes degenerate conditional on $D$, and set $Y_t(0)=0$ for every $t\in\mathcal T_0\cup\{1\}$, $Y_1(1)=1$, $Y_{t'}(1)=c^{|t'|+1}/2$, and $Y_{t_0}(1)=0$ for $t_0\in\mathcal T_0\setminus\{t'\}$. All selection biases are zero, so strong PT and bias-set stability hold. Moreover, $ATT=1$ and
\[
0<AE_{t'}=\frac{c^{|t'|+1}}{2}<c^{|t'|+1}|ATT|,
\]
while every other baseline anticipation effect is zero. Thus bounded anticipation holds but no anticipation fails. Hence
\[
Q_a\in\mathcal M(A_{\text{AE-R}})\cap\mathcal M(A_{\text{R}}),
\qquad
Q_a\notin\mathcal M(A_{\text{PT-R}})\cup\mathcal M(A_0).
\]

Second, let $Q_b$ satisfy $D\sim\operatorname{Bernoulli}(1/2)$, again with degenerate potential outcomes conditional on $D$, and set
\[
Y_{t'}(0)=\mathbbm 1\{D=1\},
\qquad
Y_{t_0}(0)=0\quad(t_0\in\mathcal T_0\setminus\{t'\}),
\qquad
Y_1(0)=\frac{\mathbbm 1\{D=1\}}{2},
\]
together with $Y_t(1)=Y_t(0)$ for every $t\in\mathcal T_0\cup\{1\}$. Then all anticipation effects and the ATT are zero, so no anticipation and bounded anticipation hold. The selection biases satisfy $SB_{t'}=1$, $SB_{t''}=0$, and $SB_1=1/2\in[0,1]$, so bias-set stability holds but strong PT fails. Hence
\[
Q_b\in\mathcal M(A_{\text{PT-R}})\cap\mathcal M(A_{\text{R}}),
\qquad
Q_b\notin\mathcal M(A_{\text{AE-R}})\cup\mathcal M(A_0).
\]
The law $Q_a$ shows that $A_{\text{AE-R}}\not\succeq A_{\text{PT-R}}$ and $A_{\text{R}}\not\succeq A_{\text{PT-R}}$. The law $Q_b$ shows that $A_{\text{PT-R}}\not\succeq A_{\text{AE-R}}$ and $A_{\text{R}}\not\succeq A_{\text{AE-R}}$. Together with the displayed implications, these relations show that $A_{\text{PT-R}}$ and $A_{\text{AE-R}}$ are incomparable and that each is strictly stronger than $A_{\text{R}}$.

Because $A_{\text{PT-R}}$ is always data-consistent and strictly stronger than $A_{\text{R}}$, $A_{\text{R}}$ cannot be a minimum relaxation. If $\Theta(A_{\text{AE-R}})=\varnothing$, only $A_{\text{PT-R}}$ remains. If $\Theta(A_{\text{AE-R}})\neq\varnothing$, the two incomparable bundles $A_{\text{PT-R}}$ and $A_{\text{AE-R}}$ are the minimum relaxations. This proves part (ii).

\subsubsection*{Step 3: The MR bound}

Definition \ref{def:mrbound} takes the union of the identified sets associated with the bundles characterized in Step 2. Combining that characterization with Step 1 gives exactly the three cases in \eqref{eq:mrbound-cases}. Nonemptiness follows because $A_0$ is data-consistent in the first regime and $A_{\text{PT-R}}$ is data-consistent in the other two. Finally, $A_0\succeq A_{\text{PT-R}}$ gives $\Theta(A_0)\subseteq\Theta(A_{\text{PT-R}})$ in the first regime, and Appendix \ref{sec:inc} gives
\[
\Theta(A_{\text{PT-R}})\cup\Theta(A_{\text{AE-R}})
\subseteq
\Theta(A_{\text{R}}).
\]
The asserted inclusions follow in all three regimes.

\clearpage

\clearpage

\counterwithin{figure}{section}

\counterwithin{table}{section}
\renewcommand\theclaim{D.\arabic{claim}} 
\setcounter{claim}{0}
\begin{center}
\vspace{2cm}
    \huge{{Online Appendix}}

\huge{From Pre-Trends to Bounds: \\ Robust Difference-in-Differences}\\
\bigskip
    \Large{Kyunghoon Ban, D\'esir\'e K\'edagni, Zhengrun Chen, and Augustine Denteh}
\end{center}
\vspace{1cm}
\pagenumbering{arabic}
\renewcommand{\thepage}{OA-\arabic{page}}

{
\raggedbottom
\setlength{\emergencystretch}{2em}
\section{\texorpdfstring{{Details of} Estimation and Inference}{Details of Estimation and Inference}}\label{app:inference-details}

For one focal post-treatment period in the no-covariate model, the identified sets are maps of a finite vector of DID contrasts, $\bm{\theta}^{{DID}}$. The algorithms below use this common first stage for identified-set and ATT inference. Covariate-adjusted or pooled post-treatment contrasts require their own identification and first-stage justification.\footnote{The application to \citet{baten} retains the original specification's covariate and fixed effects in the saturated regression. Section~\ref{sec:empirical} discusses the assumptions needed to interpret this adjusted, pooled contrast.}

Denote the unique post-treatment period by $1$, a nonempty finite baseline set by $\mathcal I_0$, and the size of the baseline set by $K=|\mathcal I_0|$, respectively. 
Furthermore, a baseline index $t_0$ runs over $\mathcal I_0$, a cluster index $i$ over $\{1,\ldots,n\}$, an observation index $\ell$ over $\{1,\ldots,N\}$, and a simulation index $b$ over $\{1,\ldots,B\}$.
The disjoint sets $\mathcal O_i$ partition the $N$ observations into $n$ independent clusters. 
Using the baseline-specific DID estimand from Lemma~\ref{lem:samesign}, write
\begin{gather*}
\theta_{t_0}^{{DID}}=\theta_1^{OLS}-\theta_{t_0}^{OLS}\quad(t_0\in\mathcal I_0),\qquad
\bm{\theta}^{{DID}}\equiv(\theta_{t_0}^{{DID}})_{t_0\in\mathcal I_0}\in\mathbb R^K,\\
g_{t_0,\sigma}(\bm{\theta}^{{DID}};c)=\frac{\theta_{t_0}^{{DID}}}{1+\sigma c_{t_0}},
\quad c_{t_0}=c^{|t_0|+1},\quad (t_0,\sigma)\in\mathcal I_0\times\{-1,+1\}.
\end{gather*}
Define $\mathcal S_{+1}=[0,\infty)$ and $\mathcal S_{-1}=(-\infty,0]$.
Throughout the section, $s,\sigma\in\{-1,+1\}$ are numerical signs so that $1+s c_{t_0}$ always has its ordinary arithmetic meaning, and subscripts $+$ and $-$ are shorthand for $+1$ and $-1$.
Unless noted otherwise, minimum, maximum, or product over $t_0$ uses all of $\mathcal I_0$, and one over $(t_0,\sigma)$ uses $\mathcal I_0\times\{-1,+1\}$.

{The identification results give the following maps.}
\begin{align}
\Gamma_{\text{PT-R}}(\bm{\theta}^{{DID}})&=[\min_{t_0}\theta_{t_0}^{{DID}},\max_{t_0}\theta_{t_0}^{{DID}}],\label{eq:Gamma-I}\\
\Gamma_{\text{AE-R},s}(\bm{\theta}^{{DID}};c)
&=[\max_{t_0}g_{t_0,s}(\bm{\theta}^{{DID}};c),\min_{t_0}g_{t_0,-s}(\bm{\theta}^{{DID}};c)]
  \cap\mathcal S_s,\quad s\in\{-1,+1\},\label{eq:Gamma-II-branch}\\
\Gamma_{\text{AE-R}}(\bm{\theta}^{{DID}};c)
&=\Gamma_{\text{AE-R},+}(\bm{\theta}^{{DID}};c)\cup\Gamma_{\text{AE-R},-}(\bm{\theta}^{{DID}};c),\label{eq:Gamma-II}\\
\Gamma_{\text{R}}(\bm{\theta}^{{DID}};c)
&=[\min_{t_0,\sigma}g_{t_0,\sigma}(\bm{\theta}^{{DID}};c),
   \max_{t_0,\sigma}g_{t_0,\sigma}(\bm{\theta}^{{DID}};c)],\label{eq:Gamma-III}
\end{align}
where a reversed interval is empty. 
Let $\Gamma_*(\bm{\theta}^{{DID}};c)$ be the MR map of Proposition \ref{prop:mrbound}, and let $\Gamma_0(\bm{\theta}^{{DID}})$ equal the common DID singleton when all coordinates of $\bm{\theta}^{{DID}}$ are equal and the empty set otherwise.
We use
\begin{equation}
\Gamma_*(\bm{\theta}^{{DID}};c)\subseteq\Gamma_{\text{PT-R}}(\bm{\theta}^{{DID}})\cup\Gamma_{\text{AE-R}}(\bm{\theta}^{{DID}};c)
\subseteq\Gamma_{\text{R}}(\bm{\theta}^{{DID}};c),
\label{eq:Gamma-inclusions}
\end{equation}
implied by Proposition \ref{prop:mrbound}.
Applying these maps to $\widehat{\bm{\theta}}^{{DID}}$ gives estimated identified sets, but it does not add sampling uncertainty. 
In particular, an empty estimated set alone does not reject the model.

For $j\in\{0,\text{PT-R},\text{AE-R},\text{R},*\}$, we denote an identified-set confidence envelope by $C^S_j$ and an ATT confidence set by $C^P_j$, respectively.
The former covers the entire population set simultaneously. The latter covers each admissible scalar ATT. The ATT interpretation maintains the corresponding identifying restrictions.
For a regular class $\mathcal P$ of observed-data laws, the respective coverage requirements\footnote{In \eqref{eq:uniform-parameter-coverage}, the infimum ranges over admissible pairs $(P,\vartheta)$ with $\vartheta\in\Gamma_j(\bm{\theta}^{{DID}}(P);c)$ so that laws with empty population identified sets therefore contribute no admissible pairs.
Coverage here is asymptotic as the number of independent sampling clusters $n$ grows.
} are
\begin{align}
\liminf_{n\to\infty}\inf_{P\in\mathcal P}
 P\{\Gamma_j(\bm{\theta}^{{DID}}(P);c)\subseteq C^S_j(c)\}&\geq1-\alpha,
 \label{eq:uniform-set-coverage}\\
\liminf_{n\to\infty}\inf_{\substack{P\in\mathcal P\\
             \vartheta\in\Gamma_j(\bm{\theta}^{{DID}}(P);c)}}
 P\{\vartheta\in C^P_j(c)\}&\geq1-\alpha.
 \label{eq:uniform-parameter-coverage}
\end{align}

We maintain fixed $K$, $0<\alpha<1/2$, and $c\in[0,\bar c]$ with $\bar c<1$ and impose three first-stage conditions: 
(i) a Gaussian approximation to $\sqrt n(\widehat{\bm{\theta}}^{{DID}}-\bm{\theta}^{{DID}}(P))$, with covariance $\bm\Sigma_P$, uniformly over axis-aligned rectangles and $P\in\mathcal P$;
(ii) uniform consistency of $n\widehat{\bm V}$ for $\bm\Sigma_P$; and
(iii) marginal variances uniformly bounded above and away from zero.
The covariance estimator is symmetric positive semidefinite, as in the cluster-score construction below. 
These general conditions permit singular primitive covariance and impose no convergence rate for covariance consistency. 
Gaussian covariance comparison and anti-concentration justify fixed-level calibration of finite signed-coordinate maxima using the estimated covariance matrix \citep[Theorems 2--3]{CCK2015anticoncentration}.\footnote{The coverage justification for Algorithm~\ref{alg:ii-set} additionally imposes Condition~P \citep{chernozhukov2013intersection} over the subclass $\mathcal P_{\mathrm{CLR}}$ defined below, including positive definiteness of the primitive covariance and the approximation requirements in \eqref{eq:CLR-P-regularity}. 
These stronger conditions are specific to that justification. The rectangle construction retains the general first-stage conditions.}

For critical-value calculations, given simulated values
$\bm v=(v_b)_{b=1}^B$ and $p\in[0,1]$, define
\[
Q_B(p;\bm v)=v_{(k_B(p))},
\qquad
k_B(p)=\min\{B,\max\{1,\lceil Bp\rceil\}\},
\]
where $v_{(k)}$ is the $k$th smallest simulated value.
When the argument is written as an expression in $b$, $Q_B$
uses all $B$ simulated values of that expression.
Algorithm~\ref{alg:first-stage} computes the joint DID estimates
and generates normal multiplier perturbations using common
multipliers across coordinates.
Write $\Phi$ for the standard normal distribution function and $z_p=\Phi^{-1}(p)$.

\begin{algorithm}[!htbp]
\small\setstretch{1.05}
\setlength{\linewidth}{\hsize}
\let\fullwidthdisplay\relax \SetAlgoNlRelativeSize{-1}
\DontPrintSemicolon
\caption{Joint DID estimation and common normal multiplier draws}
\label{alg:first-stage}
\Input{Sample of $N$ observations in $n$ independent clusters (or $n=N$ independent observations), clustering unit, $\mathcal I_0$, focal period 1, confidence level $1-\alpha$, draw count $B$, and random seed.}
\Output{Joint estimates $\widehat{\bm{\theta}}^{{DID}}$, covariance $\widehat{\bm V}$,
common draws $\bm Z_b$, marginal intervals, and joint rectangle $\mathcal C_{{DID}}$.}
Fit the saturated regression
\begin{equation}
Y_\ell=\sum_{t\in\mathcal I_0\cup\{1\}}
\mathbbm1\{t_\ell=t\}(\alpha_t+\beta_tD_\ell)+u_\ell,
\quad \ell=1,\ldots,N.
\label{eq:saturated-regression}
\end{equation}
Let $\bm X$ be its $N\times p_X$ design matrix, $p_X=2(K+1)$,
$\bm Y=(Y_\ell)_{\ell=1}^N$, and
$\bm\beta=(\alpha_t,\beta_t)_{t\in\mathcal I_0\cup\{1\}}$ the full coefficient
vector. With a $K\times p_X$ contrast matrix $\bm R$, compute
$\widehat{\bm\beta}=(\bm X'\bm X)^{-1}\bm X'\bm Y$ and
$\widehat{\bm{\theta}}^{{DID}}=\bm R\widehat{\bm\beta}$, so
$\widehat{\theta}_{t_0}^{{DID}}=\widehat\beta_1-\widehat\beta_{t_0}$ for $t_0\in\mathcal I_0$.
Set $\widehat u_\ell=Y_\ell-\bm x_\ell'\widehat{\bm\beta}$ for $\ell=1,\ldots,N$.\;
For $i=1,\ldots,n$, compute the raw-scale cluster score and covariance
\[
\bm a_i=\bm R(\bm X'\bm X)^{-1}
  \sum_{\ell\in\mathcal O_i}\bm x_\ell\widehat u_\ell,
\qquad \widehat{\bm V}=f_n\sum_{i=1}^n\bm a_i\bm a_i',
\qquad s_{t_0}=\sqrt{\widehat V_{t_0t_0}}\ (t_0\in\mathcal I_0),
\]
where $\bm x_\ell'$ is row $\ell$ of $\bm X$ and $f_n$ is defined below.\;
Draw $\xi_{ib}\sim N(0,1)$ independently over $i=1,\ldots,n$ and
$b=1,\ldots,B$, independently of the data. For each $b$, set
\[
\bm Z_b=\sqrt{f_n}\sum_{i=1}^n\xi_{ib}\bm a_i,\qquad
S_{b,t_0}=Z_{b,t_0}/s_{t_0}\ (t_0\in\mathcal I_0),\qquad
\widehat{\bm{\theta}}_b^{{DID},\#}=\widehat{\bm{\theta}}^{{DID}}+\bm Z_b.
\]
Reuse these same draws for every transformation and model.\;
For a finite index set $\mathcal R$ and known loadings $\bm w_r\in\mathbb R^K$,
define $g_r(\bm{\theta}^{{DID}})=\bm w_r'\bm{\theta}^{{DID}}$ for $r\in\mathcal R$, requiring positive variances. Compute
$\widehat g_r=\bm w_r'\widehat{\bm{\theta}}^{{DID}}$,
$s_r=(\bm w_r'\widehat{\bm V}\bm w_r)^{1/2}$, and
$Z^g_{b,r}=\bm w_r'\bm Z_b$ for $(b,r)\in\{1,\ldots,B\}\times\mathcal R$.
For $A_{\text{PT-R}}$, use $\bm w_r=\bm e_{t_0}$ for $r=t_0$, where $\bm e_{t_0}$ is the unit vector for baseline $t_0$.
For $A_{\text{R}}$, use $\bm w_r=\bm e_{t_0}/(1+\sigma c_{t_0})$ for $r=(t_0,\sigma)$.\;
For each $t_0\in\mathcal I_0$, save the marginal interval
$[\widehat{\theta}_{t_0}^{{DID}}\pm z_{1-\alpha/2}s_{t_0}]$ and form
\begin{equation}
\begin{gathered}
q_R=Q_B\bigl(1-\alpha;(\max_{t_0\in\mathcal I_0}|S_{b,t_0}|)_{b=1}^B\bigr),
\qquad \mathcal C_{{DID}}=\prod_{t_0\in\mathcal I_0}[\ell_{t_0},u_{t_0}],\\
\ell_{t_0}=\widehat{\theta}_{t_0}^{{DID}}-q_Rs_{t_0},\qquad
u_{t_0}=\widehat{\theta}_{t_0}^{{DID}}+q_Rs_{t_0}\quad(t_0\in\mathcal I_0).
\end{gathered}
\label{eq:primitive-rectangle}
\end{equation}\par
\end{algorithm}
\FloatBarrier

In Algorithm~\ref{alg:first-stage}, we use $f_n=\frac{n(N-1)}{(n-1)(N-r_X)}$ where $r_X=\operatorname{rank}(\bm X)$ based on a conventional finite-sample covariance correction \citep[equation (12)]{CameronMiller2015}.
When individual observations are the independent units ($n=N$), this reduces
to the usual correction $f_n=N/(N-r_X)$.
We require $n>1$, $N>r_X$, and a full-rank design, and the asymptotic argument requires $f_n\to1$.

The saturated regression estimates all cell-mean contrasts jointly and preserves their common post-treatment-period covariance. 
Under the stated Gaussian approximation, each marginal interval has the advertised marginal coverage and $P\{\bm{\theta}^{{DID}}(P)\in\mathcal C_{{DID}}\}\geq1-\alpha+o(1)$ uniformly. 
With normal multipliers,
$\bm Z_b\mid\text{data}\sim N(\bm 0_K,\widehat{\bm V})$ exactly, where $\bm 0_K$ is the $K$-vector of zeros, so no separate bootstrap theorem for OLS is being asserted. 
Note that the perturbation $\bm Z_b$ is on the same scale as the estimation error $\widehat{\bm{\theta}}^{{DID}}-\bm{\theta}^{{DID}}$.
Conditional quantile simulation error is required to vanish, for example by letting $B\to\infty$.
For the empirical common-DID benchmark under $A_0$, restrict $\beta_{t_0}=\beta_{\rm pre}$ for all $t_0\in\mathcal I_0$ in \eqref{eq:saturated-regression} and set $\widehat{\theta}_{\mathrm{com}}^{{DID}}=\widehat\beta_1-\widehat\beta_{\rm pre}$. Report $[\widehat{\theta}_{\mathrm{com}}^{{DID}}\pm z_{1-\alpha/2}s_{\rm com}]$ and $2\Phi(-|\widehat{\theta}_{\mathrm{com}}^{{DID}}/s_{\rm com}|)$ for a zero-effect null, using the restricted regression's cluster sandwich and design rank. These calculations require the corresponding normal approximation and consistent, nondegenerate variance estimation.

For $K>1$, the empirical primitive-DID equality diagnostic tests $\bm H\bm{\theta}^{{DID}}=0$, where $\bm H$ takes adjacent differences of the chronologically ordered DID vector. Its Wald statistic is
\[
W=(\bm H\widehat{\bm{\theta}}^{{DID}})'(\bm H\widehat{\bm V}\bm H')^{-1}\bm H\widehat{\bm{\theta}}^{{DID}},
\]
with a $\chi^2_{K-1}$ reference distribution when the eigenvalues of $\bm H\bm\Sigma_P\bm H'$ are uniformly bounded away from zero. Non-rejection does not establish $A_0$ or select the MR regime.

\subsection{Union bounds: $\Theta^{\text{PT-RDID}}$ and
$\Theta^{\text{RDID}}$}
\label{sec:union-inference}

For $j=\text{PT-R}$, set $\mathcal R_{\text{PT-R}}=\mathcal I_0$ and
$g_r=\theta_{t_0}^{{DID}}$ with $r=t_0$. For $j=\text{R}$, set
$\mathcal R_{\text{R}}=\mathcal I_0\times\{-1,+1\}$ and
$g_r=g_{t_0,\sigma}(\bm{\theta}^{{DID}};c)$ with $r=(t_0,\sigma)$.
In a calculation for model $j\in\{\text{PT-R},\text{R}\}$, every
constituent index $r$ runs over $\mathcal R_j$. In particular,
$r_L,r_U\in\mathcal R_j$. In both cases write
$L=\min_r g_r$ and $U=\max_r g_r$.

\begin{algorithm}[!htbp]
\small\setstretch{1.05}
\setlength{\linewidth}{\hsize}
\let\fullwidthdisplay\relax \SetAlgoNlRelativeSize{-1}
\DontPrintSemicolon
\caption{Constituent-normal envelope}
\label{alg:ec}
\Input{Estimates $\widehat g_r$ and standard errors $s_r$ for
$r\in\mathcal R_j$, confidence level $1-\alpha$, and null ATT value $h\in\mathbb R$.}
\Output{Identified-set envelope $C^{S,EC}_j$, also usable for conservative ATT inference, and ATT $p$-value $p_j(h)$.}
Form each constituent's equal-tailed normal interval
$[\widehat g_r-z_{1-\alpha/2}s_r,\widehat g_r+z_{1-\alpha/2}s_r]$.\;
Report
\begin{equation}
C^{S,EC}_j=
[\min_r\{\widehat g_r-z_{1-\alpha/2}s_r\},
 \max_r\{\widehat g_r+z_{1-\alpha/2}s_r\}].
\label{eq:confidencebounds}
\end{equation}
Use this same interval as a conservative ATT confidence interval.
\;
For $H_0:ATT=h$, $h\in\mathbb R$, compute
\[
T_{EC}(h)=\max\left\{0,\min_r\frac{\widehat g_r-h}{s_r},
                         \min_r\frac{h-\widehat g_r}{s_r}\right\},
\qquad p_j(h)=2\Phi\{-T_{EC}(h)\}.
\]
Then $h\in C^{S,EC}_j$ if and only if $p_j(h)\geq\alpha$.\;
\end{algorithm}
\FloatBarrier

Algorithm~\ref{alg:ec} applies the standard intersection--union
principle \citep[Section~3]{Berger1996}, with a Bonferroni allocation
across the two endpoint errors. A related construction appears in
\citet[Section~5.2]{KolesarRothe2018}. The same normal envelope is
discussed by \citet[p.~2262]{ye2023negative}, whose Supplement,
Theorem~S6 and equation~(S11), establishes uniform identified-set
coverage under their assumptions.

Coverage follows by controlling the lower error at a population minimizer $r_L$ and the upper error at a population maximizer $r_U$, each at probability at most $\alpha/2+o(1)$. Their union has probability at most $\alpha+o(1)$, regardless of the number of constituents. This argument permits ties and zero set width. Appendix~\ref{sec:inferenceproof} gives the endpoint argument, which also applies to the transformed constituents of $A_{\text{R}}$.

\subsection{Intersection bounds: $\Theta^{\text{AE-RDID}}$}\label{sec:inf2}

Whole-set and scalar inference for $A_{\text{AE-R}}$ use the same joint DID estimates but different calibrations. 
We first apply the intersection-bound construction of \citet{chernozhukov2013intersection}, abbreviated CLR, with adaptive inequality selection as implemented by \citet{CKLR2015}, abbreviated CKLR, for the whole-set inference. 
We then implement generalized moment selection for the scalar ATT target.

For each $s\in\{-1,+1\}$, write $L_s=\max_{t_0}g_{t_0,s}$, $U_s=\min_{t_0}g_{t_0,-s}$.
Also, express both endpoints as maxima by indexing four families\footnote{All four families use the same first-stage draws to preserve dependence among the DID contrasts.} with
$f=(r,s)\in\{L,U\}\times\{-1,+1\}$:
\begin{align}
h_{L,s,t_0}&=g_{t_0,s},&
h_{U,s,t_0}&=-g_{t_0,-s},\nonumber\\
H_{L,s}&=\max_{t_0}h_{L,s,t_0}=L_s,&
H_{U,s}&=\max_{t_0}h_{U,s,t_0}=-U_s,
\label{eq:CLR-families}
\end{align}
where $h$ gets a hat when we replace $\bm{\theta}^{{DID}}$ by $\widehat{\bm{\theta}}^{{DID}}$.
The associated estimated standard errors and common standardized draws are
\begin{align}
\tau_{L,s,t_0}&=\frac{s_{t_0}}{1+s c_{t_0}},&
\tau_{U,s,t_0}&=\frac{s_{t_0}}{1-s c_{t_0}},\nonumber\\
W_{b,L,s,t_0}&=S_{b,t_0},&
W_{b,U,s,t_0}&=-S_{b,t_0}.
\label{eq:CLR-draws}
\end{align}
Note that these signs account for negating the upper endpoint. 

\begin{algorithm}[!htbp]
\small\setstretch{1.05}
\setlength{\linewidth}{\hsize}
\let\fullwidthdisplay\relax
\SetAlgoNlRelativeSize{-1}
\DontPrintSemicolon
\caption{CLR confidence set for the $A_{\text{AE-R}}$ identified set}
\label{alg:ii-set}
\Input{Joint DID estimates, standard errors, and common draws from Algorithm~\ref{alg:first-stage}, $n$ independent clusters, fixed sensitivity value $c$, and confidence level $1-\alpha$.}
\Output{Whole-set confidence set $C^S_{\text{AE-R}}(c)$.}
Use the four maximum families, standard errors, and signed draws in
\eqref{eq:CLR-families}--\eqref{eq:CLR-draws}. Set
$p=1-\alpha/2$ and $\gamma_n=1-0.1/\log n$.\;
For each $f\in\{L,U\}\times\{-1,+1\}$, compute the preliminary critical value and lower bound
\[
a_f=Q_B\!\left(\gamma_n;
       (\max_{t_0\in\mathcal I_0}W_{b,f,t_0})_{b=1}^B\right),\qquad
\widetilde H_f=\max_{t_0\in\mathcal I_0}
       (\widehat h_{f,t_0}-a_f\tau_{f,t_0}).
\]
Select the potentially binding indices
\[
\widehat{\mathcal J}_f=
\{t_0\in\mathcal I_0:
  \widehat h_{f,t_0}\geq\widetilde H_f-2a_f\tau_{f,t_0}\}.
\]
If finite simulation yields an empty selected family, retain $\mathcal I_0$ for that family.\;
Calibrate each endpoint separately and form its precision-corrected bound:
\begin{equation}
q_f=Q_B\!\left(p;
       (\max_{t_0\in\widehat{\mathcal J}_f}W_{b,f,t_0})_{b=1}^B\right),
\qquad
\underline H_f=\max_{t_0\in\mathcal I_0}
       (\widehat h_{f,t_0}-q_f\tau_{f,t_0}).
\label{eq:CLR-corrected-max}
\end{equation}
Selection calibrates $q_f$. The final maximum uses all baseline indices.\;
For each sign $s$, set
\[
C^S_{\text{AE-R},s}(c)=
 [\underline H_{L,s},-\underline H_{U,s}]\cap\mathcal S_s,
\]
and report
\begin{equation}
C^S_{\text{AE-R}}(c)=C^S_{\text{AE-R},+}(c)\cup C^S_{\text{AE-R},-}(c).
\label{eq:II-set-envelope}
\end{equation}
Compute both signs. Discard reversed or sign-infeasible branches. Retain singletons and every remaining component. 
If both branches are empty, report the empty set.\;
\end{algorithm}
\FloatBarrier

For Algorithm~\ref{alg:ii-set}, we impose Condition~P of \citet[pp.~692--693]{chernozhukov2013intersection} uniformly over
$\mathcal P_{\mathrm{CLR}}\subseteq\mathcal P$, with primitive parameter $\bm\beta=\bm{\theta}^{{DID}}$ and fixed index set $V=\mathcal I_0$. 
In this linear setting, its substantive requirements are eigenvalues of $\bm\Sigma_P$ uniformly bounded above and away from zero, a Gaussian coupling, and a polynomial rate for covariance estimation:
\begin{equation}
\begin{gathered}
\bm\Sigma_P^{-1/2}\sqrt n(\widehat{\bm{\theta}}^{{DID}}-\bm{\theta}^{{DID}}(P))
 =\bm N_{n,P}+o_P(1),\qquad
 \bm N_{n,P}\sim N(\bm0_K,\bm I_K),\\
\|n\widehat{\bm V}-\bm\Sigma_P\|_{\rm op}=O_P(n^{-b_0}),
\qquad b_0>0,
\end{gathered}
\label{eq:CLR-P-regularity}
\end{equation}
uniformly over that class. 
The finite index set and known linear gradients $\pm\bm e_{t_0}/(1\pm c_{t_0})$ satisfy the continuity and smoothness requirements.\footnote{Non-singularity is imposed on the primitive DID covariance. Tied population effects and singularity of the expanded endpoint vector are allowed.}
For each family, set $\theta_{n,f}(t_0)=-h_{f,t_0}$, so CLR's target is $\theta_{n0,f}=\min_{t_0}\theta_{n,f}(t_0)=-H_f$. 
Their upper bound, after negation, is our all-index lower bound $\underline H_f$.
Theorem~4(a) therefore gives $P\{\underline H_f\leq H_f\}\geq p-o(1)$ for conditional population quantiles. 

To approximate CLR's conditional population quantiles with empirical quantiles, we assume the sufficient Monte Carlo condition
\begin{equation}
 B=B_n,\qquad B_n/(\log n)^2\longrightarrow\infty.
\label{eq:CLR-MC-rate}
\end{equation}
This sufficient condition makes empirical-CDF errors $o_P(1/\log n)$ over the finitely many signed subsets, preserving asymptotic endpoint coverage with common draws for selection and calibration.

Since $c_{t_0}<1$, every nonempty population identified set is contained entirely in at least one sign branch. 
With $p=1-\alpha/2$ as specified in Algorithm~\ref{alg:ii-set}, Bonferroni applied to that branch's two endpoints, followed by sign
clipping and unioning the reported branches, establishes the whole-set coverage requirement in \eqref{eq:uniform-set-coverage} for $j=\text{AE-R}$, with $\mathcal P$ replaced by $\mathcal P_{\mathrm{CLR}}$, at each fixed $c\in[0,\bar c]$.
Empty population sets are contained automatically.
This is the two-endpoint construction of \citet[p.~25, equation~(6)]{CKLR2015}. No additional split of $\alpha$ across signs is needed.

For the ATT inference, a whole-set confidence set is valid but may be conservative. 
Hence, we introduce Algorithm~\ref{alg:ii-att} that implements a particular generalized moment selection (GMS) procedure of \citet{AndrewsSoares2010}.

\begin{algorithm}[!htbp]
\small\setstretch{1.05}
\setlength{\linewidth}{\hsize}
\let\fullwidthdisplay\relax \SetAlgoNlRelativeSize{-1}
\DontPrintSemicolon
\caption{ATT test inversion: \citet{AndrewsSoares2010},
generalized moment selection}
\label{alg:ii-att}
\Input{DID estimates, standard errors, common draws, $c$, $\alpha$, null ATT value $h\in\mathbb R$, and $\kappa_n$
with $\kappa_n\to\infty$ and $\kappa_n/\sqrt n\to0$ (e.g., $\kappa_n=\sqrt{\log n}$).}
\Output{ATT confidence set $C^P_{\text{AE-R}}(c)$ from test inversion and ATT $p$-value $p_{\text{AE-R},B}(h)$.}
For each candidate $\vartheta\in\mathbb R$, form the $2K$ violations
\[
\widehat m_{1,t_0}(\vartheta)=\vartheta-\widehat{\theta}_{t_0}^{{DID}}-c_{t_0}|\vartheta|,
\qquad
\widehat m_{2,t_0}(\vartheta)=\widehat{\theta}_{t_0}^{{DID}}-\vartheta-c_{t_0}|\vartheta|.
\]
Here $t_0\in\mathcal I_0$. Index the moments by
$k=(q,t_0)\in\mathcal K=\{1,2\}\times\mathcal I_0$ and set
$s_k=s_{t_0}$ for both $q=1,2$. All unrestricted maxima over $k$ use
$\mathcal K$. Define
\[
T(\vartheta)=\max\{0,\max_k\widehat m_k(\vartheta)/s_k\},\qquad
\widehat{\mathcal K}(\vartheta)=
       \{k\in\mathcal K:\widehat m_k(\vartheta)/s_k\geq-\kappa_n\}.
\]\par
Use centered standardized draws
$E_{b,(1,t_0)}=-S_{b,t_0}$ and $E_{b,(2,t_0)}=S_{b,t_0}$ for $b=1,\ldots,B$ and $t_0\in\mathcal I_0$.
Let
\[
q(\vartheta)=Q_B\left(1-\alpha;
       \max\{0,\max_{k\in\widehat{\mathcal K}(\vartheta)}E_{b,k}\}\right),
\]
with $q(\vartheta)=0$ if the selected set is empty.\;
Invert the test over the whole real line:
\begin{equation}
C^P_{\text{AE-R}}(c)=\{\vartheta\in\mathbb R:T(\vartheta)\leq q(\vartheta)\}.
\label{eq:II-inverted-CI}
\end{equation}
Keep all connected components. A hull may be displayed only if identified
explicitly as a hull.\;
For $H_0:ATT=h$, report $p_{\text{AE-R},B}(h)=1$ if $T(h)=0$. If $T(h)>0$, compute
\[
p_{\text{AE-R},B}(h)=\frac{1+\sum_{b=1}^B
\mathbbm1\{\max_{k\in\widehat{\mathcal K}(h)}E_{b,k}\geq T(h)\}}{B+1}.
\]\par
\end{algorithm}
\FloatBarrier

Defining $m_k(\vartheta)$ by replacing $\widehat{\theta}_{t_0}^{{DID}}$ with $\theta_{t_0}^{{DID}}$ in $\widehat m_k(\vartheta)$, we note that the inequalities $m_k(\vartheta)\leq0$ for all $k\in\mathcal K$ characterize $A_{\text{AE-R}}$ feasibility. 
Algorithm~\ref{alg:ii-att} applies the Gaussian GMS procedure of \citet{AndrewsSoares2010} to $-m_k(\vartheta)$, using $S_3$ with $p_1=1$, no moment equalities, and the hard-threshold rule $\varphi^{(1)}$ (their equations (3.6) and (4.5)).\footnote{Our $T(\vartheta)^2$ and $q(\vartheta)^2$ are their statistic and Gaussian critical value.}
Under the maintained first-stage and tuning conditions and vanishing conditional quantile simulation error, 
the uniform coverage \eqref{eq:uniform-parameter-coverage} is achieved for $j=\text{AE-R}$ at each fixed $c$ \citep[Theorem~1(a)]{AndrewsSoares2010}.\footnote{Note that empty \textit{exact} inversion yields a specification test with uniform asymptotic size at most $\alpha$ \citep[Section~5]{AndrewsSoares2010}.}
The simulated $p$-values are asymptotically valid under the same conditions. The plus-one tail rule and numerical enlargement can prevent exact finite-$B$ duality with the reported confidence set. They do not provide exact finite-sample inference.

Algorithm~\ref{alg:ii-numerics} evaluates this inversion using the finitely many selection breakpoints of the affine moments on each sign half-line.

\begin{algorithm}[!htbp]
\small\setstretch{1.05}
\setlength{\linewidth}{\hsize}
\let\fullwidthdisplay\relax \SetAlgoNlRelativeSize{-1}
\DontPrintSemicolon
\caption{Exhaustive closed outer evaluation of the intersection ATT confidence set}
\label{alg:ii-numerics}
\Input{Inputs to Algorithm \ref{alg:ii-att} and fixed simulated draws.}
\Output{All components of a closed outer approximation to $C^P_{\text{AE-R}}(c)$, with ATT $p$-value $p_{\text{AE-R},B}(h)$ from Algorithm~\ref{alg:ii-att}.}
Use $q_R$ from \eqref{eq:primitive-rectangle}. Every accepted candidate
lies in
\[
\mathcal B_s=
\left[\max_{t_0}\frac{\widehat{\theta}_{t_0}^{{DID}}-q_Rs_{t_0}}{1+s c_{t_0}},
      \min_{t_0}\frac{\widehat{\theta}_{t_0}^{{DID}}+q_Rs_{t_0}}{1-s c_{t_0}}\right]
\cap\mathcal S_s,\qquad s\in\{+1,-1\}.
\]
This follows because $q(\vartheta)\leq q_R$ for the same draws.\;
In exact arithmetic, split each bracket at every
selection breakpoint indexed by $t_0\in\mathcal I_0$ in its sign half-line:
\[
\vartheta=\frac{\widehat{\theta}_{t_0}^{{DID}}-\kappa_n s_{t_0}}{1-s c_{t_0}}
\quad\hbox{or}\quad
\vartheta=\frac{\widehat{\theta}_{t_0}^{{DID}}+\kappa_n s_{t_0}}{1+s c_{t_0}}.
\]
Within each open cell the selected set, hence $q(\vartheta)=q$, is constant.
Compute that critical value and intersect the cell with
\[
\left[\max_{t_0}\frac{\widehat{\theta}_{t_0}^{{DID}}-q s_{t_0}}{1+s c_{t_0}},
      \min_{t_0}\frac{\widehat{\theta}_{t_0}^{{DID}}+q s_{t_0}}{1-s c_{t_0}}\right].
\]
Evaluate all cell boundaries separately using the weak selection inequality.
Union the accepted pieces. Taking the closure for reporting can only enlarge
the confidence set.\;
Apply outward numerical guards to affine roots and acceptance bounds.
Partition each sign bracket at every selection root and both guard endpoints. 
In uncertain regions, compute critical values using the union of potentially selected moments. Use weak inequalities at boundaries.
Enlarge lower acceptance bounds downward and upper bounds upward.
Retain all nonempty closed accepted pieces, including singletons. Merge only overlapping or touching pieces and report all components of the resulting closed outer approximation.\;
\end{algorithm}
\FloatBarrier

Before closure, the exact-arithmetic cell and boundary calculations recover the full finite-draw acceptance set. Closure and outward guards preserve containment if the guards contain the exact roots. Enlarging uncertain selection families can only increase critical values. Under Algorithm~\ref{alg:ii-att}'s conditions, an empty outer approximation therefore yields a conservative asymptotic level-$\alpha$ model-refutation decision. Computational failures must not be reported as empty sets.

\subsection{Inference for the MR bound and the ATT}\label{sec:mrinference}

MR coverage follows from the component procedures and the population inclusion in \eqref{eq:Gamma-inclusions}.

\begin{algorithm}[!htbp]
\small\setstretch{1.05}
\setlength{\linewidth}{\hsize}
\let\fullwidthdisplay\relax \SetAlgoNlRelativeSize{-1}
\DontPrintSemicolon
\caption{MR scalar coverage by a union of confidence sets}
\label{alg:mr-att}
\Input{Component ATT confidence sets at level $1-\alpha$ and their $p$-values, baseline count $K$, and null ATT value $h\in\mathbb R$.}
\Output{MR scalar confidence set $C^P_*(c)$ and the corresponding ATT $p$-value.}
Use the constituent-normal envelope from Algorithm~\ref{alg:ec} as the
conservative ATT interval for $A_{\text{PT-R}}$:
$C^P_{\text{PT-R}}=C^{S,EC}_{\text{PT-R}}$.\;
For $K>1$, report
\begin{equation}
C^P_*(c)=C^P_{\text{PT-R}}\cup C^P_{\text{AE-R}}(c),
\label{eq:MR-parameter-CI}
\end{equation}
using the exact inversion or the guarded outer approximation in
Algorithm~\ref{alg:ii-numerics}, under its stated guard condition, for the
second component.
Preserve all components.\;
If $K=1$ and the target is the MR bound under the minimum-relaxation
premise, use $C^P_*=C^P_{\text{PT-R}}$, since $\Gamma_*(\bm{\theta}^{{DID}};c)=\{\theta_{t_0}^{{DID}}\}$ for $\mathcal I_0=\{t_0\}$.
If coverage over the larger full-data union of $A_{\text{PT-R}}$ and
$A_{\text{AE-R}}$ is desired, retain the union even for $K=1$.\;
For $H_0:ATT=h$, the union reports
$p_{*,B}(h)=\max\{p_{\text{PT-R}}(h),p_{\text{AE-R},B}(h)\}$, using Algorithms~\ref{alg:ec} and~\ref{alg:ii-att}. For the $K=1$ MR refinement, report $p_*(h)=p_{\text{PT-R}}(h)$.\;
\end{algorithm}
\FloatBarrier

If the true law satisfies $A_{\text{PT-R}}$, the first component covers its ATT. If it satisfies $A_{\text{AE-R}}$, the second does. 
Because $A_0$ implies $A_{\text{PT-R}}$, the first component also provides coverage under $A_0$, so no separate benchmark interval is needed for MR coverage. 
Taking a union cannot reduce either coverage probability, so no Bonferroni adjustment is needed for this scalar target. 
The resulting confidence set covers each fixed scalar in $\Gamma_*(\bm{\theta}^{{DID}};c)$.

For MR whole-set coverage, the $A_{\text{PT-R}}$ and $A_{\text{AE-R}}$
components must be protected jointly. A union of separately calibrated
level-$(1-\alpha)$ component envelopes does not establish that joint event.
We therefore use the common rectangle for both components, retaining the
general first-stage assumptions.

\begin{algorithm}[!htbp]
\small\setstretch{1.05}
\setlength{\linewidth}{\hsize}
\let\fullwidthdisplay\relax
\SetAlgoNlRelativeSize{-1}
\DontPrintSemicolon
\caption{MR identified-set envelope from a common confidence rectangle}
\label{alg:mr-set}
\Input{One joint rectangle $\mathcal C_{{DID}}$ from Algorithm~\ref{alg:first-stage}, sensitivity value $c$, and baseline count $K$.}
\Output{MR identified-set envelope $C^S_*(c)$.}
Use the same $q_R$, $\ell_{t_0}$, and $u_{t_0}$ as in
\eqref{eq:primitive-rectangle}. Set
$C^{S,J}_{\text{PT-R}}=[\min_{t_0}\ell_{t_0},\max_{t_0}u_{t_0}]$.\;
For the $A_{\text{AE-R}}$ component, replace each endpoint-specific
$q_f$ in Algorithm~\ref{alg:ii-set}'s all-index final bounds by this
common $q_R$. Equivalently, compute
\[
C^{S,J}_{\text{AE-R},s}(c)=
\left[\max_{t_0}\frac{\ell_{t_0}}{1+s c_{t_0}},
      \min_{t_0}\frac{u_{t_0}}{1-s c_{t_0}}\right]\cap\mathcal S_s,
\qquad
C^{S,J}_{\text{AE-R}}(c)=\bigcup_s C^{S,J}_{\text{AE-R},s}(c).
\]
No endpoint selection is needed for this common calibration.\;
Report
\begin{equation}
C^S_*(c)=C^{S,J}_{\text{PT-R}}\cup C^{S,J}_{\text{AE-R}}(c)
\quad(K>1),\qquad C^S_*(c)=C^{S,J}_{\text{PT-R}}\quad(K=1).
\label{eq:MR-set-envelope}
\end{equation}
Retain singletons and all nonempty components. For a sensitivity curve,
hold the rectangle fixed at every $c$.\;

\end{algorithm}
\FloatBarrier

On $\bm{\theta}^{{DID}}(P)\in\mathcal C_{{DID}}$, the sets $\Gamma_{\text{PT-R}}(\bm{\theta}^{{DID}}(P))$
and $\Gamma_{\text{AE-R}}(\bm{\theta}^{{DID}}(P);c)$ are contained in
$C^{S,J}_{\text{PT-R}}$ and $C^{S,J}_{\text{AE-R}}(c)$ simultaneously.
Equation~\eqref{eq:Gamma-inclusions} gives MR whole-set coverage for $K>1$.
For $K=1$, the population MR set is the single DID covered by $C^{S,J}_{\text{PT-R}}$. 
This same event works for every $c\in[0,\bar c]$ so no further level adjustment over sensitivity values is needed.
The joint component $C^{S,J}_{\text{AE-R}}$ is a rectangle projection, distinct from the CLR envelope $C^S_{\text{AE-R}}$ in Algorithm~\ref{alg:ii-set}.
Their critical values use different probability levels and families, and no ordering between the two envelopes is asserted.

\subsection{Simultaneous sensitivity bands}\label{sec:simultaneous-c-bands}

A single joint confidence rectangle delivers simultaneous bands over the
sensitivity parameter and across models. This is a different coverage
target from Algorithm~\ref{alg:ii-set}'s fixed-$c$ CLR envelope. The MR
construction already uses this common event. The next algorithm collects
its projection with those for the other models.

\begin{algorithm}[!htbp]
\small\setstretch{1.05}
\setlength{\linewidth}{\hsize}
\let\fullwidthdisplay\relax \SetAlgoNlRelativeSize{-1}
\DontPrintSemicolon
\caption{Projection of a joint confidence rectangle}
\label{alg:projection}
\Input{One level-$(1-\alpha)$ rectangle $\mathcal C_{{DID}}$ and sensitivity range $[0,\bar c]$.}
\Output{Simultaneous identified-set bands $B^S_j(c)$ for $j\in\{0,\text{PT-R},\text{AE-R},\text{R},*\}$.}
Construct one rectangle $\mathcal C_{{DID}}$ at level $1-\alpha$ using
Algorithm \ref{alg:first-stage}. Hold this rectangle fixed for every
$c\in[0,\bar c]$.\;
For each $c\in[0,\bar c]$, calculate
\begin{align}
B^S_{\text{PT-R}}&=[\min_{t_0}\ell_{t_0},\max_{t_0}u_{t_0}],\label{eq:band-I}\\
B^S_{\text{AE-R},s}(c)
 &=\left[\max_{t_0}\frac{\ell_{t_0}}{1+s c_{t_0}},
          \min_{t_0}\frac{u_{t_0}}{1-s c_{t_0}}\right]\cap\mathcal S_s,
 \quad s\in\{-1,+1\},\nonumber\\
B^S_{\text{AE-R}}(c)&=\bigcup_{s\in\{-1,+1\}} B^S_{\text{AE-R},s}(c),
 \label{eq:band-II}\\
B^S_{\text{R}}(c)
 &=\left[\min_{t_0,\sigma}\frac{\ell_{t_0}}{1+\sigma c_{t_0}},
          \max_{t_0,\sigma}\frac{u_{t_0}}{1+\sigma c_{t_0}}\right],
 \label{eq:band-III}\\
B^S_0&=[\max_{t_0}\ell_{t_0},\min_{t_0}u_{t_0}].
\label{eq:band-A0}
\end{align}\par
For the MR curve report the outer band
\begin{equation}
B^S_*(c)=B^S_{\text{PT-R}}\cup B^S_{\text{AE-R}}(c)\quad(K>1),
\qquad B^S_*(c)=B^S_{\text{PT-R}}\quad(K=1).
\label{eq:band-MR}
\end{equation}
Retain empty sets and disconnected components as such. A plotting grid
evaluates these formulas. It does not define the coverage event.\;
\end{algorithm}
\FloatBarrier

This is the usual projection principle. On the single event $\bm{\theta}^{{DID}}(P)\in\mathcal C_{{DID}}$,
\[
\Gamma_j(\bm{\theta}^{{DID}}(P);c)\subseteq\bigcup_{\bm v\in\mathcal C_{{DID}}}\Gamma_j(\bm v;c)
\quad c\in[0,\bar c],\quad j\in\{0,\text{PT-R},\text{AE-R},\text{R},*\}.
\]

Consequently the bands jointly cover the identified-set curves with probability at least $1-\alpha+o(1)$, without another adjustment over $c$ or models.

The formulas for $A_0$, $A_{\text{PT-R}}$, $A_{\text{AE-R}}$, and $A_{\text{R}}$ are exact rectangle projections. The MR band is an outer projection obtained from \eqref{eq:Gamma-inclusions}.
By construction, $B^S_{\text{AE-R}}(c)=C^{S,J}_{\text{AE-R}}(c)$ and $B^S_*(c)=C^S_*(c)$ since Algorithm~\ref{alg:mr-set} uses the same joint rectangle as the simultaneous sensitivity bands. 
The MR band is contained in $B^S_{\text{R}}(c)$. 
At $K=1$, it equals $B^S_{\text{PT-R}}$ because the population MR set is the single DID.

\clearpage

}

\section{Sufficient Conditions for Bias Set Stability}\label{sec:sufficient-conditions}

{A natural question is which conditions on the data-generating process are sufficient for Assumption \ref{sb:bounds}. We provide one such condition for the counterfactual untreated potential outcome.}

\begin{assumption}\label{ass:sc}
$\,$
\begin{enumerate}[(i)]
\item The untreated potential outcome satisfies
\[
Y_t(0)=g_t(\varepsilon)\lambda(U)+\gamma(V)+\eta_t,
\]
where all terms are integrable, $(\varepsilon,\eta_t)\ \indep\ (U,V)$ for every $t$, and $g_t(\cdot)$, $\lambda(\cdot)$, and $\gamma(\cdot)$ are unknown measurable functions. \label{ass:sc1}
\item There exists an observed baseline period $t^{\star}\in\mathcal I_0\setminus\{0\}$ such that
\[
\mathbb E[g_1(\varepsilon)]=\mathbb E[g_{t^{\star}}(\varepsilon)].
\]
In particular, this condition holds with $t^{\star}=-1$ when $-1\in\mathcal I_0$ and $g_t(e)=g_{-t}(e)$ for all $(t,e)$. \label{ass:sc2}
\item Treatment receipt satisfies $D=h(U,V)$ and $0<\mathbb P(D=1)<1$. \label{ass:sc3}
\item Selection on the interactive factor is nondegenerate:
\[
\Delta_{\lambda}\equiv
\mathbb E[\lambda(U)\mid D=1]-\mathbb E[\lambda(U)\mid D=0]\neq0.
\]
\label{ass:sc4}
\end{enumerate}
\end{assumption}
Assumption \ref{ass:sc}.(\ref{ass:sc1}) postulates a factor, or interactive fixed-effects, structure for the untreated potential outcome. Assumption \ref{ass:sc}.(\ref{ass:sc2}) requires the period-1 factor loading in expectation to be reproduced at an observed baseline period. Time symmetry is one transparent sufficient condition, similar in spirit to the symmetric DID intuition in \citet[page 652]{AshenfelterCard1985}, but an asymmetric sequence can also qualify, for example if $\mathbb E[g_{-2}(\varepsilon)]=\mathbb E[g_1(\varepsilon)]$. Assumption \ref{ass:sc}.(\ref{ass:sc3}) permits treatment selection through the time-invariant unobservables, while Assumption \ref{ass:sc}.(\ref{ass:sc4}) ensures that such selection changes the mean of the interactive factor.

\begin{proposition}\label{prop:sc}
Suppose $\mathcal I_0=\{-T_0,-T_0+1,\ldots,0\}$ with $T_0\geq1$, $\mathbb E[g_1(\varepsilon)]\neq\mathbb E[g_0(\varepsilon)]$, and Assumption~\ref{ass:sc} holds. Then Assumption \ref{sb:bounds} holds while PT fails.\end{proposition}
{
\begin{proof}
Define
\[
\Delta_\gamma\equiv
\mathbb E[\gamma(V)\mid D=1]-\mathbb E[\gamma(V)\mid D=0].
\]
By Assumptions \ref{ass:sc}.(\ref{ass:sc1}) and \ref{ass:sc}.(\ref{ass:sc3}), and because $(\varepsilon,\eta_t)\indep(U,V)$,
\[
SB_t=\mathbb E[g_t(\varepsilon)]\Delta_\lambda+\Delta_\gamma.
\]
Assumption \ref{ass:sc}.(\ref{ass:sc2}) gives an observed $t^\star\in\mathcal I_0$ for which $SB_1=SB_{t^\star}$. Consequently,
\[
SB_1\in
\left[\inf_{t_0\in\mathcal I_0}SB_{t_0},
\sup_{t_0\in\mathcal I_0}SB_{t_0}\right],
\]
so Assumption \ref{sb:bounds} holds. On the other hand,
\[
SB_1-SB_0=
\big\{\mathbb E[g_1(\varepsilon)]-\mathbb E[g_0(\varepsilon)]\big\}\Delta_\lambda\neq0
\]
by the proposition's loading condition and Assumption \ref{ass:sc}.(\ref{ass:sc4}). Hence PT fails.
\end{proof}
}

Our main assumption (Assumption \ref{sb:bounds}) can hold in settings where there exist common life-cycle factors affecting untreated potential outcomes. These factors can generate time symmetry or periodicity that reproduces the period-1 mean loading in a selected baseline period. A complementary nonlinear sufficient condition is provided below. It is not a literal nesting of Assumption \ref{ass:sc}; Assumption \ref{ass:sc} permits equality only after averaging the factor loading, whereas Assumption \ref{ass:scgen} below imposes the stronger pointwise matching condition.

\begin{assumption}\label{ass:scgen}
$\,$
\begin{enumerate}[(i)]
\item The untreated potential outcome satisfies
\[
Y_t(0)=\varphi(t,U,\varepsilon)+\eta_t,
\]
where all terms are integrable and $U$ may be vector-valued. \label{ass:scgen1}
\item There exists an observed period $t^{\star}\in\mathcal I_0$ such that
\[
\varphi(1,u,e)=\varphi(t^{\star},u,e)
\quad\text{for all }(u,e).
\]
In particular, one may take $t^{\star}=-1$ when $-1\in\mathcal I_0$ and $\varphi(t,u,e)$ is even in $t$. \label{ass:scgen2}
\item Treatment receipt satisfies $D=h(U,V)$ and $0<\mathbb P(D=1)<1$. \label{ass:scgen3}
\item $(\varepsilon,\eta_t)\ \indep\ (U,V)$ for every $t$. \label{ass:scgen4}
\end{enumerate}
\end{assumption}
Because the $\eta_t$ term has the same conditional mean across treatment groups and the nonlinear component is identical at periods $1$ and $t^{\star}$, Assumption \ref{ass:scgen} implies $SB_1=SB_{t^{\star}}$ and therefore Assumption \ref{sb:bounds}.

For example, let $D=\mathbbm{1}\{U\geq1\}$, $U\sim N(0,\sigma^2)$, and $\eta_t\indep U$. Assumption \ref{ass:scgen} holds for
\[
Y_t(0)=\sqrt{t^2+\exp(U)}+\eta_t,
\]
which satisfies $\varphi(1,u)=\varphi(-1,u)$, and also for
\[
Y_t(0)=\sqrt{(t+2)(t-1)+3+\exp(U)}+\eta_t,
\]
which satisfies $\varphi(1,u)=\varphi(-2,u)$.
\section{Identification with Covariates}
This appendix derives conditional and treated-population ATT bounds for one focal post-treatment outcome under conditional no anticipation in the selected baseline cells.

\begin{definition}
The baseline information set is a {nonempty} measurable subset $\mathcal I_0\subseteq\mathcal I_{\mathcal X}$, where
\[
\mathcal I_{\mathcal X}\equiv\bigcup_{t_0\in\mathcal T_0}\{t_0\}\times\mathcal X_{t_0},
\]
$\mathcal X_{t_0}$ is the support of a baseline covariate $X_{t_0}$, and $\mathcal T_0$ is the set of pre-treatment periods. Write $\mathcal I_{0,t_0}\equiv\{x:(t_0,x)\in\mathcal I_0\}$. {For every period $t_0$ represented in $\mathcal I_0$, require
\[
\Pr\!\left(X_{t_0}\in\mathcal I_{0,t_0}\right)>0.
\]
Together with overlap, this ensures that both treatment groups have positive probability within each represented baseline covariate region and that the conditional comparisons and essential extrema below are defined almost surely on non-null sets.}
\end{definition}

For $t\in\mathcal T_0\cup\{1\}$, define, wherever the conditional means are well defined,
\begin{align*}
\theta_t^{OLS}(x)&\equiv \mathbb E[Y_t\mid D=1,X_t=x]-\mathbb E[Y_t\mid D=0,X_t=x],\\
SB_t(x)&\equiv \mathbb E[Y_t(0)\mid D=1,X_t=x]-\mathbb E[Y_t(0)\mid D=0,X_t=x],\\
AE_t(x)&\equiv \mathbb E[Y_t(1)-Y_t(0)\mid D=1,X_t=x],\\
ATT(x)&\equiv \mathbb E[Y_1(1)-Y_1(0)\mid D=1,X_1=x].
\end{align*}

Under overlap, consider a measurable comparison region $C\subseteq\mathcal X_0\cap\mathcal X_1$ of positive probability in both periods on which the restrictions of $F_{X_0}$ and $F_{X_1}$ are mutually absolutely continuous. Conditional PT between periods $0$ and $1$ on $C$ means $SB_1(x)=SB_0(x)$ for $(F_{X_0}+F_{X_1})$-almost every $x\in C$.
Define the baseline-bias endpoints directly by
\begin{align*}
\underline{SB}(\mathcal I_0)
&\equiv \inf_{t_0:\,\mathcal I_{0,t_0}\neq\varnothing}
\operatorname*{ess\,inf}_{x\in\mathcal I_{0,t_0}}SB_{t_0}(x),\\
\overline{SB}(\mathcal I_0)
&\equiv \sup_{t_0:\,\mathcal I_{0,t_0}\neq\varnothing}
\operatorname*{ess\,sup}_{x\in\mathcal I_{0,t_0}}SB_{t_0}(x),
\end{align*}
{where the essential extrema are taken with respect to the distribution of $X_{t_0}$ on the indicated region.} They are invariant to changes in a version of a conditional mean on a null set. In the no-covariate case, these definitions reduce to
\[
\underline{SB}(\mathcal I_0)=\inf_{t_0\in\mathcal I_0}SB_{t_0},
\qquad
\overline{SB}(\mathcal I_0)=\sup_{t_0\in\mathcal I_0}SB_{t_0}.
\]

\begin{example}
If $\mathcal T_0=\{-1,0\}$, $\mathcal X_{-1}=\{a,b\}$, and $\mathcal X_0=\{a,c\}$, then
\[
\mathcal I_{\mathcal X}=\{(-1,a),(-1,b),(0,a),(0,c)\},
\]
and {$\mathcal I_0$ may be any nonempty subset whose represented covariate regions satisfy the preceding positive-probability requirement}.
\end{example}

\begin{assumption}[Conditional No Anticipation Effects]\label{ass:NoAE_x}
For every $t_0$ represented in $\mathcal I_0$,
\[
AE_{t_0}(x)=0
\]
for $F_{X_{t_0}\mid D=1}$-almost every $x\in\mathcal I_{0,t_0}$.
\end{assumption}

\begin{assumption}[Conditional Bias Set Stability]\label{sb:bounds_x}
For $F_{X_1\mid D=1}$-almost every $x_1$,
\[
SB_1(x_1)\in
\big[\underline{SB}(\mathcal I_0),\overline{SB}(\mathcal I_0)\big]
\equiv\Delta_{SB(\mathcal I_0)}.
\]
\end{assumption}

Under Assumption \ref{ass:NoAE_x}, each baseline bias is identified almost surely as $SB_{t_0}(x)=\theta_{t_0}^{OLS}(x)$. Assumption~\ref{sb:bounds_x} therefore uses observed baseline heterogeneity to restrict the possible post-treatment conditional selection bias. It requires the range of $SB_1(X_1)$ to be \emph{contained in}, not necessarily equal to, the closed convex hull of the baseline conditional biases. Essential extrema avoid making the bounds depend on arbitrary versions of conditional expectations at probability-zero covariate values.

\begin{assumption}[Overlap]\label{overlap}
$0<\mathbb P(D=1\mid X_t)<1$ almost surely for every period $t$ used in the conditional analysis.
\end{assumption}

\begin{proposition}\label{prop1x1}
Suppose model \eqref{seq1} and Assumptions \ref{ass:NoAE_x}, \ref{sb:bounds_x}, and \ref{overlap} hold, and suppose $\underline{SB}(\mathcal I_0)$ and $\overline{SB}(\mathcal I_0)$ are finite. Then, for $F_{X_1\mid D=1}$-almost every $x_1$,
\[
ATT(x_1)\in
\left[
\theta_1^{OLS}(x_1)-\overline{SB}(\mathcal I_0),
\theta_1^{OLS}(x_1)-\underline{SB}(\mathcal I_0)
\right]
\equiv\Theta^{\text{PT-R}}(x_1).
\]
{With unrestricted real-valued potential outcomes, the bounds are functionally sharp. Under this support condition, every integrable measurable function $a(x_1)$ satisfying $a(x_1)\in\Theta^{\text{PT-R}}(x_1)$ almost surely is attained by a single joint distribution satisfying the maintained assumptions. Hence, $\Theta^{\text{PT-R}}(x_1)$ is the pointwise identified set for $ATT(x_1)$ almost surely.}
\end{proposition}
{
\begin{proof}
\textbf{Validity of the bounds}

Under overlap,
\[
\theta_1^{OLS}(x_1)=ATT(x_1)+SB_1(x_1)
\]
for $F_{X_1\mid D=1}$-almost every $x_1$. The claimed bounds follow immediately from Assumption \ref{sb:bounds_x}.

\textbf{Sharpness of the bounds}

Allow unrestricted real-valued potential outcomes.
Fix an arbitrary integrable measurable candidate function $a(x_1)$ satisfying $a(x_1)\in\Theta^{\text{PT-R}}(x_1)$ for $F_{X_1\mid D=1}$-almost every $x_1$, and define its implied conditional period-1 selection bias by
\begin{equation}\label{eq:prop1x1-sharp-s}
s(x_1)\equiv\theta_1^{OLS}(x_1)-a(x_1).
\end{equation}
By the definition of $\Theta^{\text{PT-R}}(x_1)$, candidate membership is equivalent to
\begin{equation}\label{eq:prop1x1-sharp-feasibility}
s(x_1)\in
[\underline{SB}(\mathcal I_0),\overline{SB}(\mathcal I_0)]
\quad\text{for }F_{X_1\mid D=1}\text{-almost every }x_1.
\end{equation}
Choose measurable versions of the period-1 observed conditional means,
\begin{equation}\label{eq:prop1x1-sharp-means}
m_1(x)\equiv\mathbb E[Y_1\mid D=1,X_1=x],
\qquad
m_0(x)\equiv\mathbb E[Y_1\mid D=0,X_1=x].
\end{equation}

Construct all potential outcomes on the original observed-data probability space. For every observed period $t\in\mathcal T_0\cup\{1\}$, retain the observed cells:
\begin{equation}\label{eq:prop1x1-sharp-observed}
\widetilde Y_t(1)=Y_t\quad\text{on }\{D=1\},
\qquad
\widetilde Y_t(0)=Y_t\quad\text{on }\{D=0\}.
\end{equation}
For every pre-treatment period $t_0\in\mathcal T_0$, define the treated units' missing untreated baseline outcome by
\begin{equation}\label{eq:prop1x1-sharp-baseline-completion}
\widetilde Y_{t_0}(0)=Y_{t_0}
\quad\text{on }\{D=1\}.
\end{equation}
For treated units in period 1, define
\begin{equation}\label{eq:prop1x1-sharp-followup-completion}
\widetilde Y_1(0)=m_1(X_1)-a(X_1)
\quad\text{on }\{D=1\}.
\end{equation}
Set the remaining missing treated-arm cells, $\widetilde Y_t(1)$ on $\{D=0\}$, equal to zero. The assignments in \eqref{eq:prop1x1-sharp-baseline-completion}--\eqref{eq:prop1x1-sharp-followup-completion} are jointly measurable and integrable. Equation \eqref{eq:prop1x1-sharp-observed} gives
\begin{equation}\label{eq:prop1x1-sharp-consistency}
\widetilde Y_t(1)D+\widetilde Y_t(0)(1-D)=Y_t
\qquad\text{almost surely},
\end{equation}
so consistency and the complete observed-data distribution are preserved.

We now verify the conditional restrictions. From \eqref{eq:prop1x1-sharp-observed} and \eqref{eq:prop1x1-sharp-baseline-completion}, for each represented $t_0$ and for $F_{X_{t_0}\mid D=1}$-almost every $x\in\mathcal I_{0,t_0}$,
\begin{align}
AE_{t_0}(x)
&=\mathbb E[\widetilde Y_{t_0}(1)-\widetilde Y_{t_0}(0)
\mid D=1,X_{t_0}=x]\nonumber\\
&=\mathbb E[Y_{t_0}-Y_{t_0}\mid D=1,X_{t_0}=x]
=0.
\label{eq:prop1x1-sharp-noae}
\end{align}
Thus Assumption \ref{ass:NoAE_x} holds. The corresponding baseline selection bias is
\begin{align}
SB_{t_0}(x)
&=\mathbb E[\widetilde Y_{t_0}(0)\mid D=1,X_{t_0}=x]
 -\mathbb E[\widetilde Y_{t_0}(0)\mid D=0,X_{t_0}=x]\nonumber\\
&=\mathbb E[Y_{t_0}\mid D=1,X_{t_0}=x]
 -\mathbb E[Y_{t_0}\mid D=0,X_{t_0}=x]
=\theta_{t_0}^{OLS}(x).
\label{eq:prop1x1-sharp-baseline-bias}
\end{align}
In period 1, equations \eqref{eq:prop1x1-sharp-observed}--\eqref{eq:prop1x1-sharp-followup-completion} imply, for $F_{X_1\mid D=1}$-almost every $x_1$,
\begin{align}
SB_1(x_1)
&=\mathbb E[\widetilde Y_1(0)\mid D=1,X_1=x_1]
 -\mathbb E[\widetilde Y_1(0)\mid D=0,X_1=x_1]\nonumber\\
&=m_1(x_1)-a(x_1)-m_0(x_1)\nonumber\\
&=\theta_1^{OLS}(x_1)-a(x_1)
=s(x_1).
\label{eq:prop1x1-sharp-post-bias}
\end{align}
Because \eqref{eq:prop1x1-sharp-baseline-bias} makes the constructed baseline-bias endpoints equal to $\underline{SB}(\mathcal I_0)$ and $\overline{SB}(\mathcal I_0)$, equations \eqref{eq:prop1x1-sharp-feasibility} and \eqref{eq:prop1x1-sharp-post-bias} verify Assumption \ref{sb:bounds_x}. Finally,
\begin{align}
ATT(x_1)
&=\mathbb E[\widetilde Y_1(1)-\widetilde Y_1(0)
\mid D=1,X_1=x_1]\nonumber\\
&=m_1(x_1)-\{m_1(x_1)-a(x_1)\}
=a(x_1)
\label{eq:prop1x1-sharp-att}
\end{align}
for $F_{X_1\mid D=1}$-almost every $x_1$. Thus a single joint completion attains the entire candidate function, establishing functional sharpness.

For completeness, consider any scalar candidate $\vartheta$ in the integrated interval displayed after Proposition \ref{prop1x1}. 
The conditional gap $\theta_1^{OLS}(X_1)$ is integrable under the treated law because it equals $ATT(X_1)+SB_1(X_1)$, where the conditional ATT is integrable and the bias is bounded by the finite baseline endpoints.
Let
\begin{equation}\label{eq:prop1x1-integrated-s}
s_{\vartheta}
\equiv
\int\theta_1^{OLS}(x_1)dF_{X_1\mid D=1}(x_1)-\vartheta.
\end{equation}
Membership of $\vartheta$ in that interval implies $s_{\vartheta}\in[\underline{SB}(\mathcal I_0),\overline{SB}(\mathcal I_0)]$. Hence the function
\[
a_{\vartheta}(x_1)\equiv\theta_1^{OLS}(x_1)-s_{\vartheta}
\]
satisfies the pointwise feasibility condition \eqref{eq:prop1x1-sharp-feasibility}. Applying the construction above with $a=a_{\vartheta}$ yields $ATT(x_1)=a_{\vartheta}(x_1)$ almost surely by \eqref{eq:prop1x1-sharp-att}, and therefore
\begin{align}
ATT
&=\int a_{\vartheta}(x_1)dF_{X_1\mid D=1}(x_1)\nonumber\\
&=\int\theta_1^{OLS}(x_1)dF_{X_1\mid D=1}(x_1)-s_{\vartheta}
=\vartheta,
\label{eq:prop1x1-integrated-att}
\end{align}
where the last equality follows from \eqref{eq:prop1x1-integrated-s}. Every scalar in the integrated interval, including its endpoints, is therefore attained.
\end{proof}
}

Integrating the conditional bounds over $F_{X_1\mid D=1}$ yields the sharp identified set
\[
ATT\in\left[
\int\theta_1^{OLS}(x_1)dF_{X_1\mid D=1}(x_1)-\overline{SB}(\mathcal I_0),
\int\theta_1^{OLS}(x_1)dF_{X_1\mid D=1}(x_1)-\underline{SB}(\mathcal I_0)
\right].
\]

The following Proposition \ref{propdr} provides a doubly robust estimand for $\int\theta_1^{OLS}(x_1)d F_{X_1\vert D=1}(x_1)$. This result is probably achieved in the literature, but since we could not find a closed-form expression of a doubly robust estimand for this quantity, we provide an estimand along with its proof for completeness.  

\begin{proposition}\label{propdr}
Let $e(X_1)\equiv\mathbb E[D\mid X_1]$ and consider
\[
\tau^{DR}\equiv
\frac{1}{\mathbb E[D]}
\mathbb E\left[
\frac{D-P(X_1)}{1-P(X_1)}\big(Y_1-\mu_0(X_1)\big)
\right],
\]
where $P(X_1)$ and $\mu_0(X_1)$ are postulated models for $e(X_1)$ and $\mathbb E[Y_1\mid D=0,X_1]$, respectively. Suppose $0<\mathbb E[D]<1$, $0<e(X_1)<1$ almost surely, $0\leq P(X_1)<1$ almost surely, and the displayed expectation is absolutely integrable. Then
\[
\tau^{DR}=\int\theta_1^{OLS}(x_1)dF_{X_1\mid D=1}(x_1)
\]
if either, but not necessarily both, $P(X_1)=e(X_1)$ almost surely or $\mu_0(X_1)=\mathbb E[Y_1\mid D=0,X_1]$ almost surely.
\end{proposition}

{
\begin{proof}
We have
\begin{eqnarray*}
	\int\theta_1^{OLS}(x_1)d F_{X_1\vert D=1}(x_1) &=& \mathbb E[Y_1 \vert D=1] - \int \mathbb E[Y_1 \vert D=0, X_1] d F_{X_1\vert D=1}(x_1), \\
	&=& \frac{1}{\mathbb E[D]} \mathbb E[DY_1] - \frac{1}{\mathbb E[D]}\mathbb E \big[ D \cdot \mathbb E[Y_1 \vert D=0, X_1] \big], \\
	&=& \frac{1}{\mathbb E[D]} \mathbb E \Big[ D \big( Y_1 - \mathbb E[Y_1 \vert D=0, X_1] \big) \Big].
\end{eqnarray*}
Then,
\begin{eqnarray*}
	&& \tau^{DR} - \int\theta_1^{OLS}(x_1)d F_{X_1\vert D=1}(x_1) \\
	&& \qquad \qquad = \frac{1}{\mathbb E[D]} \mathbb E \bigg[ \frac{D-P(X_1)}{1-P(X_1)} \big(Y_1 - \mu_0(X_1) \big) - D \big( Y_1 - \mathbb E[Y_1 \vert D=0, X_1] \big) \bigg], \\
	&& \qquad \qquad = \frac{1}{\mathbb E[D]} \mathbb E \bigg[ \frac{D-P(X_1)}{1-P(X_1)} \big(Y_1 - \mu_0(X_1) \big) \\
	&&  \qquad \qquad \qquad \qquad \qquad \qquad  - D \big( Y_1  - \mu_0(X_1) + \mu_0(X_1) - \mathbb E[Y_1 \vert D=0, X_1] \big) \bigg], \\
	&& \qquad \qquad = \frac{1}{\mathbb E[D]} \mathbb E \bigg[ \Big( \frac{D-P(X_1)}{1-P(X_1)} - D \Big) \big(Y_1 - \mu_0(X_1) \big) - D \big( \mu_0(X_1)  - \mathbb E[Y_1 \vert D=0, X_1] \big) \bigg], \\
	&& \qquad \qquad = \frac{1}{\mathbb E[D]} \mathbb E \bigg[ \frac{P(X_1) ( 1 - D )}{1-P(X_1)} \big( \mu_0(X_1) - Y_1 \big)  - D \big( \mu_0(X_1)  - \mathbb E[Y_1 \vert D=0, X_1] \big) \bigg].
\end{eqnarray*}
By the law of iterated expectations, this implies
\begin{eqnarray*}
	&& \tau^{DR} - \int\theta_1^{OLS}(x_1)d F_{X_1\vert D=1}(x_1) \\
	&& \qquad \qquad = \frac{1}{\mathbb E[D]} \mathbb E \Bigg[ \mathbb E \bigg[ \frac{P(X_1) ( 1 - D )}{1-P(X_1)} \big( \mu_0(X_1) - Y_1 \big)  - D \big( \mu_0(X_1)  - \mathbb E[Y_1 \vert D=0, X_1] \big) \bigg \vert X_1 \bigg] \Bigg], \\
	&& \qquad \qquad = \frac{1}{\mathbb E[D]} \mathbb E \bigg[ \frac{P(X_1)}{1-P(X_1)} \big(  \mathbb E [1 - D \vert X_1 ]\cdot \mu_0(X_1) -  \mathbb E [( 1 - D )Y_1\vert X_1 ] \big) \\
	&& \qquad \qquad \qquad \qquad \qquad \qquad \qquad \qquad \qquad \qquad -  \mathbb E[D \vert X_1]\cdot \big( \mu_0(X_1)  - \mathbb E[Y_1 \vert D=0, X_1] \big) \bigg].
\end{eqnarray*}
Finally, using the identity $\mathbb E[Y_1 \vert D=0, X_1] = \frac{\mathbb E[(1-D) Y_1\vert X_1]}{1 - \mathbb E[D \vert X_1]}$, we have
\begin{eqnarray*}
	&& \tau^{DR} - \int\theta_1^{OLS}(x_1)d F_{X_1\vert D=1}(x_1) \\
	&& \qquad \qquad = \frac{1}{\mathbb E[D]} \mathbb E \bigg[ \frac{P(X_1)}{1-P(X_1)} \big(  \mathbb E [1 - D \vert X_1 ]\cdot \mu_0(X_1) - \mathbb E [1 - D \vert X_1 ]\cdot \mathbb E[Y_1 \vert D=0, X_1] \big) \\
	&& \qquad \qquad \qquad \qquad \qquad \qquad \qquad \qquad \qquad \qquad  -  \mathbb E[D \vert X_1]\cdot  \big( \mu_0(X_1)  - \mathbb E[Y_1 \vert D=0, X_1] \big) \bigg] , \\
	&& \qquad \qquad = \frac{1}{\mathbb E[D]} \mathbb E \bigg[ \frac{P(X_1) \cdot \mathbb E [1 - D \vert X_1 ]}{1-P(X_1)} \big( \mu_0(X_1) - \mathbb E[Y_1 \vert D=0, X_1] \big) \\
	&& \qquad \qquad \qquad \qquad \qquad \qquad \qquad \qquad \qquad \qquad  -  \mathbb E[D \vert X_1]\cdot  \big( \mu_0(X_1)  - \mathbb E[Y_1 \vert D=0, X_1] \big) \bigg] , \\
	&& \qquad \qquad = \frac{1}{\mathbb E[D]} \mathbb E \bigg[ \frac{P(X_1)- \mathbb E [D \vert X_1 ]}{1-P(X_1)} \big( \mu_0(X_1) - \mathbb E[Y_1 \vert D=0, X_1] \big) \bigg] , \\
	&& \qquad \qquad = 0,
\end{eqnarray*}
if either $P(X_1) = \mathbb E[D \vert X_1]$ a.s. or $\mu_0(X_1) = \mathbb E[Y_1 \vert D=0, X_1]$ a.s.
\end{proof}
}
Note that the proposed estimand $\tau^{DR}$ is equal to the desired quantity $\int\theta_1^{OLS}(x_1)d F_{X_1\vert D=1}(x_1)$ even if either the propensity score function or the conditional outcome mean function is misspecified. However, if both functions are misspecified, $\tau^{DR}$ is generally different from $\int\theta_1^{OLS}(x_1)d F_{X_1\vert D=1}(x_1)$.  

	\begin{example}
		Consider another version model (\ref{seq1}) where
		\begin{eqnarray*}
			\left\{ \begin{array}{lcl}
				Y_t&=&(1+ X_t)U+\theta X_t D*t\mathbbm{1}\{t\geq 0\} \\ \\
				D&=&\mathbbm{1}\{U\geq 1\}\\ \\
				U &\sim& N(0,1),\text{ } X_t \sim \mathcal U_{\left[0,\frac{1}{1+t^2}\right]}, \text{  and  } X_t\ \indep\ U
			\end{array} \right.
		\end{eqnarray*}  and $\mathcal I_0=\{0\}\times\mathcal X_0$, where $\mathcal X_0=[0,1]$. In this model, $SB_t(x_t)=(1+ x_t)(\alpha_1-\alpha_0)$ where $\alpha_1=\frac{\phi(1)}{1-\Phi(1)}\approx 1.53$ and $\alpha_0=-\frac{\phi(1)}{\Phi(1)} \approx -0.29$. We have $SB_0(x_0) \in [\alpha_1-\alpha_0, 2(\alpha_1-\alpha_0)]$ and $  SB_1(x_1) \in \left[\alpha_1-\alpha_0, 1.5(\alpha_1-\alpha_0)\right] \subseteq [\alpha_1-\alpha_0, 2(\alpha_1-\alpha_0)]\equiv \Delta_{SB_{0X}}$. Thus, the standard unconditional PT condition fails because the unconditional selection bias changes over time. However, the selection bias $SB_1(x_1) $ in period~1 belongs to the convex hull of all selection biases in period 0, i.e., $SB_1(x_1) \in \Delta_{SB_{0X}}$. Hence, our identifying assumption holds. We have $Y_1(1)=(1+X_1)U+\theta X_1$. Then, $\theta_1^{OLS}(x_1)=(1+ x_1)(\alpha_1-\alpha_0)+\theta x_1$, which implies the bounds $ATT(x_1) \in [(x_1-1)(\alpha_1-\alpha_0)+\theta x_1,x_1(\alpha_1-\alpha_0)+\theta x_1]$. The actual conditional ATT function is $ATT(x_1)=\theta x_1$. 
        For the numerical illustration, set the treatment-effect slope to $\theta=8$. Figure \ref{Fig:ex_withX} plots the bounds over the post-treatment covariate support $[0,0.5]$. Writing $\Delta=\alpha_1-\alpha_0$, the lower bound is $(8+\Delta)x_1-\Delta$. Thus the conditional ATT is identified as strictly positive almost everywhere on $(\Delta/(8+\Delta),0.5]$. The cutoff is approximately $0.1847$, so this region covers about $63\%$ of the post-treatment covariate distribution.

\begin{figure}[h]
    \centering
        \pgfmathsetmacro{\biasgap}{1.812735247}
    \pgfmathsetmacro{\effectslope}{8}
    \pgfmathsetmacro{\signcutoff}{\biasgap/(\effectslope+\biasgap)}
    \begin{tikzpicture}
        \pgfmathsetmacro{\biasgap}{1.812735247}
        \pgfmathsetmacro{\effectslope}{8}
        \pgfmathsetmacro{\signcutoff}{\biasgap/(\effectslope+\biasgap)}
        \begin{axis}[
            width=0.66\textwidth,
            height=0.40\textwidth,
            axis lines=box,
            xmin=0, xmax=0.5,
            ymin=-2, ymax=5.2,
            domain=0:0.5,
            samples=2,
            xtick={0,0.1,0.2,0.3,0.4,0.5},
            ytick={-2,-1,0,1,2,3,4,5},
            xlabel={$x_1$},
            ylabel={$ATT(x_1)$},
            legend style={
                at={(0.03,0.97)},
                anchor=north west,
                draw=none,
                fill=none,
                font=\small
            },
            legend cell align=left
        ]
            \addplot[black, thick, name path=lower]
                {(\effectslope+\biasgap)*x-\biasgap};
            \addlegendentry{RDID bounds}
    
            \addplot[black, thick, name path=upper, forget plot]
                {(\effectslope+\biasgap)*x};
            \addplot[fill=black!10, draw=none, forget plot]
                fill between[of=lower and upper];
    
            \addplot[black!55, dotted, forget plot] {0};
            \addplot[black!55, dotted, forget plot]
                coordinates {(\signcutoff,-2) (\signcutoff,0)};
    
            \addplot[black, thick, dashed] {\effectslope*x};
            \addlegendentry{True $ATT(x_1)$}
        \end{axis}
    \end{tikzpicture}
    \caption{Conditional ATT bounds for $\theta=8$ on $x_1\in[0,0.5]$. The vertical guide marks the cutoff above which the lower bound is positive.}
    \label{Fig:ex_withX}
\end{figure}

	\end{example}

\subsection{A sufficient condition for Assumption \ref{sb:bounds_x}}
The following factor structure separates containment of the post-treatment bias range from equality of the conditional bias functions.

\begin{assumption}\label{ass:scx}
$\,$
\begin{enumerate}[(i)]
\item For $t\in\{0,1\}$,
\[
Y_t(0)=g_t(X_t)\lambda(U)+\gamma(V)+\varepsilon_t,
\]
where all terms are integrable, $(X_t,\varepsilon_t)\ \indep\ (U,V)$, and $g_t(\cdot)$, $\lambda(\cdot)$, and $\gamma(\cdot)$ are unknown measurable functions. \label{ass:scx1}
\item The post-treatment loading lies in the closed convex hull of the baseline loading almost surely:
\[
g_1(X_1)\in
\left[
\operatorname*{ess\,inf}_{x_0\in\mathcal X_0}g_0(x_0),
\operatorname*{ess\,sup}_{x_0\in\mathcal X_0}g_0(x_0)
\right].
\]
\label{ass:scx2}
\item Treatment receipt satisfies $D=h(U,V)$, $0<\mathbb P(D=1)<1$, and
\[
\Delta_{\lambda}\equiv
\mathbb E[\lambda(U)\mid D=1]-\mathbb E[\lambda(U)\mid D=0]\neq0.
\]
\label{ass:scx3}
\item Let $\nu\equiv F_{X_0}+F_{X_1}$ on $\mathcal X_0\cap\mathcal X_1$. There exists a measurable set $B\subseteq\mathcal X_0\cap\mathcal X_1$ with $F_{X_0}(B)>0$ and $F_{X_1}(B)>0$ such that the restrictions $F_{X_0}|_B$ and $F_{X_1}|_B$ are mutually absolutely continuous and $g_1(x)\neq g_0(x)$ for $\nu$-almost every $x\in B$. \label{ass:scx4}
\end{enumerate}
\end{assumption}

Assumption \ref{ass:scx}.(\ref{ass:scx2}) directly restricts the loading ranges and therefore does not require monotonicity of either loading. Assumptions \ref{ass:scx}.(\ref{ass:scx3})--(\ref{ass:scx4}) separately ensure that time variation in the loading generates a genuine failure of conditional PT on the comparison region $B$.

\begin{proposition}\label{prop:scx}
Suppose $\mathcal I_0=\{0\}\times\mathcal X_0$ and Assumption \ref{ass:scx} holds. Then Assumption \ref{sb:bounds_x} holds, and for any versions of the conditional biases, $SB_1(x)\neq SB_0(x)$ for $\nu$-almost every $x\in B$.
\end{proposition}
{
\begin{proof}
Define
\[
\Delta_\gamma\equiv
\mathbb E[\gamma(V)\mid D=1]-\mathbb E[\gamma(V)\mid D=0].
\]
For $t\in\{0,1\}$, Assumptions \ref{ass:scx}.(\ref{ass:scx1}) and \ref{ass:scx}.(\ref{ass:scx3}) imply
\[
SB_t(x)=g_t(x)\Delta_\lambda+\Delta_\gamma
\]
for $F_{X_t}$-almost every $x\in\mathcal X_t$, since $(X_t,\varepsilon_t)$ is independent of $D=h(U,V)$.. Let
\[
\ell_0\equiv\operatorname*{ess\,inf}_{x_0\in\mathcal X_0}g_0(x_0),
\qquad
\nu_0\equiv\operatorname*{ess\,sup}_{x_0\in\mathcal X_0}g_0(x_0).
\]
By Assumption \ref{ass:scx}.(\ref{ass:scx2}), $g_1(X_1)\in[\ell_0,\nu_0]$ almost surely. Applying the affine map $z\mapsto z\Delta_\lambda+\Delta_\gamma$ shows, regardless of the sign of $\Delta_\lambda$, that
\[
SB_1(X_1)\in
\left[
\operatorname*{ess\,inf}_{x_0\in\mathcal X_0}SB_0(x_0),
\operatorname*{ess\,sup}_{x_0\in\mathcal X_0}SB_0(x_0)
\right]
\]
almost surely. 
Since $X_1$ is independent of $D$, the inclusion also holds $F_{X_1\mid D=1}$-almost everywhere.
Since $\mathcal I_0=\{0\}\times\mathcal X_0$, Assumption \ref{sb:bounds_x} holds.

By Assumption \ref{ass:scx}.(\ref{ass:scx4}), any subset of $B$ that is null under either period's covariate law is also $\nu$-null. Thus both factor-model identities hold outside a common $\nu$-null subset of $B$, for any versions of the conditional biases.
For $\nu$-almost every $x\in B$,
\[
SB_1(x)-SB_0(x)=\{g_1(x)-g_0(x)\}\Delta_\lambda\neq0
\]
by Assumptions \ref{ass:scx}.(\ref{ass:scx3})--(\ref{ass:scx4}). Thus conditional PT fails for $\nu$-almost every $x\in B$.
\end{proof}
}

For example, let $X_0=X_1=X$ have support $[0,1]$, $g_0(x)=1+x$, and $g_1(x)=1+\alpha x$ for some $\alpha\in(0,1)$. Then $g_1(X)\in[1,1+\alpha]\subset[1,2]$, but $g_1(x)\neq g_0(x)$ for every $x>0$. 
The two covariate laws are identical, so $B=(0,1]$ satisfies Assumption \ref{ass:scx}.(\ref{ass:scx4}).
Thus the conditional bias-set restriction holds while conditional PT fails whenever $\Delta_{\lambda}\neq0$.

\section{Additional examples}
\begin{example}
		Consider a modified version of the previous models where
		\begin{eqnarray*}
			\left\{ \begin{array}{lcl}
				Y_t&=&(1+ 0.25^t*X_t)U+\theta X_t D*t\mathbbm{1}\{t\geq 0\} \\ \\
				D&=&\mathbbm{1}\{U\geq 1\}\\ \\
				U &\sim& N(0,1),\text{ } X_t \sim \mathcal U_{\left[0,1+t^2\right]}, \text{  and  } X_t\ \indep\ U
			\end{array} \right.
		\end{eqnarray*}  and $\mathcal I_0=\{0\}\times\mathcal X_0$, where $\mathcal X_0=[0,1]$.
In this model, $SB_t(x_t)=(1+0.25^t* x_t)(\alpha_1-\alpha_0)$ where $\alpha_1=\frac{\phi(1)}{1-\Phi(1)}\approx 1.53$ and $\alpha_0=-\frac{\phi(1)}{\Phi(1)} \approx -0.29$. We have $SB_0(x_0) \in [\alpha_1-\alpha_0, 2(\alpha_1-\alpha_0)]$ and $  SB_1(x_1) \in \left[\alpha_1-\alpha_0, 1.5(\alpha_1-\alpha_0)\right] \subseteq [\alpha_1-\alpha_0, 2(\alpha_1-\alpha_0)]\equiv \Delta_{SB_{0X}}$. Thus, the standard unconditional PT condition fails because the unconditional selection bias changes over time. However, the selection bias $SB_1(x_1) $ in period 1 belongs to the convex hull of all selection biases in period 0, i.e., $SB_1(x_1) \in \Delta_{SB_{0X}}$. Hence, our identifying assumption holds.
\end{example}

\begin{example}\label{ex:ife}
		Consider the following model where
		\begin{eqnarray*}
			\left\{ \begin{array}{lcl}
				Y_t&=&\eta_t + V+ U*t+\theta D*t\mathbbm{1}\{t\geq 0\} \\ \\
				D&=&\mathbbm{1}\{\vert U \vert \geq 1\}\\ \\
				U &\sim& N(0,\sigma^2),\text{ } \eta_t\ \indep\ (U,V)
			\end{array} \right.
		\end{eqnarray*}  and $\mathcal I_0=[-T_0,0]$.
 Our bias set stability assumption holds in this example.
\end{example}

\begin{example}[Multiple treatment periods with staggered adoption where PT holds]\label{ex:staggeredTWFE}
We consider a DGP in which there is selection on a time-invariant unobservable and there are no instrumental variables available.
\begin{eqnarray*}
\left\{ \begin{array}{lcl}
          Y_{t} &=& U+\varepsilon_t +(\sum_{s=1}^T\theta_s D_s) \mathbbm{1}\{t>0\}\ \text{ for } t=0,\ldots,T\\ \\
     D_t &=& \mathbbm{1}\{U \geq 2-\frac{t}{T}\}    
     \end{array} \right.
     \end{eqnarray*} 
where $U\ \indep\ (\varepsilon_0,\{\varepsilon_t,\theta_t\}_{t=1}^T)$, $\theta_t\sim \mathcal{U}_{[0,1+t^2]}$, $\mathcal I_0=\{0\}$, $\varepsilon_t \sim \mathcal N(t^2,1)$, and $U \sim \mathcal U_{[0,2]}$.
In this DGP, 
\begin{eqnarray*}
&& \mathbb E[Y_t(\bm d)-Y_0(0) \vert \bm D=\bm d']\\
&& \qquad \qquad =\mathbb E[Y_t(\bm d)-Y_0(0) \vert \bm D=\bm d].
\end{eqnarray*}
Therefore, PT holds.

\begin{table}[!htbp] \centering 
  \caption{Summary of the TWFE estimation results (B=500)} 
  \label{tab:TWFEMC} 
\setlength{\tabcolsep}{5pt}\begin{tabular}{@{}ccccccccccc@{}} 
\hline 
\hline \\
 & \multirow{2}{*}{True Value} & \multicolumn{3}{c}{$N=1,000$}  & \multicolumn{3}{c}{$N=5,000$} & \multicolumn{3}{c}{$N=10,000$} \\
 & & Est. & MAE & RMSE & Est. & MAE & RMSE & Est. & MAE & RMSE \\
\hline \\
$\theta_{1}^{1}$ & $8.500$ & $8.512$ & $0.229$ & $0.291$ & $8.497$ & $0.095$ & $0.120$ & $8.504$ & $0.069$ & $0.087$ \\ 
$\theta_{1}^{2}$ & $7.500$ & $7.500$ & $0.230$ & $0.282$ & $7.498$ & $0.104$ & $0.129$ & $7.501$ & $0.074$ & $0.092$ \\ 
$\theta_{1}^{3}$ & $5$ & $5.005$ & $0.188$ & $0.236$ & $5.001$ & $0.089$ & $0.111$ & $5.003$ & $0.064$ & $0.079$ \\ 
$\theta_{2}^{1}$ & $8.500$ & $8.507$ & $0.237$ & $0.297$ & $8.498$ & $0.098$ & $0.123$ & $8.502$ & $0.068$ & $0.086$ \\ 
$\theta_{2}^{2}$ & $7.500$ & $7.506$ & $0.231$ & $0.281$ & $7.498$ & $0.104$ & $0.130$ & $7.503$ & $0.074$ & $0.092$ \\ 
$\theta_{2}^{3}$ & $5$ & $5.007$ & $0.196$ & $0.244$ & $4.999$ & $0.092$ & $0.114$ & $5.003$ & $0.065$ & $0.080$ \\ 
$\theta_{3}^{1}$ & $8.500$ & $8.510$ & $0.228$ & $0.285$ & $8.498$ & $0.096$ & $0.122$ & $8.500$ & $0.069$ & $0.087$ \\ 
$\theta_{3}^{2}$ & $7.500$ & $7.502$ & $0.230$ & $0.284$ & $7.496$ & $0.103$ & $0.128$ & $7.501$ & $0.074$ & $0.092$ \\ 
$\theta_{3}^{3}$ & $5$ & $5.007$ & $0.190$ & $0.238$ & $4.998$ & $0.093$ & $0.115$ & $5.001$ & $0.066$ & $0.081$ \\ 
\hline \\
\end{tabular} 

{\raggedright\footnotesize \textit{
	Note: Est. is the mean estimate over $B=500$ Monte Carlo replications. MAE is the mean absolute error, $B^{-1}\sum_{b=1}^B|\widehat\theta_b-\theta|$, and RMSE is the root mean squared error. $N$ denotes the sample size. 
}\par}
\end{table} 
In Table~\ref{tab:TWFEMC}, the Monte Carlo means are within $0.012$ of their targets, and both MAE and RMSE decrease with sample size for every coefficient.

\end{example}

\begin{example}[Multiple treatment periods with non-staggered adoption where PT holds]
We consider a DGP in which there is selection on a time-invariant unobservable and there are no instrumental variables available.
\begin{eqnarray*}
\left\{ \begin{array}{lcl}
          Y_{t} &=& U+\varepsilon_t +(\sum_{s=1}^T\theta_s D_s) \mathbbm{1}\{t>0\}\ \text{ for } t=0,\ldots,T\\ \\
     D_t &=& \mathbbm{1}\{U \geq 2-\frac{V_t}{T}\}    
     \end{array} \right.
     \end{eqnarray*} 
where $(U,\{V_t\}_{t=1}^T)\ \indep\ (\varepsilon_0,\{\varepsilon_t,\theta_t\}_{t=1}^T)$, $\theta_t\sim \mathcal{U}_{[0,1+t^2]}$, $\mathcal I_0=\{0\}$, $\varepsilon_t \sim \mathcal N(t^2,1)$, $V_t \sim \mathcal U_{[0,t]}$, and $U \sim \mathcal U_{[0,2]}$.
In this DGP, 
\begin{eqnarray*}
&& \mathbb E[Y_t(\bm d)-Y_0(0) \vert \bm D=\bm d']\\
&& \qquad \qquad =\mathbb E[Y_t(\bm d)-Y_0(0) \vert \bm D=\bm d].
\end{eqnarray*}
Therefore, PT holds. 
\end{example}

\subsection{Simple example where Assumption \ref{ass:expoAE} holds}
\begin{example}\label{AE_example}
Suppose for each individual $i\in\{1,2,...,N\}$ and $t\in\mathcal{T}=\{-T_0,-T_0+1,...,-2,-1,0,1\}$,
\[
\begin{cases}
Y_{it}=\delta^{|t-1|}U_{it}D_{i}+V_{i}+\varepsilon_{it}\\
D_{i}=\mathds{1}\{V_{i}>0\}
\end{cases}
\]
where $\delta\in(0,1)$, $U_{it}\sim\mathcal{U}(0,4)$, $V_{i}\sim\mathcal{N}(0,1)$,
$\varepsilon_{it}\sim\mathcal{N}(0,2)$ and $\forall t$, $U_{it}\perp\!\!\!\!\perp V_{i}$.

\end{example}

In Example \ref{AE_example}, 
{ $Y_{it}$ is the observed outcome, $D_i$ is the treatment indicator, and $\varepsilon_{it}$ is an independent error term for individual $i$ and time $t$.}
The individual-specific variable $V_{i}$ drives selection into treatment, and $U_{it}$ is the treatment effect which is uniformly distributed and heterogeneous across individuals and time, and the true ATT is equal to 2. However, the magnitude of anticipation effect is governed by a positive constant $\delta$ since 
 $AE_t=\delta^{|t-1|}\mathbb E[U_{it}\mid D_i=1]=2\delta^{|t-1|}$ and $ATT=\mathbb E[U_{i1}\mid D_i=1]=2$.

\begin{figure}[H]

\begin{centering}
\includegraphics[scale=0.55]{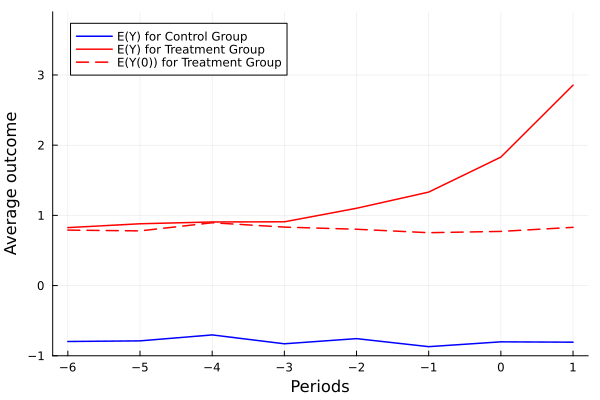}
\par\end{centering}
\caption{Average outcomes over time for two groups}
\label{AEexample_figure}

\vspace{0.5cm}
\footnotesize
 Notes: Treatment happens between period 0 and period 1. The red solid line represents the observed average outcome for the treatment group, and the red dashed line is the potential outcome for the treatment group if treatment never happens. The blue solid line represents the observed outcome for the control group. In period 1, the difference between the red solid line and the red dashed line is ATT, and the difference between the red dashed line and the blue solid line is $SB_1$. In period 0, the difference between the red solid line and the red dashed line is $AE_0$, and the difference between the red dashed line and the blue solid line is $SB_0$. Similarly, during the pre-treatment periods, the difference between the red solid line and the blue solid line can be decomposed into corresponding $AE$ and $SB$.

\end{figure}

Let $\delta=0.5,N=2000,T_0=6$. We can easily verify that this DGP satisfies Assumption \ref{ass:strongPT}. Furthermore, if we set the bounding parameter $c\geq\delta$, then Assumption \ref{ass:expoAE} will hold. The average observed outcomes over time for the two groups and the potential outcomes for the treatment group are illustrated in Figure \ref{AEexample_figure}, and our model is correctly specified when we set $c\geq 0.5$.

\subsection{A Motivating model for Assumption \ref{ass:expoAE}} \label{ap:motivating}
To motivate Assumption \ref{ass:expoAE}, consider the forward-looking potential-outcome model (\citealp{chow1989, malani2015interpreting})
\begin{equation}\label{forwardmodel}
Y_t(\bm D)=Y_t(\bm 0)+\tau_tD_t+\sum_{j=1}^{\infty}\beta^j\mathbb E_t(\tau_{t+j}D_{t+j}),
\end{equation}
where $\bm D\equiv(\ldots,D_{t-1},D_t,D_{t+1},\ldots)$ is the treatment path, $\tau_t\equiv Y_t(\ldots,0,D_t=1,0,\ldots)-Y_t(\bm0)$ is the contemporaneous individual treatment effect, and $\beta\in[0,1)$ is an individual-specific stochastic discount factor. Use the conventional forecast-error decomposition
\[
\tau_{t+j}D_{t+j}=\mathbb E_t(\tau_{t+j}D_{t+j})+e_{t,t+j},
\]
so that the forecast equals the realization minus the forecast error.

{The model accommodates heterogeneous degrees of forward-looking behavior; an individual is a non-anticipator when $\beta=0$ and an anticipator when $\beta>0$.}

Suppose treatment can occur only in period 1, and write $\bm d^1=(\ldots,0,D_1=1,0,\ldots)$. Then
\[
ATT_1=\mathbb E[\tau_1\mid\bm D=\bm d^1]
\]
and, for $t_0\in\mathcal T_0=\{\ldots,-1,0\}$,
\[
AE_{t_0}=\mathbb E[Y_{t_0}(\bm d^1)-Y_{t_0}(\bm0)\mid\bm D=\bm d^1].
\]

\begin{proposition}\label{prop_motivation}
Suppose treatment can occur only in period 1 and, for every $t_0\in\mathcal T_0$,
\begin{equation}\label{eq:forecast-orthogonality}
\mathbb E\!\left[\beta^{|t_0|+1}e_{t_0,1}\mid\bm D=\bm d^1\right]=0.
\end{equation}
Under model \eqref{forwardmodel}, exponential $c$-bounded anticipation holds if either
\begin{enumerate}[(i)]
\item $\beta=b$ almost surely for some $b\in[0,1)$; or
\item $\beta\ \indep\ \tau_1\mid\bm D=\bm d^1$ and $0\leq\beta\leq\bar\beta<1$ almost surely.
\end{enumerate}
In case (i), one may take $c=b$; in case (ii), one may take $c=\bar\beta$.\end{proposition}

\begin{proof}
Let $h\equiv|t_0|+1$. Along the treated path $\bm d^1$, the only nonzero future-treatment term at time $t_0$ is the period-1 term. Since the forecast-error convention gives
\[
\mathbb E_{t_0}(\tau_1D_1)=\tau_1-e_{t_0,1}
\quad\text{on }\{\bm D=\bm d^1\},
\]
model \eqref{forwardmodel} and \eqref{eq:forecast-orthogonality} imply
\begin{align*}
AE_{t_0}
&=\mathbb E[\beta^h(\tau_1-e_{t_0,1})\mid\bm D=\bm d^1]\\
&=\mathbb E[\beta^h\tau_1\mid\bm D=\bm d^1].
\end{align*}
In case (i), $\beta=b$ almost surely, so
\[
AE_{t_0}=b^h ATT_1
\]
and the result holds with $c=b$. In case (ii), conditional independence gives
\[
AE_{t_0}
=\mathbb E[\beta^h\mid\bm D=\bm d^1]ATT_1.
\]
Because $0\leq\beta\leq\bar\beta<1$ almost surely,
\[
|AE_{t_0}|
\leq\bar\beta^h|ATT_1|,
\]
so the result holds with $c=\bar\beta$.
\end{proof}

For two periods, under the weighted forecast-error orthogonality condition and conditional independence of $\beta$ and $\tau_1$ for treated units, $|AE_0|=\mathbb E[\beta\mid D=1]|ATT_1|$; hence one may take $c=\mathbb E[\beta\mid D=1]$.

\clearpage
\section{Policy-oriented RDID and projection-based forecasting}\label{app:point-approaches}

Building on the framework in Section~\ref{sec:setting} and maintaining the no-anticipation assumption in Section~\ref{subsec:did-relax-pt}, this appendix develops two approaches to determining the period-1 selection bias: a policy-oriented decision based on a loss function and projection-based forecasting over ordered baseline periods.

Both approaches use only pre-treatment information to choose or forecast the period-1 selection bias. PO-RDID can use selected periods or covariate-indexed baseline cells. Projection-based forecasting uses ordered pre-treatment periods.

{

\subsection{PO-RDID: Loss-based choice of the period-1 selection bias}\label{subsec:po-rdid}
Section~\ref{subsec:did-relax-pt} delivers an identified set for the ATT. For policy communication or internal decision-making, a researcher may want a single summary of that set. We consider three losses ($L_1$, $L_2$, and $L_\infty$) that use the distribution of selected baseline biases to choose an adjustment for the period-1 outcome gap.

First consider a period-only information set $\mathcal I_0\subseteq\mathcal T_0$. Let $I_0$ be a random period in $\mathcal I_0$ with known law $\mu_{I_0}$, and write $SB(I_0)=SB_{I_0}$. Under Assumption~\ref{ass:NoAE}, the selected baseline biases equal their observed outcome gaps. For each minimizing choice $SB_1^{opt}$, the ATT decomposition in \eqref{eq:att} motivates reporting $\theta_1^{OLS}-SB_1^{opt}$ as a \textit{policy-oriented robust difference-in-differences (PO-RDID)} summary. Proposition~\ref{prop:po-att} characterizes when this summary recovers the ATT, and Section~\ref{sec:po-rdid-covariates} extends the loss rules to covariate-indexed baseline cells.

\begin{assumption}\label{ass:optimalDID}
Let $\mathcal L(b;SB(I_0),\mu_{I_0})$ be the decision maker's loss from choosing $b$. The set of optimal choices is
\[
\mathcal B^{opt}(\mu_{I_0})
\equiv
\arg\min_{b\in\mathbb R}\mathcal L(b;SB(I_0),\mu_{I_0}),
\]
and the decision maker selects some $SB_1^{opt}\in\mathcal B^{opt}(\mu_{I_0})$.
\end{assumption}

For $1\leq p<\infty$, consider
\[
\mathcal L_p(b;SB(I_0),\mu_{I_0})
=
\left(\mathbb E_{\mu_{I_0}}\left[|b-SB(I_0)|^p\right]\right)^{1/p}.
\]
Assume the corresponding moment is finite; for the $L_\infty$ problem below, assume $SB(I_0)$ is essentially bounded. We derive the optimal choices for $p\in\{1,2,\infty\}$. These losses have closed-form minimizers, while other loss functions can be treated analogously.

\subsubsection{$L1$ loss: Mean absolute error (MAE)}
Let
\[
\operatorname{Med}_{\mu_{I_0}}\{SB(I_0)\}
\equiv
\left\{m:
\mathbb P_{\mu_{I_0}}\{SB(I_0)\leq m\}\geq\tfrac12,
\ \mathbb P_{\mu_{I_0}}\{SB(I_0)\geq m\}\geq\tfrac12
\right\}
\]
denote the possibly set-valued median. Since
\[
\mathcal L_1(b;SB(I_0),\mu_{I_0})
=\mathbb E_{\mu_{I_0}}|b-SB(I_0)|,
\]
the set of optimal decisions is
\[
\mathcal B^{opt}(\mu_{I_0})
=\operatorname{Med}_{\mu_{I_0}}\{SB(I_0)\}.
\]
Accordingly, absent a tie-breaking rule, the $L_1$ PO-RDID decision is set-valued:
\[
\Theta_{PO\text{-}RDID}^{\mathcal L_1}
=\left\{\theta_1^{OLS}-m:m\in\operatorname{Med}_{\mu_{I_0}}\{SB(I_0)\}\right\}.
\]

\subsubsection{$L2$ loss: Root mean square error (RMSE)}
Provided $\mathbb E_{\mu_{I_0}}[SB(I_0)^2]<\infty$,
\[
\mathcal L_2(b;SB(I_0),\mu_{I_0})
=\left(\mathbb E_{\mu_{I_0}}[(b-SB(I_0))^2]\right)^{1/2}.
\]
Minimizing RMSE is equivalent to minimizing MSE, whose unique minimizer is
\[
SB_1^{opt}=\mathbb E_{\mu_{I_0}}[SB(I_0)].
\]
Hence
\[
\theta_{PO\text{-}RDID}^{\mathcal L_2}
=\theta_1^{OLS}-\mathbb E_{\mu_{I_0}}[SB(I_0)].
\]

\subsubsection{$L\infty$ loss: Maximal regret}
The maximal-regret loss is
\begin{align*}
\mathcal L_\infty(b;SB(I_0),\mu_{I_0})
&=\operatorname*{ess\,sup}_{\mu_{I_0}}|b-SB(I_0)|\\
&=\inf\left\{M\geq0:
\mathbb P_{\mu_{I_0}}\{|b-SB(I_0)|\leq M\}=1
\right\}.
\end{align*}
Let
\[
\underline s_\mu
\equiv\operatorname*{ess\,inf}_{\mu_{I_0}}SB(I_0),
\qquad
\overline s_\mu
\equiv\operatorname*{ess\,sup}_{\mu_{I_0}}SB(I_0).
\]
The $L_\infty$ problem has the unique minimizer
\[
SB_1^{opt}=\frac{\underline s_\mu+\overline s_\mu}{2},
\]
and therefore
\[
\theta_{PO\text{-}RDID}^{\mathcal L_\infty}
=\theta_1^{OLS}-\frac{\underline s_\mu+\overline s_\mu}{2}.
\]
When $\mu_{I_0}$ gives positive probability to every selected baseline period, these essential endpoints equal the infimum and supremum of the selected period-specific biases.

\subsubsection{The distributionally robust envelope and recovery of the ATT}
If $\mathcal I_0$ is a singleton, $SB(I_0)$ is degenerate and each of the three losses selects the sole baseline bias, yielding the DID contrast for that baseline period. Unlike the $L_1$ and $L_2$ decisions, the $L_\infty$ decision depends only on the essential endpoints, not on probability weights within the essential support.

For these three losses, every minimizing adjustment lies between the infimum and supremum of the selected baseline biases. Under the assumptions of Proposition~\ref{prop1}, each resulting PO-RDID summary therefore lies in the PT-RDID interval.

For the distributionally robust envelope, vary the law of the baseline bias over all Borel probability measures supported on the interval $\Delta_{SB(\mathcal I_0)}$ in Assumption~\ref{sb:bounds}, and denote this class of laws by $\Lambda$. Assume that the interval has finite endpoints. For $\lambda\in\Lambda$, define $\mathcal L_p(z;\lambda)$ by replacing the law of $SB(I_0)$ in the preceding $L_p$ criterion with $\lambda$. For $p\in\{1,2,\infty\}$, let
\[
\mathcal A_p
\equiv
\left\{
\theta_1^{OLS}-b:
\lambda\in\Lambda,\quad
b\in\arg\min_{z\in\mathbb R}\mathcal L_p(z;\lambda)
\right\}.
\]
\begin{definition}
The distributionally robust PO-RDID envelope under loss $p$ is
\[
[\inf\mathcal A_p,\sup\mathcal A_p].
\]
\end{definition}

\begin{lemma}\label{lem1}
{Suppose Assumption \ref{ass:NoAE} holds for the selected pre-treatment periods $\mathcal I_0\subseteq\mathcal T_0$ and $\underline{SB}(\mathcal I_0)$ and $\overline{SB}(\mathcal I_0)$ are finite.} Then, for each $p\in\{1,2,\infty\}$, the distributionally robust PO-RDID envelope coincides with the bounds in Proposition \ref{prop1}.\end{lemma}
{
\begin{proof}
Write $\Delta_{SB(\mathcal I_0)}=[m,M]$. For every $\lambda\in\Lambda$, every relevant minimizer lies in $[m,M]$: under $L_2$ the minimizer is the mean; under $L_1$ the minimizer set is the set of medians; and under $L_\infty$ the minimizer is the midpoint of the essential lower and upper endpoints of the support. Hence
\[
\mathcal A_p\subseteq[\theta_1^{OLS}-M,\theta_1^{OLS}-m]
\]
for each $p\in\{1,2,\infty\}$.

Conversely, the degenerate measures $\delta_M$ and $\delta_m$ belong to $\Lambda$. Under $\delta_M$, all three losses are uniquely minimized at $M$, yielding $\theta_1^{OLS}-M$; under $\delta_m$, all three are uniquely minimized at $m$, yielding $\theta_1^{OLS}-m$. Therefore
\[
[\inf\mathcal A_p,\sup\mathcal A_p]
=[\theta_1^{OLS}-M,\theta_1^{OLS}-m].
\]
Under Assumption \ref{ass:NoAE}, $SB_{t_0}=\theta_{t_0}^{OLS}$ for every $t_0\in\mathcal I_0$. Hence $m=\inf_{t_0\in\mathcal I_0}\theta_{t_0}^{OLS}$ and $M=\sup_{t_0\in\mathcal I_0}\theta_{t_0}^{OLS}$, so this is the interval in Proposition \ref{prop1}.
\end{proof}
}

{\begin{proposition}[When PO-RDID recovers the ATT]\label{prop:po-att}
Suppose Assumption \ref{ass:optimalDID} and the moment conditions stated for the relevant loss hold. Then:
\begin{enumerate}[(i)]
\item $ATT\in\Theta_{PO\text{-}RDID}^{\mathcal L_1}$ if and only if
\[
SB_1\in\operatorname{Med}_{\mu_{I_0}}\{SB(I_0)\}.
\]
If a tie-breaking rule selects a single median, the resulting $L_1$ PO-RDID estimand equals the ATT if and only if that rule selects $SB_1$.
\item $\theta_{PO\text{-}RDID}^{\mathcal L_2}=ATT$ if and only if
\[
SB_1=\mathbb E_{\mu_{I_0}}[SB(I_0)].
\]
\item $\theta_{PO\text{-}RDID}^{\mathcal L_\infty}=ATT$ if and only if
\[
SB_1=\frac{\underline s_\mu+\overline s_\mu}{2}.
\]
\end{enumerate}
A common sufficient condition for all three criteria is that
\[
U\equiv SB(I_0)-SB_1
\]
is essentially bounded, is symmetrically distributed about zero, $U\overset{d}{=}-U$, and has zero as its unique median. Under this condition,
\[
\Theta_{PO\text{-}RDID}^{\mathcal L_1}
=\{ATT\},
\qquad
\theta_{PO\text{-}RDID}^{\mathcal L_2}
=\theta_{PO\text{-}RDID}^{\mathcal L_\infty}
=ATT.
\]
\end{proposition}
{
\begin{proof}
Equation \eqref{eq:att} gives
\begin{equation}\label{eq:po-att-basic}
ATT=\theta_1^{OLS}-SB_1.
\end{equation}
For the $L_1$ criterion, a value belongs to $\Theta_{PO\text{-}RDID}^{\mathcal L_1}$ precisely when it can be written as $\theta_1^{OLS}-m$ for some median $m$ of $SB(I_0)$. By \eqref{eq:po-att-basic},
\[
\theta_1^{OLS}-m=ATT
\quad\Longleftrightarrow\quad
m=SB_1.
\]
Therefore $ATT\in\Theta_{PO\text{-}RDID}^{\mathcal L_1}$ if and only if $SB_1$ is a median; after a tie-breaking rule selects one median, the resulting point equals the ATT if and only if the selected median is $SB_1$.

For the $L_2$ criterion, the unique optimal bias is $\mathbb E_{\mu_{I_0}}[SB(I_0)]$. Hence, again by \eqref{eq:po-att-basic},
\[
\theta_{PO\text{-}RDID}^{\mathcal L_2}=ATT
\quad\Longleftrightarrow\quad
\mathbb E_{\mu_{I_0}}[SB(I_0)]=SB_1.
\]
Likewise, the $L_\infty$ criterion selects $(\underline s_\mu+\overline s_\mu)/2$, so
\[
\theta_{PO\text{-}RDID}^{\mathcal L_\infty}=ATT
\quad\Longleftrightarrow\quad
\frac{\underline s_\mu+\overline s_\mu}{2}=SB_1.
\]

For the common sufficient condition, write $SB(I_0)=SB_1+U$. Symmetry and essential boundedness imply $\mathbb E_{\mu_{I_0}}[U]=0$, so
\[
\mathbb E_{\mu_{I_0}}[SB(I_0)]=SB_1.
\]
Because zero is the unique median of $U$, $SB_1$ is the unique median of $SB(I_0)$. Finally, if $\underline u$ and $\overline u$ are the essential lower and upper endpoints of $U$, symmetry gives $\underline u=-\overline u$. Therefore
\[
\frac{\underline s_\mu+\overline s_\mu}{2}
=
\frac{(SB_1+\underline u)+(SB_1+\overline u)}{2}
=SB_1.
\]
All three conclusions follow from the preceding characterizations.
\end{proof}
}

For example, if $SB(I_0)$ takes the values $SB_1-\delta$, $SB_1$, and $SB_1+\delta$ with probabilities $1/4$, $1/2$, and $1/4$, respectively, the common sufficient condition holds and all three criteria select $SB_1$.
}

\subsubsection{Covariate-indexed baseline cells}\label{sec:po-rdid-covariates}

Now let $\mathcal I_0$ consist of cells $(t_0,x)$ as in Appendix~D, and write $SB(t_0,x)=SB_{t_0}(x)$. Let $I_0$ have a known law $\mu_{I_0}$ on $\mathcal I_0$. For every selected period $t_0$, require the measure $A\mapsto\mu_{I_0}(\{t_0\}\times A)$ on covariate sets $A$ to be absolutely continuous with respect to $F_{X_{t_0}}$. Under conditional no anticipation and overlap (Assumptions~\ref{ass:NoAE_x} and \ref{overlap}), $SB_{t_0}(x)=\theta_{t_0}^{OLS}(x)$ for $F_{X_{t_0}}$-almost every $x$ in the selected region. The weights give these exceptional sets zero probability, so the induced law of $SB(I_0)$ is identified and independent of the versions of the conditional biases.

Use the same median, mean, and midrange rules to select a scalar bias $b$, under the corresponding moment conditions. Assume $\theta_1^{OLS}(X_1)$ is integrable under the treated law, as it is under the assumptions of Proposition~\ref{prop1x1}. Define the post-treatment gap and bias standardized to the treated covariate distribution by
\[
\bar\theta_1^{OLS}\equiv\int\theta_1^{OLS}(x)\,dF_{X_1\mid D=1}(x),
\qquad
\bar s_1\equiv\int SB_1(x)\,dF_{X_1\mid D=1}(x).
\]
The PO-RDID summary is $\bar\theta_1^{OLS}-b$ for each minimizing $b$. Since $ATT=\bar\theta_1^{OLS}-\bar s_1$, the summary recovers the ATT exactly when $b=\bar s_1$. Under the assumptions of Proposition~\ref{prop1x1}, the induced bias law is supported on the interval $\Delta_{SB(\mathcal I_0)}$ in Assumption~\ref{sb:bounds_x}. Each of the three rules then yields a summary within the integrated bounds.

For the distributionally robust envelope, allow any probability law supported on this interval of bias values. Using $\bar\theta_1^{OLS}$ in place of $\theta_1^{OLS}$, the same endpoint argument as in Lemma~\ref{lem1} yields exactly the integrated bounds following Proposition~\ref{prop1x1}.

{\subsection{Projection-based forecasting of $SB_1$ when baseline periods are ordered}\label{sec:orderedinfo}
For forecasting, let $\mathcal I_0\subseteq\mathcal T_0$ consist of ordered pre-treatment periods.
Let $\mu_{I_0}$ be a known, analyst-chosen probability measure with support $\mathcal I_0$, and let $p_K(t)$ be a $K\times1$ vector of known basis functions defined on $\mathcal I_0\cup\{1\}$. Suppose
\[
\mathbb E_{\mu_{I_0}}[SB(I_0)^2]<\infty,
\qquad
G_K\equiv
\mathbb E_{\mu_{I_0}}[p_K(I_0)p_K(I_0)']
\]
is finite and positive definite. Define the population projection coefficient and its period-1 extrapolation by
\begin{align}
\beta_K
&\equiv
\arg\min_{b\in\mathbb R^K}
\mathbb E_{\mu_{I_0}}
\left[(SB(I_0)-p_K(I_0)'b)^2\right]
\nonumber\\
&=G_K^{-1}
\mathbb E_{\mu_{I_0}}[p_K(I_0)SB(I_0)],
\label{eq:projection-beta}\\
SB_1^{\mathrm{proj}}
&\equiv p_K(1)'\beta_K.
\label{eq:projection-bias}
\end{align}

\begin{assumption}[Projection extrapolation]\label{ass:projection-extrapolation}
The period-1 selection bias equals the extrapolated population projection:
\[
SB_1=SB_1^{\mathrm{proj}}=p_K(1)'\beta_K.
\]
\end{assumption}

\begin{proposition}[Projection-based RDID]\label{prop:projection-rdid}
Suppose model \eqref{seq1}, Assumption \ref{ass:NoAE}, and Assumption \ref{ass:projection-extrapolation} hold. Then the ATT is point identified by
\begin{align}
\theta^{\mathrm{Proj\text{-}RDID}}
&\equiv
\theta_1^{OLS}-p_K(1)'\beta_K
\nonumber\\
&=
\theta_1^{OLS}
-p_K(1)'G_K^{-1}
\mathbb E_{\mu_{I_0}}
\left[p_K(I_0)\theta_{I_0}^{OLS}\right],
\label{eq:projection-rdid}
\end{align}
where $\theta_{I_0}^{OLS}$ denotes the baseline difference in means evaluated at the random index $I_0$.
\end{proposition}
{
\begin{proof}
Because $G_K$ is positive definite, the projection criterion in \eqref{eq:projection-beta} is strictly convex in $b$. Its first-order condition is
\[
G_K b
=
\mathbb E_{\mu_{I_0}}[p_K(I_0)SB(I_0)],
\]
so its unique minimizer is
\begin{equation}\label{eq:projection-proof-beta}
\beta_K
=G_K^{-1}
\mathbb E_{\mu_{I_0}}[p_K(I_0)SB(I_0)].
\end{equation}
Under Assumption \ref{ass:NoAE},
\begin{equation}\label{eq:projection-proof-observed-bias}
SB_{t_0}=\theta_{t_0}^{OLS}
\qquad\text{for every }t_0\in\mathcal I_0.
\end{equation}
Substituting \eqref{eq:projection-proof-observed-bias} into \eqref{eq:projection-proof-beta} identifies
\[
\beta_K
=G_K^{-1}
\mathbb E_{\mu_{I_0}}[p_K(I_0)\theta_{I_0}^{OLS}].
\]
Assumption \ref{ass:projection-extrapolation} and \eqref{eq:att} then give
\begin{align*}
ATT
&=\theta_1^{OLS}-SB_1\\
&=\theta_1^{OLS}-p_K(1)'\beta_K\\
&=\theta_1^{OLS}
-p_K(1)'G_K^{-1}
\mathbb E_{\mu_{I_0}}[p_K(I_0)\theta_{I_0}^{OLS}],
\end{align*}
which is \eqref{eq:projection-rdid}.
\end{proof}
}

With finitely many pre-treatment periods, choosing $\mu_{I_0}$ to be uniform and $p_K(t_0)=(1,t_0,\ldots,t_0^{K-1})'$ makes \eqref{eq:projection-beta} the ordinary least-squares projection of the period-specific baseline biases on a polynomial in time; \eqref{eq:projection-bias} evaluates that polynomial at period $1$. This extrapolation can lie outside the convex hull of the observed baseline biases. Thus, unlike Proposition \ref{prop1}, the projection-based estimand obtains point identification by imposing the functional-form extrapolation in Assumption \ref{ass:projection-extrapolation}; it is not a bound that is robust to misspecification of that extrapolation.
}
}

\clearpage
\section{Supporting empirical diagnostics and sensitivity analyses}\label{app:empirical-diagnostics}

\begin{figure}[p]
\centering
\captionsetup{font={small,stretch=1},skip=3pt}
\captionsetup[subfigure]{labelfont=normalfont,font={small,stretch=1}}
\begin{subfigure}{.76\textwidth}
\centering
\includegraphics[width=\linewidth,height=.23\textheight,keepaspectratio,trim=0 66bp 0 6bp,clip]{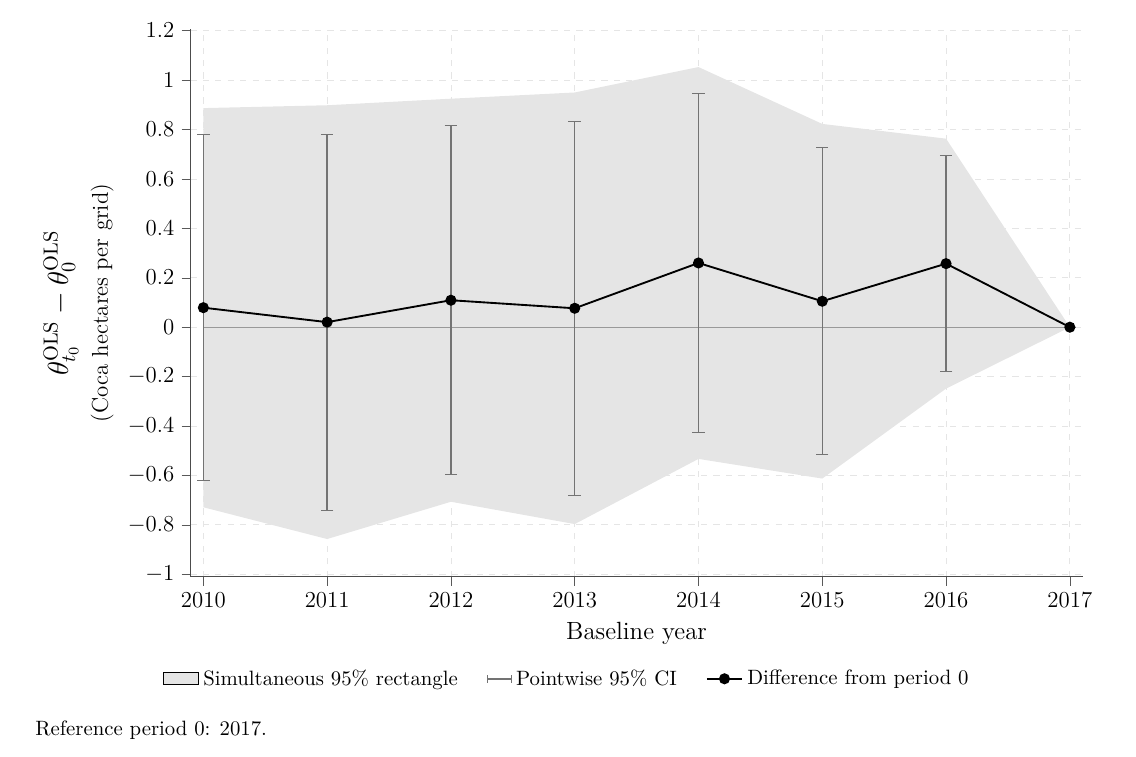}
\caption{Colombia's coca substitution program (baseline year)}
\end{subfigure}
\par
\begin{subfigure}{.76\textwidth}
\centering
\includegraphics[width=\linewidth,height=.23\textheight,keepaspectratio,trim=0 66bp 0 6bp,clip]{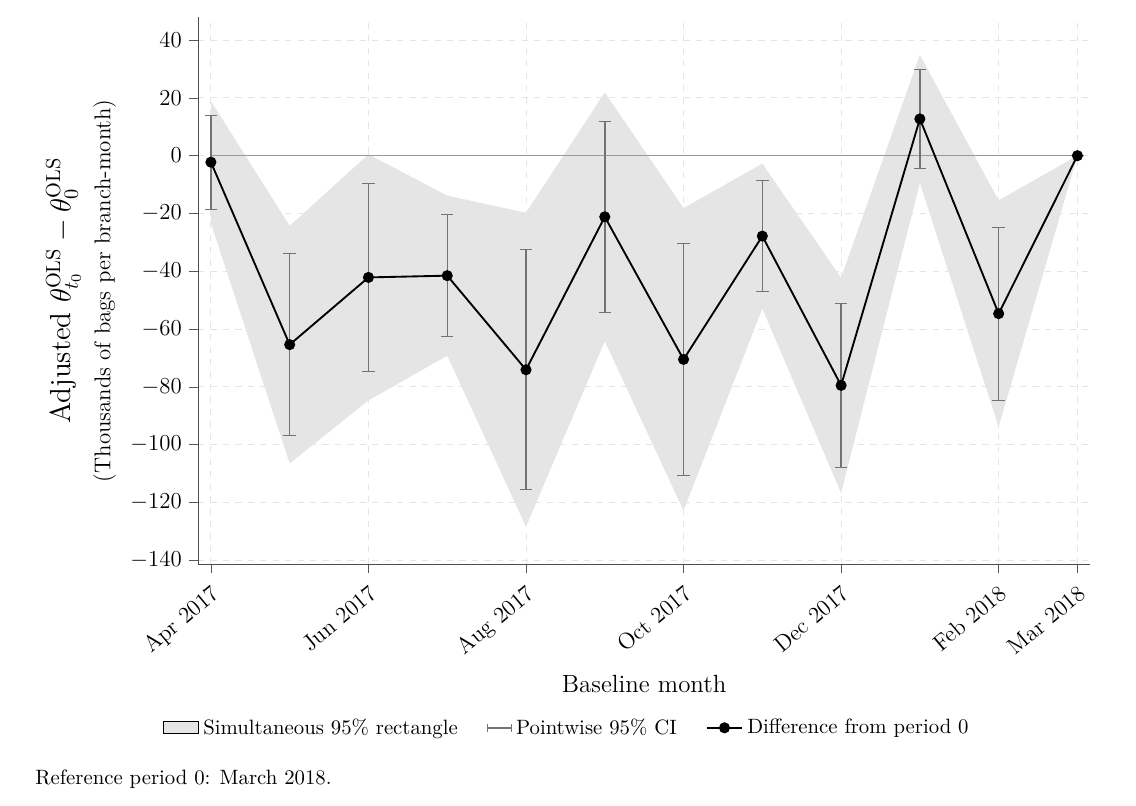}
\caption{Plastic-bag charges in Uruguay (baseline month)}
\end{subfigure}
\par
\begin{subfigure}{.76\textwidth}
\centering
\includegraphics[width=\linewidth,height=.23\textheight,keepaspectratio,trim=0 66bp 0 6bp,clip]{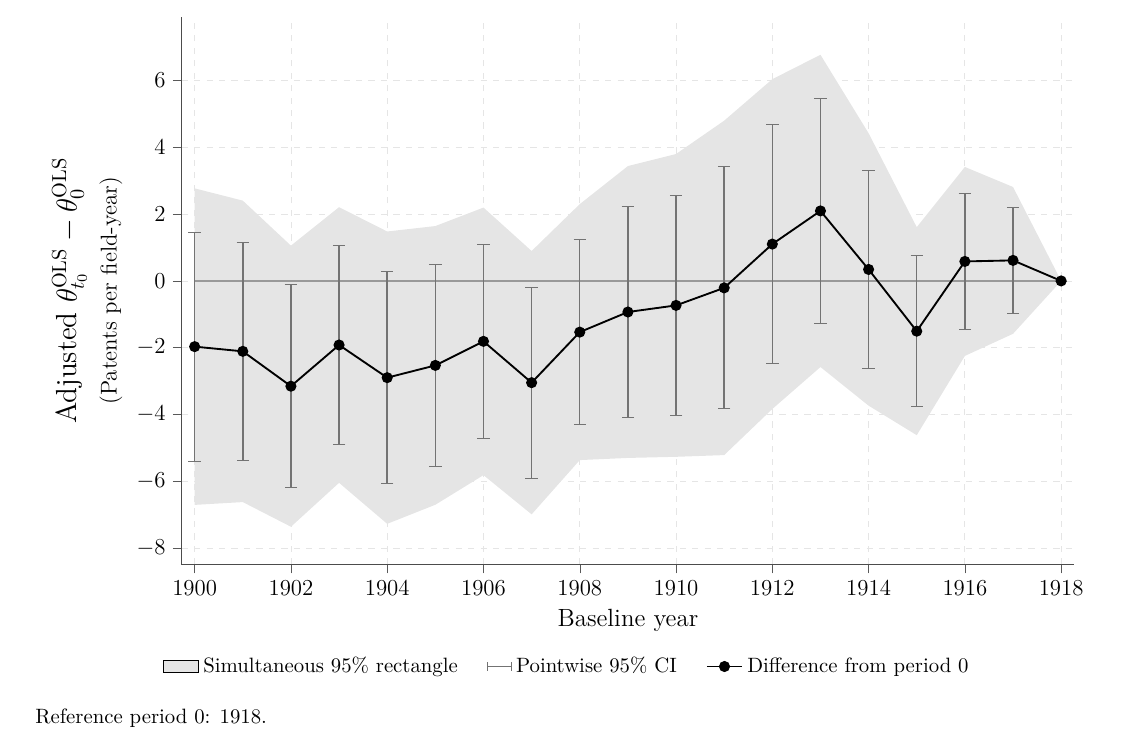}
\caption{Compulsory licensing in Germany (baseline year)}
\end{subfigure}
\caption{Pre-treatment placebo comparisons relative to the final baseline period. Each point reports $\theta_{t_0}^{OLS}-\theta_0^{OLS}$; the reference periods are 2017 (a), March 2018 (b), and 1918 (c). The reference-period zero is a normalization. Vertical bars show pointwise 95\% confidence intervals; shading shows simultaneous 95\% bounds. These panels report pre-treatment coefficients only. Units: coca hectares per grid (a), thousands of bags per branch-month (b), and patents per field-year (c).}
\label{fig:application-eventstudy}
\end{figure}

This appendix supplements the three applications in Section~\ref{sec:empirical}. For each application, we report confidence envelopes for the entire identified set, full pre-treatment equality diagnostics, observed group means, and sensitivity to the anticipation bound under $A_{\text{AE-R}}$. The appendix tables use the full baseline information sets. Additional $A_{\text{PT-R}}$ results using selected baseline periods appear in the main application tables. The baseline-specific DID comparisons are shown in Figure~\ref{fig:application-baselines}; Figure~\ref{fig:application-eventstudy} below reports pre-treatment placebo comparisons relative to a reference period.

The ATT confidence sets in the main tables concern a fixed true effect; whole-set coverage concerns simultaneous containment of the entire population identified set.  The $A_{\text{PT-R}}$ and $A_{\text{R}}$ intervals coincide for these two targets by construction. {$A_{\text{AE-R}}$ uses exhaustive scalar inversion for ATT and CLR/CKLR adaptive-selection, Bonferroni endpoint bounds for set coverage; MR uses the scalar union for ATT and projection of the joint primitive rectangle for set coverage.} These different procedures need not be nested in a realized sample. The tables use $c=0.5$, and $\varnothing$ denotes an empty set.

The sensitivity figures vary $c$ under $A_{\text{AE-R}}$ while maintaining strong parallel trends. Their ATT summaries are hulls when scalar inversion yields multiple components; the pointwise confidence results and simultaneous bands have distinct coverage targets. The displays assess this particular assumption bundle, with time indexed in years for Colombia and Germany and in months for Uruguay.

\FloatBarrier

\subsection{Colombia's coca substitution program}\label{app:ladino-diagnostics}

\begin{figure}[H]
\centering
\includegraphics[width=.88\textwidth]{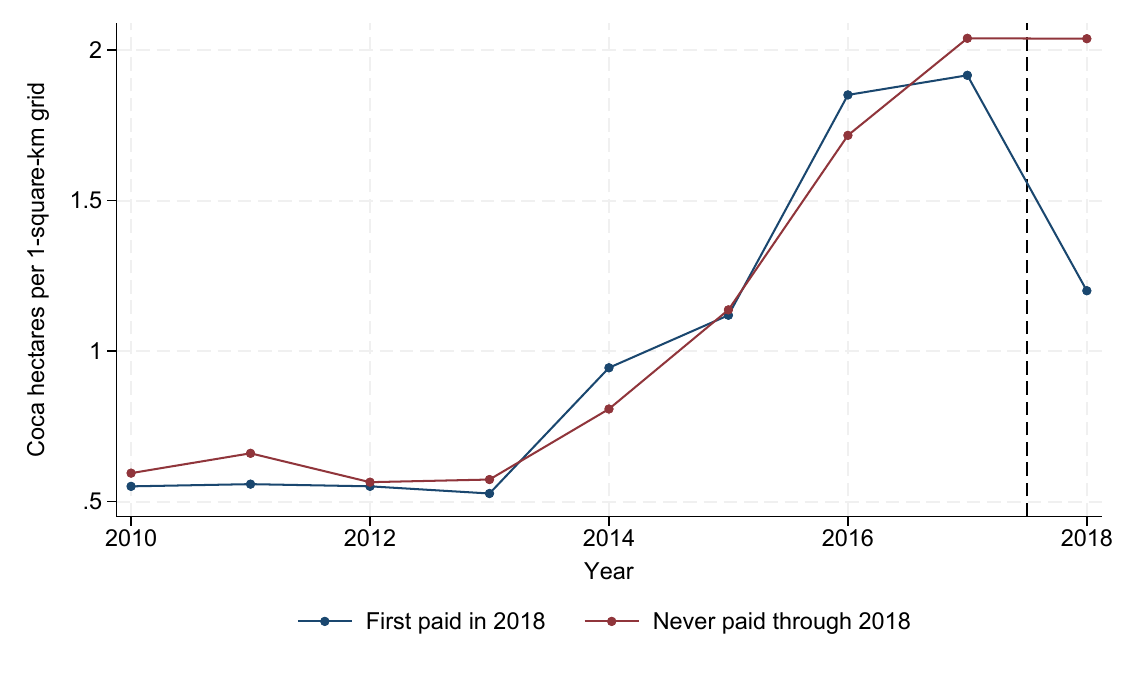}
\caption{Colombia selected sample: observed coca hectares per grid in treated and comparison grids.}
\label{fig:ladino-diagnostics}
\end{figure}

\begin{figure}[H]\centering
\includegraphics[width=.98\textwidth,height=.73\textheight,keepaspectratio]{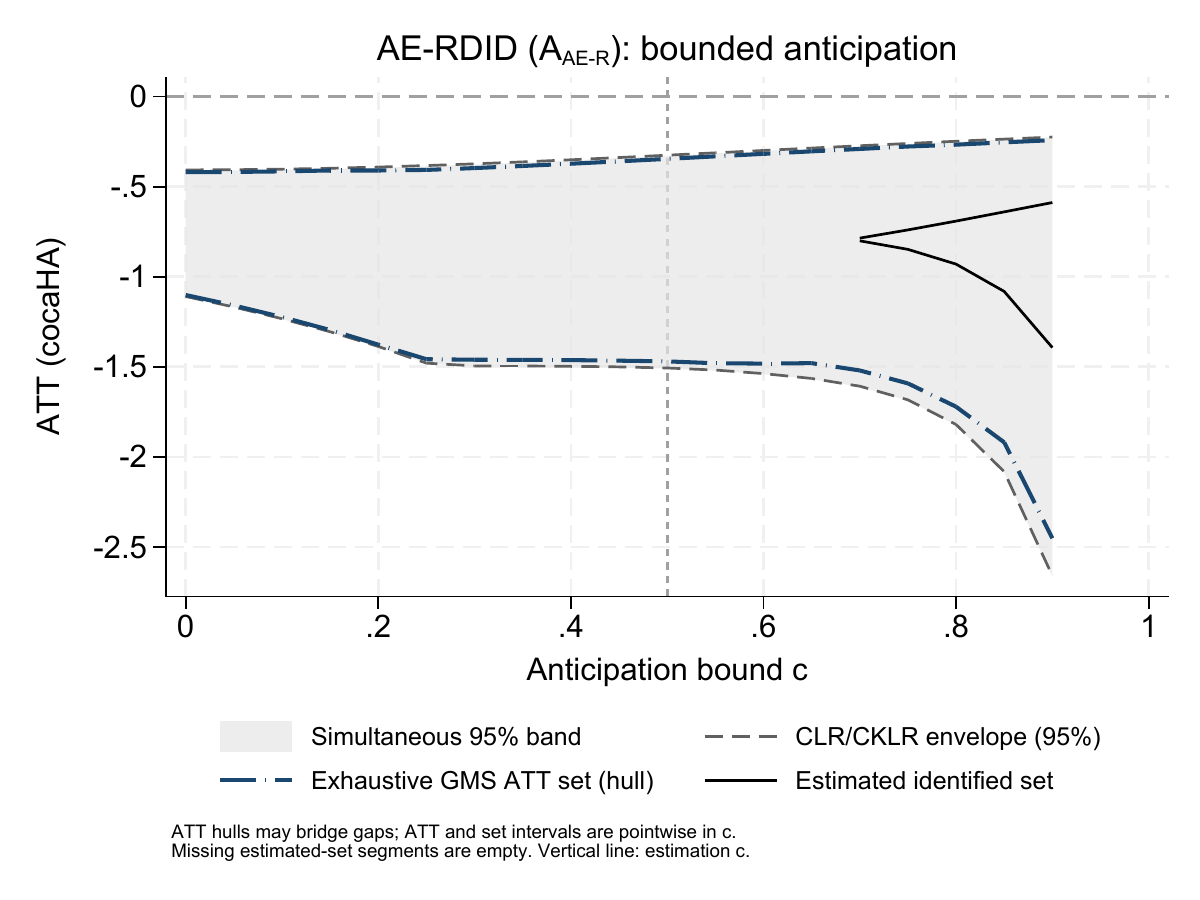}
\caption{Ladino: $A_{\text{AE-R}}$ sensitivity to the anticipation bound. The displayed ATT summary is a hull if multiple scalar components exist. }
\label{fig:ladino-sensitivity-aii}
\end{figure}

Tables~\ref{tab:ladino-set-coverage} and~\ref{tab:ladino-pretests} report coverage envelopes and pre-treatment diagnostics; Figures~\ref{fig:ladino-diagnostics} and~\ref{fig:ladino-sensitivity-aii} show observed group means and $A_{\text{AE-R}}$ sensitivity.

\begin{table}[H]
\centering\footnotesize
\caption{Ladino: confidence intervals for ATT and for the entire identified set.}
\label{tab:ladino-set-coverage}
\begin{tabular}{lccc}
\toprule
Model & \shortstack{Estimated\\identified set} & \shortstack{95\% ATT\\confidence set} & \shortstack{95\% identified-set\\confidence envelope} \\
\midrule
\(A_{\text{PT-R}}\), baseline 2010--2017 & \([-0.974, -0.714]\) & \([-1.587, -0.121]\) & \([-1.587, -0.121]\) \\
\(A_{\text{AE-R}}\) & \(\varnothing\) & \([-1.470, -0.346]\) & \([-1.506, -0.326]\) \\
\(A_{\text{R}}\) & \([-1.429, -0.476]\) & \([-2.067, -0.120]\) & \([-2.067, -0.120]\) \\
MR & \([-0.974, -0.714]\) & \([-1.587, -0.121]\) & \([-1.736, 0.028]\) \\
\bottomrule
\end{tabular}

\end{table}

\begin{table}[H]
\centering\small
\caption{Ladino joint pre-treatment diagnostics on the selected sample.}
\label{tab:ladino-pretests}
\begin{tabular}{lrrr}
\toprule
Preperiod equality diagnostic & Statistic & Degrees of freedom & \(p\)-value \\
\midrule
Original-style regression & \(F=\)  2.508 & (  7,  273) & 0.0163 \\
RDID primitive DIDs & \(\chi^2=\) 17.557 &   7 & 0.0141 \\
\bottomrule
\end{tabular}

\end{table}

\FloatBarrier

\subsection{Pricing plastic bags in Uruguay}\label{app:ccc-diagnostics}

\begin{figure}[H]
\centering
\includegraphics[width=.88\textwidth]{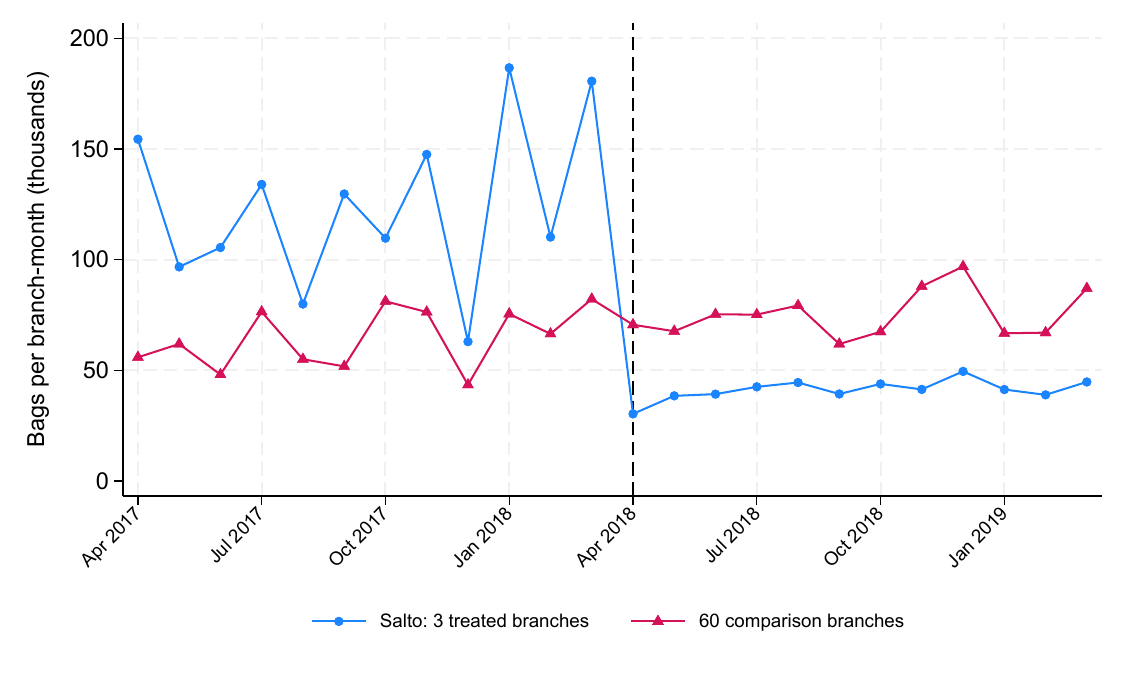}
\caption{Uruguay: observed disposable-bag use in treated and comparison branches. The group means use available records; their plotting aggregation does not change the estimation data.}
\label{fig:ccc0923design}
\end{figure}

\begin{figure}[H]\centering
\includegraphics[width=.98\textwidth,height=.73\textheight,keepaspectratio]{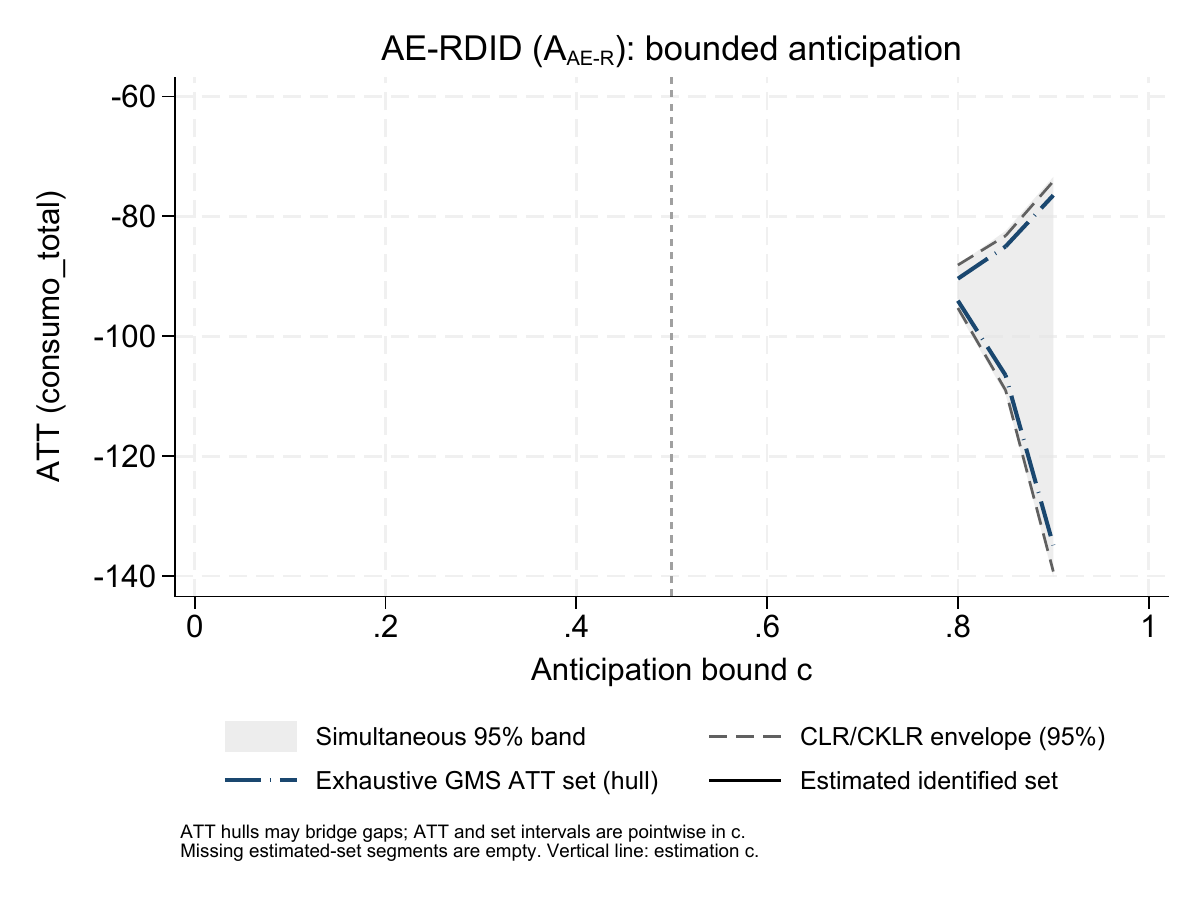}
\caption{CCC: $A_{\text{AE-R}}$ sensitivity to the anticipation bound. The displayed ATT summary is a hull if multiple scalar components exist. }
\label{fig:ccc-sensitivity-aii}
\end{figure}

Tables~\ref{tab:ccc-set-coverage} and~\ref{tab:ccc0923pre} report coverage envelopes and pre-treatment diagnostics; Figures~\ref{fig:ccc0923design} and~\ref{fig:ccc-sensitivity-aii} show observed group means and $A_{\text{AE-R}}$ sensitivity.

\begin{table}[H]
\centering\footnotesize
\caption{CCC: confidence intervals for ATT and for the entire identified set.}
\label{tab:ccc-set-coverage}
\begin{tabular}{lccc}
\toprule
Model & \shortstack{Estimated\\identified set} & \shortstack{95\% ATT\\confidence set} & \shortstack{95\% identified-set\\confidence envelope} \\
\midrule
\(A_{\text{PT-R}}\), baseline Apr 2017--Mar 2018 & \([-145.115, -52.870]\) & \([-170.923, -31.320]\) & \([-170.923, -31.320]\) \\
\(A_{\text{AE-R}}\) & \(\varnothing\) & \(\varnothing\) & \(\varnothing\) \\
\(A_{\text{R}}\) & \([-264.762, -49.760]\) & \([-330.498, -30.838]\) & \([-330.498, -30.838]\) \\
MR & \([-145.115, -52.870]\) & \([-170.923, -31.320]\) & \([-179.891, -20.704]\) \\
\bottomrule
\end{tabular}

\end{table}

\begin{table}[H]
\centering\small
\caption{CCC pre-treatment diagnostics on the common sample.}
\label{tab:ccc0923pre}
\begin{tabular}{lrrr}
\toprule
Preperiod equality diagnostic & Statistic & Degrees of freedom & \(p\)-value \\
\midrule
Original-style regression & \(F=\) 25.833 & ( 11,   62) & \(\leq 0.0001\) \\
RDID primitive DIDs & \(\chi^2=\)286.532 &  11 & \(\leq 0.0001\) \\
\bottomrule
\end{tabular}

\end{table}

\FloatBarrier

\subsection{Compulsory licensing and German invention}\label{app:baten-diagnostics}

\begin{figure}[H]
\centering
\includegraphics[width=0.88\textwidth]{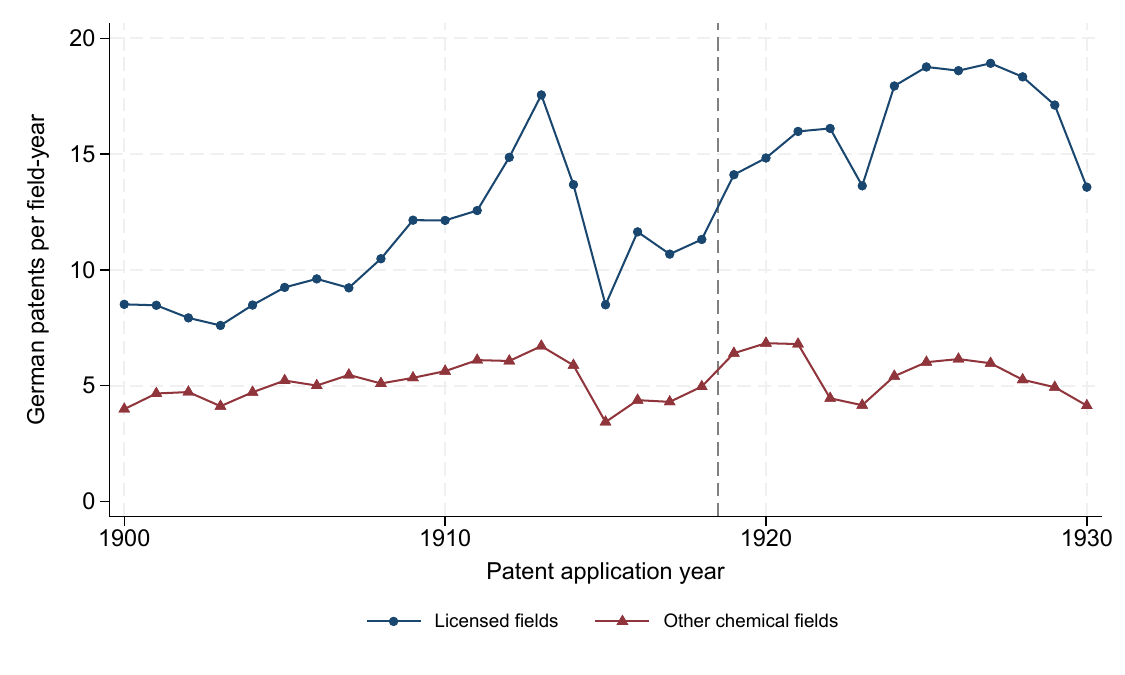}
\caption{Baten: observed, unadjusted patent counts in the same treated and comparison fields. The dashed line marks the start of the postperiod.}
\label{fig:baten0923-means}
\end{figure}

\begin{figure}[H]\centering
\includegraphics[width=.98\textwidth,height=.73\textheight,keepaspectratio]{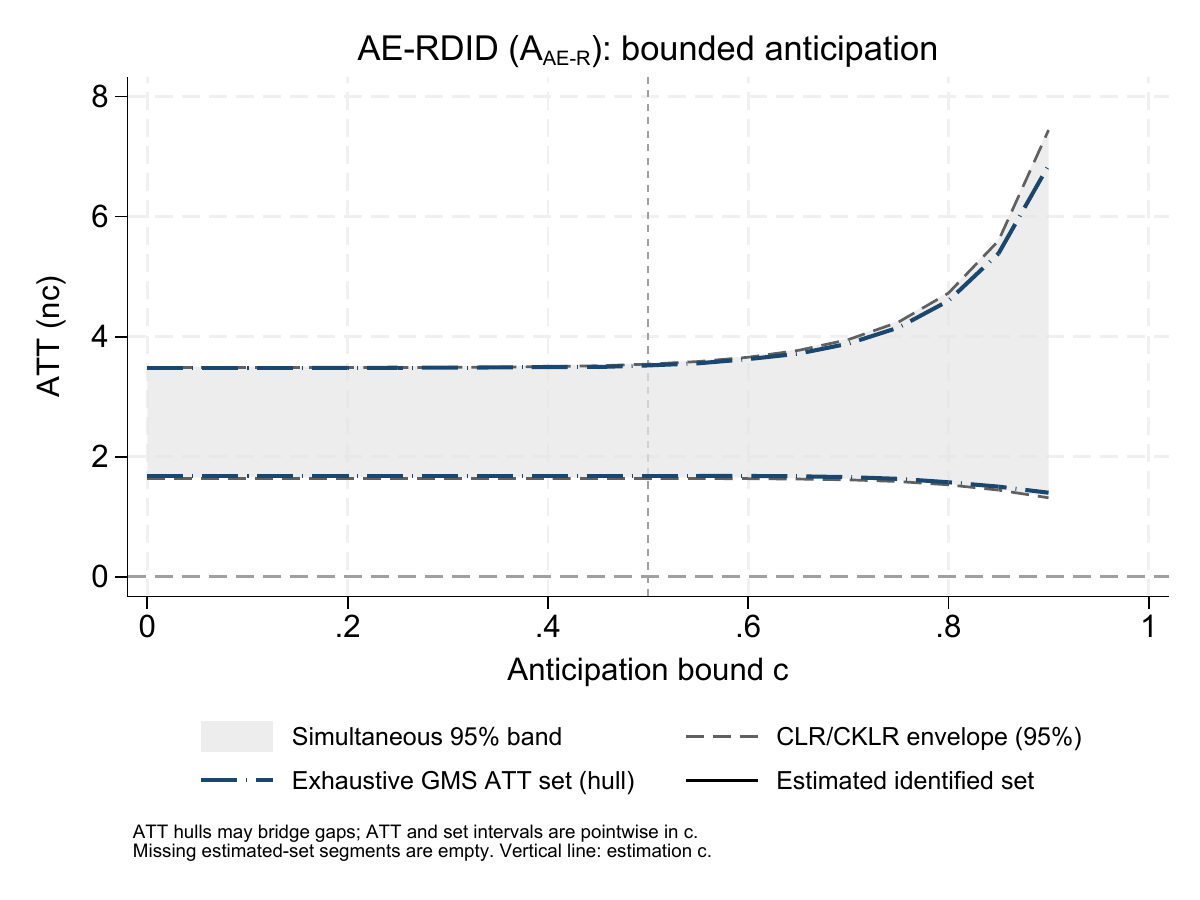}
\caption{Baten: $A_{\text{AE-R}}$ sensitivity to the anticipation bound. The displayed ATT summary is a hull if multiple scalar components exist. }
\label{fig:baten-sensitivity-aii}
\end{figure}

Tables~\ref{tab:baten-set-coverage} and~\ref{tab:baten0923-pretests} report coverage envelopes and pre-treatment diagnostics; Figures~\ref{fig:baten0923-means} and~\ref{fig:baten-sensitivity-aii} show observed group means and $A_{\text{AE-R}}$ sensitivity.

\begin{table}[H]
\centering\footnotesize
\caption{Baten: confidence intervals for ATT and for the entire identified set.}
\label{tab:baten-set-coverage}
\begin{tabular}{lccc}
\toprule
Model & \shortstack{Estimated\\identified set} & \shortstack{95\% ATT\\confidence set} & \shortstack{95\% identified-set\\confidence envelope} \\
\midrule
\(A_{\text{PT-R}}\), baseline 1900--1918 & \([0.073, 5.325]\) & \([-2.302, 7.934]\) & \([-2.302, 7.934]\) \\
\(A_{\text{AE-R}}\) & \(\varnothing\) & \([1.683, 3.521]\) & \([1.637, 3.542]\) \\
\(A_{\text{R}}\) & \([0.072, 5.325]\) & \([-2.339, 9.234]\) & \([-2.339, 9.234]\) \\
MR & \([0.073, 5.325]\) & \([-2.302, 7.934]\) & \([-3.356, 9.093]\) \\
\bottomrule
\end{tabular}

\end{table}

\begin{table}[H]
\centering\small
\caption{Baten: preperiod diagnostics for the selected design}
\label{tab:baten0923-pretests}
\begin{tabular}{lrrr}
\toprule
Preperiod equality diagnostic & Statistic & Degrees of freedom & \(p\)-value \\
\midrule
Original-style regression & \(F=\)  1.775 & ( 18,  211) & 0.0300 \\
RDID primitive DIDs & \(\chi^2=\) 30.946 &  18 & 0.0292 \\
\bottomrule
\end{tabular}

\end{table}

\FloatBarrier

\end{document}